\documentclass[12pt,a4paper]{article}

\DeclareMathSizes{12}{12}{7}{5}
\usepackage{amsfonts}
\usepackage[dvips]{graphicx}
\usepackage{latexsym,amsmath,amssymb,graphics,stmaryrd}
\usepackage{cite}
\usepackage{array}
\usepackage{subfigure}
\usepackage{multirow}
\usepackage{epsfig}
\usepackage[dvips]{graphicx}
\usepackage[dvips]{color}
\usepackage{pstricks}
\usepackage{pst-node}
\usepackage{pst-plot}
\usepackage{dsfont}
\usepackage{amsthm}
\usepackage{graphics}
\usepackage{subfigure}
\usepackage{fancyhdr}

\usepackage{rotating}
\newlength{\arlength}

\usepackage{feynmp}

\usepackage{Dalgebra}

\usepackage{feynmp}

\fancypagestyle{plain}{
\fancyhf{}
\newcommand{\fpage}{\iffloatpage{}{\thepage}}
\fancyfoot[C]{\fpage}

}

\newcommand{\col}{~,}
\newcommand{\pnt}{~.}
\newcommand{\AdS}{\text{AdS}}
\newcommand{\CFT}{\text{CFT}}

\newcommand{\twob}{{\text{II}\,\text{B}}}

\newcommand{\YM}{\text{YM}}

\newcommand{\unitmatrix}{\mathds{1}}

\newcommand{\comm}[2]{[#1\smash[b]{\,\mathbin{,}\,}#2]}
\newcommand{\acomm}[2]{\{#1\smash[b]{\,\mathbin{,}\,}#2\}}

\newcommand{\de}{\operatorname{d}\!}

\newcommand{\e}{\operatorname{e}}

\newcommand{\fun}{\Box\protect\rule[0pt]{0pt}{2.5mm}}
\newcommand{\afun}{\bar{\fun}}

\newlength{\neglength}
\newlength{\diameter}

\newcommand{\svertex}[3][0.5]{%
\fmfiequ{#2}{point #1*length(#3) of #3}
}
\newcommand{\dvertex}[3]{%
\fmfiequ{#1}{point length(#3)/3 of #3}
\fmfiequ{#2}{point 2length(#3)/3 of #3}
}
\newcommand{\vvertex}[3]{%
\fmfipath{px}
\fmfiequ{px}{(0,ypart(#2))..(100,ypart(#2))}
\fmfiequ{#1}{point xpart(#3 intersectiontimes px) of #3}
}

\newcommand{\wline}[6][0]{%
\fmfipath{pi[]}
\fmfiset{pi1}{#3 ..controls (xpart(#3)-(#1),ypart(#3)) and (xpart(vloc(__#4))-(#1)-0.25w,-0.15w) .. (xpart(vloc(__#4)),-0.15w)}
\fmfiset{pi2}{(xpart(vloc(__#4)),-0.15w) ..(xpart(vloc(__#5)),-0.15w)}
\fmfiset{pi3}{(xpart(vloc(__#5)),-0.15w) ..controls (xpart(vloc(__#5))+(#1)+0.25w,-0.15w) and (xpart(#6)+(#1)+0,ypart(#6)) .. #6}
\fmfi{#2}{pi1}
\fmfi{#2}{pi2}
\fmfi{#2}{pi3}
}

\newcommand{\wigglywrap}[4]{%
\fmfipath{pi[]}
\fmfiset{pi1}{#1 ..controls (xpart(#1)-0.25w,ypart(#1)) and (xpart(vloc(__#2))-0.25w,-0.15w) .. (xpart(vloc(__#2)),-0.15w)}
\fmfiset{pi2}{(xpart(vloc(__#2)),-0.15w) ..(xpart(vloc(__#3)),-0.15w)}
\fmfiset{pi3}{(xpart(vloc(__#3)),-0.15w) ..controls (xpart(vloc(__#3))+0.25w,-0.15w) and (xpart(#4)+0.25w,ypart(#4)) .. #4}
\fmfi{photon}{pi1}
\fmfi{photon}{pi2}
\fmfi{photon}{pi3}
}

\newcommand{\idonei}[6][plain]{%
\fmftop{v1}
\fmfbottom{v3}
\fmfforce{(0.125w,h)}{v1}
\fmfforce{(0.125w,0)}{v3}
\fmffixed{(0.25w,0)}{v1,v2}
\fmffixed{(0.25w,0)}{v3,v4}
\fmf{phantom}{v1,v3}
\fmf{#5}{v1,vc1}
\fmf{#2}{v3,vc1}
\fmf{phantom}{v2,v4}
\fmf{#6}{v2,vc2}
\fmf{#3}{v4,vc2}
\fmf{#1,tension=0.5,right=0,width=1mm}{v3,v4}
\fmffreeze
\fmfposition
\fmf{#4}{vc1,vc2}
\fmfipath{p[]}
\fmfiset{p1}{vpath(__v1,__v3)}
\fmfiset{p2}{vpath(__v2,__v4)}
}

\newcommand{\chionei}[6][plain]{%
\fmftop{v1}
\fmfbottom{v3}
\fmfforce{(0.125w,h)}{v1}
\fmfforce{(0.125w,0)}{v3}
\fmffixed{(0.25w,0)}{v1,v2}
\fmffixed{(0.25w,0)}{v3,v4}
\fmf{#5,tension=0.5,right=0.25}{v1,vc1}
\fmf{#6,tension=0.5,left=0.25}{v2,vc1}
\fmf{#4,tension=1.25}{vc1,vc2}
\fmf{#2,tension=0.5,left=0.25}{v3,vc2}
\fmf{#3,tension=0.5,right=0.25}{v4,vc2}
\fmf{#1,tension=0.5,right=0,width=1mm}{v3,v4}
\fmfposition
\fmfipath{p[]}
\fmfipair{vd[],vm[],vu[]}
\fmfiset{p1}{vpath(__v1,__vc1)}
\fmfiset{p2}{vpath(__v2,__vc1)}
\fmfiset{p3}{vpath(__vc1,__vc2)}
\fmfiset{p4}{vpath(__v3,__vc2)}
\fmfiset{p5}{vpath(__v4,__vc2)}
\svertex{vm1}{p1}
\dvertex{vu1}{vd1}{p1}
\svertex{vm2}{p2}
\dvertex{vu2}{vd2}{p2}
\svertex{vm3}{p3}
\dvertex{vu3}{vd3}{p3}
\svertex{vm4}{p4}
\dvertex{vd4}{vu4}{p4}
\svertex{vm5}{p5}
\dvertex{vd5}{vu5}{p5}
}

\newcommand{\chionegi}[6]{%
\fmftop{v1}
\fmfbottom{v4}
\fmfforce{(0.125w,h)}{v1}
\fmfforce{(0.125w,0)}{v4}
\fmffixed{(0.25w,0)}{v1,v2}
\fmffixed{(0.25w,0)}{v2,v3}
\fmffixed{(0.25w,0)}{v4,v5}
\fmffixed{(0.25w,0)}{v5,v6}
\fmf{#4,tension=0.5,right=0.25}{v1,vc1}
\fmf{#5,tension=0.5,left=0.25}{v2,vc1}
  \fmf{#3,tension=1.25}{vc1,vc2}
\fmf{#1,tension=0.5,left=0.25}{v4,vc2}
\fmf{#2,tension=0.5,right=0.25}{v5,vc2}
\fmf{#6}{v3,v6}
\fmf{plain,tension=0.5,right=0,width=1mm}{v4,v6}
\fmfposition
\fmfipath{p[],pg}
\fmfipair{vd[],vm[],vu[],vgd[],vgm[],vgu[],vg[]}
\fmfiset{p1}{vpath(__v1,__vc1)}
\fmfiset{p2}{vpath(__v2,__vc1)}
\fmfiset{p3}{vpath(__vc1,__vc2)}
\fmfiset{p5}{vpath(__v5,__vc2)}
\fmfiset{p4}{vpath(__v4,__vc2)}
\fmfiset{pg}{vpath(__v3,__v6)}
\svertex{vm1}{p1}
\dvertex{vu1}{vd1}{p1}
\svertex{vm2}{p2}
\dvertex{vu2}{vd2}{p2}
\svertex{vm3}{p3}
\dvertex{vu3}{vd3}{p3}
\svertex{vm4}{p4}
\dvertex{vd4}{vu4}{p4}
\svertex{vm5}{p5}
\dvertex{vd5}{vu5}{p5}
\vvertex{vgu2}{vu2}{pg}
\vvertex{vgm2}{vm2}{pg}
\vvertex{vgd2}{vd2}{pg}
\vvertex{vgu3}{vu3}{pg}
\vvertex{vgm3}{vm3}{pg}
\vvertex{vgd3}{vd3}{pg}
\vvertex{vgu5}{vu5}{pg}
\vvertex{vgm5}{vm5}{pg}
\vvertex{vgd5}{vd5}{pg}
\vvertex{vg1}{vloc(__vc1)}{pg}
\vvertex{vg2}{vloc(__vc2)}{pg}
}

\newcommand{\gchionei}[6]{%
\fmftop{v1}
\fmfbottom{v4}
\fmfforce{(0.125w,h)}{v1}
\fmfforce{(0.125w,0)}{v4}
\fmffixed{(0.25w,0)}{v1,v2}
\fmffixed{(0.25w,0)}{v2,v3}
\fmffixed{(0.25w,0)}{v4,v5}
\fmffixed{(0.25w,0)}{v5,v6}
\fmf{#5,tension=0.5,right=0.25}{v2,vc1}
\fmf{#6,tension=0.5,left=0.25}{v3,vc1}
  \fmf{#4,tension=1.25}{vc1,vc2}
\fmf{#2,tension=0.5,left=0.25}{v5,vc2}
\fmf{#3,tension=0.5,right=0.25}{v6,vc2}
\fmf{#1}{v1,v4}
\fmf{plain,tension=0.5,right=0,width=1mm}{v4,v6}
\fmfposition
\fmfipath{p[],pg}
\fmfipair{vd[],vm[],vu[],vgd[],vgm[],vgu[],vg[]}
\fmfiset{p1}{vpath(__v2,__vc1)}
\fmfiset{p2}{vpath(__v3,__vc1)}
\fmfiset{p3}{vpath(__vc1,__vc2)}
\fmfiset{p5}{vpath(__v6,__vc2)}
\fmfiset{p4}{vpath(__v5,__vc2)}
\fmfiset{pg}{vpath(__v1,__v4)}
\svertex{vm1}{p1}
\dvertex{vu1}{vd1}{p1}
\svertex{vm2}{p2}
\dvertex{vu2}{vd2}{p2}
\svertex{vm3}{p3}
\dvertex{vu3}{vd3}{p3}
\svertex{vm4}{p4}
\dvertex{vd4}{vu4}{p4}
\svertex{vm5}{p5}
\dvertex{vd5}{vu5}{p5}
\vvertex{vgu2}{vu2}{pg}
\vvertex{vgm2}{vm2}{pg}
\vvertex{vgd2}{vd2}{pg}
\vvertex{vgu3}{vu3}{pg}
\vvertex{vgm3}{vm3}{pg}
\vvertex{vgd3}{vd3}{pg}
\vvertex{vgu5}{vu5}{pg}
\vvertex{vgm5}{vm5}{pg}
\vvertex{vgd5}{vd5}{pg}
\vvertex{vg1}{vloc(__vc1)}{pg}
\vvertex{vg2}{vloc(__vc2)}{pg}
}

\newcommand{\chionetwoi}[9]{%
\fmftop{v1}
\fmfbottom{v4}
\fmfforce{(0.125w,h)}{v1}
\fmfforce{(0.125w,0)}{v4}
\fmffixed{(0.25w,0)}{v1,v2}
\fmffixed{(0.25w,0)}{v2,v3}
\fmffixed{(0.25w,0)}{v4,v5}
\fmffixed{(0.25w,0)}{v5,v6}
\fmffixed{(0,whatever)}{vc1,vc3}
\fmffixed{(0,whatever)}{vc2,vc4}
\fmf{#7,tension=0.5,right=0.25}{v1,vc1}
\fmf{#8,tension=0.5,left=0.25}{v2,vc1}
\fmf{phantom,tension=0.5,right=0.25}{v2,vc2}
\fmf{#9,tension=0.5,left=0.25}{v3,vc2}
\fmf{#1,tension=0.5,left=0.25}{v4,vc3}
\fmf{phantom,tension=0.5,right=0.25}{v5,vc3}
\fmf{#2,tension=0.5,left=0.25}{v5,vc4}
\fmf{#3,tension=0.5,right=0.25}{v6,vc4}
\fmf{#4,tension=1.25,left=0}{vc1,vc3}
\fmf{#6,tension=1.25,left=0}{vc2,vc4}
\fmffreeze
\fmf{#5,tension=1,left=0}{vc2,vc3}
\fmf{plain,tension=0.5,right=0,width=1mm}{v4,v6}
\fmffreeze
\fmfposition
\fmfipath{p[]}
\fmfipair{vd[],vm[],vu[]}
\fmfiset{p1}{vpath(__v1,__vc1)}
\fmfiset{p2}{vpath(__v2,__vc1)}
\fmfiset{p6}{vpath(__v3,__vc2)}
\fmfiset{p4}{vpath(__v4,__vc3)}
\fmfiset{p8}{vpath(__v5,__vc4)}
\fmfiset{p9}{vpath(__v6,__vc4)}
\fmfiset{p3}{vpath(__vc1,__vc3)}
\fmfiset{p7}{vpath(__vc2,__vc4)}
\fmfiset{p5}{vpath(__vc2,__vc3)}
\svertex{vm1}{p1}
\svertex{vm2}{p2}
\svertex{vm3}{p3}
\svertex{vm4}{p4}
\svertex{vm5}{p5}
\svertex{vm6}{p6}
\svertex{vm7}{p7}
\svertex{vm8}{p8}
\svertex{vm9}{p9}
}

\newcommand{\chitwoonei}[9]{%
\fmftop{v1}
\fmfbottom{v4}
\fmfforce{(0.125w,h)}{v1}
\fmfforce{(0.125w,0)}{v4}
\fmffixed{(0.25w,0)}{v1,v2}
\fmffixed{(0.25w,0)}{v2,v3}
\fmffixed{(0.25w,0)}{v4,v5}
\fmffixed{(0.25w,0)}{v5,v6}
\fmffixed{(0,whatever)}{vc1,vc3}
\fmffixed{(0,whatever)}{vc2,vc4}
\fmf{#9,tension=0.5,left=0.25}{v3,vc1}
\fmf{#8,tension=0.5,right=0.25}{v2,vc1}
\fmf{phantom,tension=0.5,left=0.25}{v2,vc2}
\fmf{#7,tension=0.5,right=0.25}{v1,vc2}
\fmf{#3,tension=0.5,right=0.25}{v6,vc3}
\fmf{phantom,tension=0.5,left=0.25}{v5,vc3}
\fmf{#2,tension=0.5,right=0.25}{v5,vc4}
\fmf{#1,tension=0.5,left=0.25}{v4,vc4}
\fmf{#6,tension=1.25,left=0}{vc1,vc3}
\fmf{#4,tension=1.25,left=0}{vc2,vc4}
\fmffreeze
\fmf{#5,tension=1,left=0}{vc2,vc3}
\fmf{plain,tension=0.5,right=0,width=1mm}{v4,v6}
\fmffreeze
\fmfposition
\fmfipath{p[]}
\fmfipair{vd[],vm[],vu[]}
\fmfiset{p1}{vpath(__v3,__vc1)}
\fmfiset{p2}{vpath(__v2,__vc1)}
\fmfiset{p6}{vpath(__v1,__vc2)}
\fmfiset{p4}{vpath(__v6,__vc3)}
\fmfiset{p8}{vpath(__v5,__vc4)}
\fmfiset{p9}{vpath(__v4,__vc4)}
\fmfiset{p3}{vpath(__vc1,__vc3)}
\fmfiset{p7}{vpath(__vc2,__vc4)}
\fmfiset{p5}{vpath(__vc2,__vc3)}
\svertex{vm1}{p1}
\svertex{vm2}{p2}
\svertex{vm3}{p3}
\svertex{vm4}{p4}
\svertex{vm5}{p5}
\svertex{vm6}{p6}
\svertex{vm7}{p7}
\svertex{vm8}{p8}
\svertex{vm9}{p9}
}

\newcommand{\chionetwothreei}[9][plain]{%
\fmftop{v1}
\fmfbottom{v5}
\fmfforce{(0.125w,h)}{v1}
\fmfforce{(0.125w,0)}{v5}
\fmffixed{(0.25w,0)}{v1,v2}
\fmffixed{(0.25w,0)}{v2,v3}
\fmffixed{(0.25w,0)}{v3,v4}
\fmffixed{(0.25w,0)}{v5,v6}
\fmffixed{(0.25w,0)}{v6,v7}
\fmffixed{(0.25w,0)}{v7,v8}
\fmffixed{(0,whatever)}{vc1,vc4}
\fmffixed{(0,whatever)}{vc2,vc5}
\fmffixed{(0,whatever)}{vc3,vc6}

\fmf{#6,tension=0.5,right=0.25}{v1,vc1}
\fmf{#7,tension=0.5,left=0.25}{v2,vc1}
\fmf{phantom,tension=0.5,right=0.25}{v2,vc2}
\fmf{#8,tension=0.5,left=0.25}{v3,vc2}
\fmf{phantom,tension=0.5,right=0.25}{v3,vc3}
\fmf{#9,tension=0.5,left=0.25}{v4,vc3}
\fmf{#2,tension=0.5,left=0.25}{v5,vc4}
\fmf{phantom,tension=0.5,right=0.25}{v6,vc4}
\fmf{#3,tension=0.5,left=0.25}{v6,vc5}
\fmf{phantom,tension=0.5,right=0.25}{v7,vc5}
\fmf{#4,tension=0.5,left=0.25}{v7,vc6}
\fmf{#5,tension=0.5,right=0.25}{v8,vc6}
\fmf{dots,tension=1.25,left=0}{vc1,vc4}
\fmf{dots,tension=1.25,left=0}{vc2,vc5}
\fmf{dots,tension=1.25,left=0}{vc3,vc6}
\fmffreeze
\fmf{dots,tension=1,left=0}{vc4,vc2}
\fmf{dots,tension=1,left=0}{vc5,vc3}
\fmf{#1,tension=0.5,right=0,width=1mm}{v5,v8}
\fmffreeze
\fmfposition
}

\newcommand{\chithreetwoonei}[9][plain]{%
\fmftop{v1}
\fmfbottom{v5}
\fmfforce{(0.125w,h)}{v1}
\fmfforce{(0.125w,0)}{v5}
\fmffixed{(0.25w,0)}{v1,v2}
\fmffixed{(0.25w,0)}{v2,v3}
\fmffixed{(0.25w,0)}{v3,v4}
\fmffixed{(0.25w,0)}{v5,v6}
\fmffixed{(0.25w,0)}{v6,v7}
\fmffixed{(0.25w,0)}{v7,v8}
\fmffixed{(0,whatever)}{vc1,vc4}
\fmffixed{(0,whatever)}{vc2,vc5}
\fmffixed{(0,whatever)}{vc3,vc6}

\fmf{#9,tension=0.5,left=0.25}{v4,vc3}
\fmf{#8,tension=0.5,right=0.25}{v3,vc3}
\fmf{phantom,tension=0.5,left=0.25}{v3,vc2}
\fmf{#7,tension=0.5,right=0.25}{v2,vc2}
\fmf{phantom,tension=0.5,left=0.25}{v2,vc1}
\fmf{#6,tension=0.5,right=0.25}{v1,vc1}
\fmf{#5,tension=0.5,right=0.25}{v8,vc6}
\fmf{phantom,tension=0.5,left=0.25}{v7,vc6}
\fmf{#4,tension=0.5,right=0.25}{v7,vc5}
\fmf{phantom,tension=0.5,left=0.25}{v6,vc5}
\fmf{#3,tension=0.5,right=0.25}{v6,vc4}
\fmf{#2,tension=0.5,left=0.25}{v5,vc4}
\fmf{dots,tension=1.25,left=0}{vc1,vc4}
\fmf{dots,tension=1.25,left=0}{vc2,vc5}
\fmf{dots,tension=1.25,left=0}{vc3,vc6}
\fmffreeze
\fmf{dots,tension=1,left=0}{vc1,vc5}
\fmf{dots,tension=1,left=0}{vc2,vc6}
\fmf{#1,tension=0.5,right=0,width=1mm}{v5,v8}
\fmffreeze
\fmfposition
}

\newcommand{\chionetwoonei}{%
\fmftop{v1}
\fmfbottom{v4}
\fmfforce{(0.125w,h)}{v1}
\fmfforce{(0.125w,0)}{v4}
\fmffixed{(0.25w,0)}{v1,v2}
\fmffixed{(0.25w,0)}{v2,v3}
\fmffixed{(0.25w,0)}{v4,v5}
\fmffixed{(0.25w,0)}{v5,v6}
\fmffixed{(0,whatever)}{vc1,vc3}
\fmffixed{(0,whatever)}{vb2,vb4}
\fmffixed{(0,whatever)}{vc1,vb1}
\fmffixed{(0,whatever)}{vc1,vb3}
\fmffixed{(whatever,0)}{vb1,vb2}
\fmffixed{(whatever,0)}{vb3,vb4}
\fmf{dots,tension=0.5,right=0.25}{v1,vc1}
\fmf{dots,tension=0.5,left=0.25}{v2,vc1}
\fmf{phantom,tension=0.5,right=0.25}{v2,vb2}
\fmf{dots,tension=0.5,left=0.25}{v3,vb2}
\fmf{dots,tension=0.5,left=0.25}{v4,vc3}
\fmf{dots,tension=0.5,right=0.25}{v5,vc3}
\fmf{phantom,tension=0.5,left=0.25}{v5,vb4}
\fmf{dots,tension=0.5,right=0.25}{v6,vb4}
\fmf{dots,tension=1.25,left=0}{vc1,vb1}
\fmf{dots,tension=1.25,left=0}{vb1,vb3}
\fmf{dots,tension=1.25,left=0}{vb3,vc3}
\fmf{dots,tension=1.25,left=0}{vb2,vb4}
\fmffreeze
\fmf{dots,tension=1,left=0}{vb1,vb2}
\fmf{dots,tension=1,left=0}{vb3,vb4}
\fmf{plain,tension=0.5,right=0,width=1mm}{v4,v6}
\fmffreeze
\fmfposition
}

\newcommand{\chitwoonetwoi}{%
\fmftop{v1}
\fmfbottom{v4}
\fmfforce{(0.125w,h)}{v1}
\fmfforce{(0.125w,0)}{v4}
\fmffixed{(0.25w,0)}{v1,v2}
\fmffixed{(0.25w,0)}{v2,v3}
\fmffixed{(0.25w,0)}{v4,v5}
\fmffixed{(0.25w,0)}{v5,v6}
\fmffixed{(0,whatever)}{vc1,vc3}
\fmffixed{(0,whatever)}{vb2,vb4}
\fmffixed{(0,whatever)}{vc1,vb1}
\fmffixed{(0,whatever)}{vc1,vb3}
\fmffixed{(whatever,0)}{vb1,vb2}
\fmffixed{(whatever,0)}{vb3,vb4}
\fmf{dots,tension=0.5,left=0.25}{v3,vc1}
\fmf{dots,tension=0.5,right=0.25}{v2,vc1}
\fmf{phantom,tension=0.5,left=0.25}{v2,vb2}
\fmf{dots,tension=0.5,right=0.25}{v1,vb2}
\fmf{dots,tension=0.5,right=0.25}{v6,vc3}
\fmf{dots,tension=0.5,left=0.25}{v5,vc3}
\fmf{phantom,tension=0.5,right=0.25}{v5,vb4}
\fmf{dots,tension=0.5,left=0.25}{v4,vb4}
\fmf{dots,tension=1.25,left=0}{vc1,vb1}
\fmf{dots,tension=1.25,left=0}{vb1,vb3}
\fmf{dots,tension=1.25,left=0}{vb3,vc3}
\fmf{dots,tension=1.25,left=0}{vb2,vb4}
\fmffreeze
\fmf{dots,tension=1,left=0}{vb1,vb2}
\fmf{dots,tension=1,left=0}{vb3,vb4}
\fmf{plain,tension=0.5,right=0,width=1mm}{v4,v6}
\fmffreeze
\fmfposition
}

\newcommand{\chionethreei}[8]{%
\fmftop{v1}
\fmfbottom{v5}
\fmfforce{(0.125w,h)}{v1}
\fmfforce{(0.125w,0)}{v5}
\fmffixed{(0.25w,0)}{v1,v2}
\fmffixed{(0.25w,0)}{v2,v3}
\fmffixed{(0.25w,0)}{v3,v4}
\fmffixed{(0.25w,0)}{v5,v6}
\fmffixed{(0.25w,0)}{v6,v7}
\fmffixed{(0.25w,0)}{v7,v8}
\fmf{#5,tension=0.5,right=0.25}{v1,vc1}
\fmf{#6,tension=0.5,left=0.25}{v2,vc1}
  \fmf{dots,tension=1.25}{vc1,vc2}
\fmf{#7,tension=0.5,left=0.25}{v5,vc2}
\fmf{#8,tension=0.5,right=0.25}{v6,vc2}
\fmf{#1,tension=0.5,right=0.25}{v3,vc3}
\fmf{#2,tension=0.5,left=0.25}{v4,vc3}
\fmf{dots,tension=1.25}{vc3,vc4}
\fmf{#3,tension=0.5,left=0.25}{v7,vc4}
\fmf{#4,tension=0.5,right=0.25}{v8,vc4}
\fmf{plain,tension=0.5,right=0,width=1mm}{v5,v8}
\fmfposition
\fmfipath{pl[],pr[]}
\fmfipair{vld[],vlm[],vlu[],vrd[],vrm[],vru[]}
\fmfiset{pl1}{vpath(__v1,__vc1)}
\fmfiset{pl2}{vpath(__v2,__vc1)}
\fmfiset{pl3}{vpath(__vc1,__vc2)}
\fmfiset{pl5}{vpath(__v6,__vc2)}
\fmfiset{pl4}{vpath(__v5,__vc2)}
\fmfiset{pr1}{vpath(__v3,__vc3)}
\fmfiset{pr2}{vpath(__v4,__vc3)}
\fmfiset{pr3}{vpath(__vc3,__vc4)}
\fmfiset{pr5}{vpath(__v8,__vc4)}
\fmfiset{pr4}{vpath(__v7,__vc4)}
\svertex{vlm1}{pl1}
\dvertex{vlu1}{vld1}{pl1}
\svertex{vlm2}{pl2}
\dvertex{vlu2}{vld2}{pl2}
\svertex{vlm3}{pl3}
\dvertex{vlu3}{vld3}{pl3}
\svertex{vlm4}{pl4}
\dvertex{vld4}{vlu4}{pl4}
\svertex{vlm5}{pl5}
\dvertex{vld5}{vlu5}{pl5}
\svertex{vrm1}{pr1}
\dvertex{vru1}{vrd1}{pr1}
\svertex{vrm2}{pr2}
\dvertex{vru2}{vrd2}{pr2}
\svertex{vrm3}{pr3}
\dvertex{vru3}{vrd3}{pr3}
\svertex{vrm4}{pr4}
\dvertex{vrd4}{vru4}{pr4}
\svertex{vrm5}{pr5}
\dvertex{vrd5}{vru5}{pr5}

}

\newcommand{\chitwoonethreei}[8]{%
\fmftop{v1}
\fmfbottom{v5}
\fmfforce{(0.125w,h)}{v1}
\fmfforce{(0.125w,0)}{v5}
\fmffixed{(0.25w,0)}{v1,v2}
\fmffixed{(0.25w,0)}{v2,v3}
\fmffixed{(0.25w,0)}{v3,v4}
\fmffixed{(0.25w,0)}{v5,v6}
\fmffixed{(0.25w,0)}{v6,v7}
\fmffixed{(0.25w,0)}{v7,v8}
\fmffixed{(whatever,0.5h)}{v5,vc1}
\fmffixed{(0,whatever)}{vc1,vc4}
\fmffixed{(0,whatever)}{vc2,vc5}
\fmffixed{(0,whatever)}{vc3,vc6}
\fmffixed{(whatever,0)}{vc1,vc3}
\fmffixed{(whatever,0)}{vc3,vc5}

\fmf{#5,tension=0.5,right=0.125}{v1,vc1}
\fmf{phantom,tension=0.5,left=0.25}{v2,vc1}
\fmf{#6,tension=0.5,right=0.25}{v2,vc2}
\fmf{#7,tension=0.5,left=0.25}{v3,vc2}
\fmf{phantom,tension=0.5,right=0.25}{v3,vc3}
\fmf{#8,tension=0.5,left=0.125}{v4,vc3}
\fmf{#1,tension=0.5,left=0.25}{v5,vc4}
\fmf{#2,tension=0.5,right=0.25}{v6,vc4}
\fmf{phantom,tension=0.5,left=0.25}{v6,vc5}
\fmf{phantom,tension=0.5,right=0.25}{v7,vc5}
\fmf{#3,tension=0.5,left=0.25}{v7,vc6}
\fmf{#4,tension=0.5,right=0.25}{v8,vc6}
\fmf{dots,tension=1.25,left=0}{vc1,vc4}
\fmf{dots,tension=1.25,left=0}{vc2,vc5}
\fmf{dots,tension=1.25,left=0}{vc3,vc6}
\fmffreeze
\fmf{dots,tension=1,left=0}{vc1,vc5}
\fmf{dots,tension=1,left=0}{vc5,vc3}
\fmf{plain,tension=0.5,right=0,width=1mm}{v5,v8}
\fmffreeze
\fmfposition
}

\newcommand{\chionethreetwoi}[8]{%
\fmftop{v1}
\fmfbottom{v5}
\fmfforce{(0.125w,h)}{v1}
\fmfforce{(0.125w,0)}{v5}
\fmffixed{(0.25w,0)}{v1,v2}
\fmffixed{(0.25w,0)}{v2,v3}
\fmffixed{(0.25w,0)}{v3,v4}
\fmffixed{(0.25w,0)}{v5,v6}
\fmffixed{(0.25w,0)}{v6,v7}
\fmffixed{(0.25w,0)}{v7,v8}
\fmffixed{(whatever,0.5h)}{v5,vc2}
\fmffixed{(0,whatever)}{vc1,vc4}
\fmffixed{(0,whatever)}{vc2,vc5}
\fmffixed{(0,whatever)}{vc3,vc6}
\fmffixed{(whatever,0)}{vc2,vc4}
\fmffixed{(whatever,0)}{vc4,vc6}
\fmf{#5,tension=0.5,right=0.25}{v1,vc1}
\fmf{#6,tension=0.5,left=0.25}{v2,vc1}
\fmf{phantom,tension=0.5,right=0.25}{v2,vc2}
\fmf{phantom,tension=0.5,left=0.25}{v3,vc2}
\fmf{#7,tension=0.5,right=0.25}{v3,vc3}
\fmf{#8,tension=0.5,left=0.25}{v4,vc3}
\fmf{#1,tension=0.5,left=0.125}{v5,vc4}
\fmf{phantom,tension=0.5,right=0.25}{v6,vc4}
\fmf{#2,tension=0.5,left=0.25}{v6,vc5}
\fmf{#3,tension=0.5,right=0.25}{v7,vc5}
\fmf{phantom,tension=0.5,left=0.25}{v7,vc6}
\fmf{#4,tension=0.5,right=0.125}{v8,vc6}
\fmf{dots,tension=1.25,left=0}{vc1,vc4}
\fmf{dots,tension=1.25,left=0}{vc2,vc5}
\fmf{dots,tension=1.25,left=0}{vc3,vc6}
\fmffreeze
\fmf{dots,tension=1,left=0}{vc2,vc4}
\fmf{dots,tension=1,left=0}{vc2,vc6}
\fmf{plain,tension=0.5,right=0,width=1mm}{v5,v8}
\fmffreeze
\fmfposition
}

\newcommand{\chionetwo}[1][black]{%
\fmftop{v1}
\fmfbottom{v4}
\fmfforce{(0.125w,h)}{v1}
\fmfforce{(0.125w,0)}{v4}
\fmffixed{(0.25w,0)}{v1,v2}
\fmffixed{(0.25w,0)}{v2,v3}
\fmffixed{(0.25w,0)}{v4,v5}
\fmffixed{(0.25w,0)}{v5,v6}
\fmffixed{(0,whatever)}{vc1,vc3}
\fmffixed{(0,whatever)}{vc2,vc4}
\fmf{plain,tension=0.5,right=0.25}{v1,vc1}
\fmf{plain,tension=0.5,left=0.25}{v2,vc1}
\fmf{phantom,tension=0.5,right=0.25}{v2,vc2}
\fmf{plain,tension=0.5,left=0.25}{v3,vc2}
\fmf{plain,tension=0.5,left=0.25}{v4,vc3}
\fmf{phantom,tension=0.5,right=0.25}{v5,vc3}
\fmf{plain,tension=0.5,left=0.25}{v5,vc4}
\fmf{plain,tension=0.5,right=0.25}{v6,vc4}
\fmf{plain,tension=1.25,left=0}{vc1,vc3}
\fmf{plain,tension=1.25,left=0}{vc2,vc4}
\fmffreeze
\fmf{plain,tension=1,left=0}{vc2,vc3}
\fmf{plain,tension=0.5,right=0,width=1mm}{v4,v6}
\fmffreeze
\fmfposition
\fmfipath{p[]}
\fmfipair{vd[],vm[],vu[]}
\fmfiset{p1}{vpath(__v1,__vc1)}
\fmfiset{p2}{vpath(__v2,__vc1)}
\fmfiset{p6}{vpath(__v3,__vc2)}
\fmfiset{p4}{vpath(__v4,__vc3)}
\fmfiset{p8}{vpath(__v5,__vc4)}
\fmfiset{p9}{vpath(__v6,__vc4)}
\fmfiset{p3}{vpath(__vc1,__vc3)}
\fmfiset{p7}{vpath(__vc2,__vc4)}
\fmfiset{p5}{vpath(__vc2,__vc3)}
\svertex{vm1}{p1}
\svertex{vm2}{p2}
\svertex{vm3}{p3}
\svertex{vm4}{p4}
\svertex{vm5}{p5}
\svertex{vm6}{p6}
\svertex{vm7}{p7}
\svertex{vm8}{p8}
\svertex{vm9}{p9}
}

\newcommand{\chionetwoone}[1][black]{%
\fmftop{v1}
\fmfbottom{v4}
\fmfforce{(0.125w,h)}{v1}
\fmfforce{(0.125w,0)}{v4}
\fmffixed{(0.25w,0)}{v1,v2}
\fmffixed{(0.25w,0)}{v2,v3}
\fmffixed{(0.25w,0)}{v4,v5}
\fmffixed{(0.25w,0)}{v5,v6}
\fmffixed{(0,whatever)}{vc1,vc3}
\fmffixed{(0,whatever)}{vb2,vb4}
\fmffixed{(0,whatever)}{vc1,vb1}
\fmffixed{(0,whatever)}{vc1,vb3}
\fmffixed{(whatever,0)}{vb1,vb2}
\fmffixed{(whatever,0)}{vb3,vb4}
\fmf{plain,tension=0.5,right=0.25}{v1,vc1}
\fmf{plain,tension=0.5,left=0.25}{v2,vc1}
\fmf{phantom,tension=0.5,right=0.25}{v2,vb2}
\fmf{plain,tension=0.5,left=0.25}{v3,vb2}
\fmf{plain,tension=0.5,left=0.25}{v4,vc3}
\fmf{plain,tension=0.5,right=0.25}{v5,vc3}
\fmf{phantom,tension=0.5,left=0.25}{v5,vb4}
\fmf{plain,tension=0.5,right=0.25}{v6,vb4}
\fmf{plain,tension=1.25,left=0}{vc1,vb1}
\fmf{plain,tension=1.25,left=0}{vb1,vb3}
\fmf{plain,tension=1.25,left=0}{vb3,vc3}
\fmf{plain,tension=1.25,left=0}{vb2,vb4}
\fmffreeze
\fmf{plain,tension=1,left=0}{vb1,vb2}
\fmf{plain,tension=1,left=0}{vb3,vb4}
\fmf{plain,tension=0.5,right=0,width=1mm}{v4,v6}
\fmffreeze
\fmfposition
}

\newcommand{\chionetwothree}[1][black]{%
\fmftop{v1}
\fmfbottom{v5}
\fmfforce{(0.125w,h)}{v1}
\fmfforce{(0.125w,0)}{v5}
\fmffixed{(0.25w,0)}{v1,v2}
\fmffixed{(0.25w,0)}{v2,v3}
\fmffixed{(0.25w,0)}{v3,v4}
\fmffixed{(0.25w,0)}{v5,v6}
\fmffixed{(0.25w,0)}{v6,v7}
\fmffixed{(0.25w,0)}{v7,v8}
\fmffixed{(0,whatever)}{vc1,vc4}
\fmffixed{(0,whatever)}{vc2,vc5}
\fmffixed{(0,whatever)}{vc3,vc6}

\fmf{plain,tension=0.5,right=0.25}{v1,vc1}
\fmf{plain,tension=0.5,left=0.25}{v2,vc1}
\fmf{phantom,tension=0.5,right=0.25}{v2,vc2}
\fmf{plain,tension=0.5,left=0.25}{v3,vc2}
\fmf{phantom,tension=0.5,right=0.25}{v3,vc3}
\fmf{plain,tension=0.5,left=0.25}{v4,vc3}
\fmf{plain,tension=0.5,left=0.25}{v5,vc4}
\fmf{phantom,tension=0.5,right=0.25}{v6,vc4}
\fmf{plain,tension=0.5,left=0.25}{v6,vc5}
\fmf{phantom,tension=0.5,right=0.25}{v7,vc5}
\fmf{plain,tension=0.5,left=0.25}{v7,vc6}
\fmf{plain,tension=0.5,right=0.25}{v8,vc6}
\fmf{plain,tension=1.25,left=0}{vc1,vc4}
\fmf{plain,tension=1.25,left=0}{vc2,vc5}
\fmf{plain,tension=1.25,left=0}{vc3,vc6}
\fmffreeze
\fmf{plain,tension=1,left=0}{vc4,vc2}
\fmf{plain,tension=1,left=0}{vc5,vc3}
\fmf{plain,tension=0.5,right=0,width=1mm}{v5,v8}
\fmffreeze
\fmfposition
}

\newcommand{\chitwoonethree}[1][black]{%
\fmftop{v1}
\fmfbottom{v5}
\fmfforce{(0.125w,h)}{v1}
\fmfforce{(0.125w,0)}{v5}
\fmffixed{(0.25w,0)}{v1,v2}
\fmffixed{(0.25w,0)}{v2,v3}
\fmffixed{(0.25w,0)}{v3,v4}
\fmffixed{(0.25w,0)}{v5,v6}
\fmffixed{(0.25w,0)}{v6,v7}
\fmffixed{(0.25w,0)}{v7,v8}
\fmffixed{(whatever,0.5h)}{v5,vc1}
\fmffixed{(0,whatever)}{vc1,vc4}
\fmffixed{(0,whatever)}{vc2,vc5}
\fmffixed{(0,whatever)}{vc3,vc6}
\fmffixed{(whatever,0)}{vc1,vc3}
\fmffixed{(whatever,0)}{vc3,vc5}

\fmf{plain,tension=0.5,right=0.125}{v1,vc1}
\fmf{phantom,tension=0.5,left=0.25}{v2,vc1}
\fmf{plain,tension=0.5,right=0.25}{v2,vc2}
\fmf{plain,tension=0.5,left=0.25}{v3,vc2}
\fmf{phantom,tension=0.5,right=0.25}{v3,vc3}
\fmf{plain,tension=0.5,left=0.125}{v4,vc3}
\fmf{plain,tension=0.5,left=0.25}{v5,vc4}
\fmf{plain,tension=0.5,right=0.25}{v6,vc4}
\fmf{phantom,tension=0.5,left=0.25}{v6,vc5}
\fmf{phantom,tension=0.5,right=0.25}{v7,vc5}
\fmf{plain,tension=0.5,left=0.25}{v7,vc6}
\fmf{plain,tension=0.5,right=0.25}{v8,vc6}
\fmf{plain,tension=1.25,left=0}{vc1,vc4}
\fmf{plain,tension=1.25,left=0}{vc2,vc5}
\fmf{plain,tension=1.25,left=0}{vc3,vc6}
\fmffreeze
\fmf{plain,tension=1,left=0}{vc1,vc5}
\fmf{plain,tension=1,left=0}{vc5,vc3}
\fmf{plain,tension=0.5,right=0,width=1mm}{v5,v8}
\fmffreeze
\fmfposition
}

\newcommand{\chionethreetwo}[1][black]{%
\fmftop{v1}
\fmfbottom{v5}
\fmfforce{(0.125w,h)}{v1}
\fmfforce{(0.125w,0)}{v5}
\fmffixed{(0.25w,0)}{v1,v2}
\fmffixed{(0.25w,0)}{v2,v3}
\fmffixed{(0.25w,0)}{v3,v4}
\fmffixed{(0.25w,0)}{v5,v6}
\fmffixed{(0.25w,0)}{v6,v7}
\fmffixed{(0.25w,0)}{v7,v8}
\fmffixed{(whatever,0.5h)}{v5,vc2}
\fmffixed{(0,whatever)}{vc1,vc4}
\fmffixed{(0,whatever)}{vc2,vc5}
\fmffixed{(0,whatever)}{vc3,vc6}
\fmffixed{(whatever,0)}{vc2,vc4}
\fmffixed{(whatever,0)}{vc4,vc6}
\fmf{plain,tension=0.5,right=0.25}{v1,vc1}
\fmf{plain,tension=0.5,left=0.25}{v2,vc1}
\fmf{phantom,tension=0.5,right=0.25}{v2,vc2}
\fmf{phantom,tension=0.5,left=0.25}{v3,vc2}
\fmf{plain,tension=0.5,right=0.25}{v3,vc3}
\fmf{plain,tension=0.5,left=0.25}{v4,vc3}
\fmf{plain,tension=0.5,left=0.125}{v5,vc4}
\fmf{phantom,tension=0.5,right=0.25}{v6,vc4}
\fmf{plain,tension=0.5,left=0.25}{v6,vc5}
\fmf{plain,tension=0.5,right=0.25}{v7,vc5}
\fmf{phantom,tension=0.5,left=0.25}{v7,vc6}
\fmf{plain,tension=0.5,right=0.125}{v8,vc6}
\fmf{plain,tension=1.25,left=0}{vc1,vc4}
\fmf{plain,tension=1.25,left=0}{vc2,vc5}
\fmf{plain,tension=1.25,left=0}{vc3,vc6}
\fmffreeze
\fmf{plain,tension=1,left=0}{vc2,vc4}
\fmf{plain,tension=1,left=0}{vc2,vc6}
\fmf{plain,tension=0.5,right=0,width=1mm}{v5,v8}
\fmffreeze
\fmfposition
}

\newcommand{\cvert}[7][]{%
\settoheight{\eqoff}{$\times$}%
\setlength{\eqoff}{0.5\eqoff}%
\addtolength{\eqoff}{-8\unitlength}%
\raisebox{\eqoff}{%
\fmfframe(1,2)(1,2){%
\begin{fmfchar*}(12,12)
\fmfright{v3,v2}
\fmfpoly{phantom}{v1,v3,v2}
\fmf{#2}{v1,vc1}
\fmf{#3}{vc1,v2}
\fmf{#4}{vc1,v3}
\fmffreeze
\fmfposition
\fmfipath{p[]}
\fmfiset{p1}{vpath(__v1,__vc1)}
\fmfiset{p2}{vpath(__vc1,__v2)}
\fmfiset{p3}{vpath(__vc1,__v3)}
{#1}
\fmfis{phantom,ptext.clen=6,ptext.hout=3,ptext.oout=12,ptext.out=#5,ptext.sep=;}{p1}
\fmfis{phantom,ptext.clen=6,ptext.hin=3,ptext.oin=14,ptext.in=#6,ptext.sep=;}{p2}
\fmfis{phantom,ptext.side=right,ptext.clen=6,ptext.hin=-3,ptext.oin=14,ptext.in=#7,ptext.sep=;}{p3}
\end{fmfchar*}}}
}

\newcommand{\qvert}[8]{%
\settoheight{\eqoff}{$\times$}%
\setlength{\eqoff}{0.5\eqoff}%
\addtolength{\eqoff}{-8\unitlength}%
\raisebox{\eqoff}{%
\fmfframe(1,2)(1,2){%
\begin{fmfchar*}(12,12)
\fmfleft{v2,v1}
\fmfright{v3,v4}
\fmfforce{(0,0)}{v1}
\fmfforce{(0,h)}{v2}
\fmfforce{(w,h)}{v3}
\fmfforce{(w,0)}{v4}
\fmf{#1}{v1,vc1}
\fmf{#2}{v2,vc1}
\fmf{#3}{vc1,v3}
\fmf{#4}{vc1,v4}
\fmffreeze
\fmfposition
\fmfipath{p[]}
\fmfiset{p1}{vpath(__v1,__vc1)}
\fmfiset{p2}{vpath(__v2,__vc1)}
\fmfiset{p3}{vpath(__vc1,__v3)}
\fmfiset{p4}{vpath(__vc1,__v4)}
\fmfis{phantom,ptext.clen=6,ptext.hout=3,ptext.oout=12,ptext.out=#6,ptext.sep=;}{p2}
\fmfis{phantom,ptext.clen=6,ptext.hout=3,ptext.oout=12,ptext.out=#5,ptext.sep=;}{p1}
\fmfis{phantom,ptext.side=right,ptext.clen=6,ptext.hin=-3,ptext.oin=12,ptext.in=#8,ptext.sep=;}{p4}
\fmfis{phantom,ptext.clen=6,ptext.hin=3,ptext.oin=12,ptext.in=#7,ptext.sep=;}{p3}
\end{fmfchar*}}}
}

\newcommand{\cVat}[7][]{%
\settoheight{\eqoff}{$\times$}%
\setlength{\eqoff}{0.5\eqoff}%
\addtolength{\eqoff}{-13\unitlength}%
\raisebox{\eqoff}{%
\fmfframe(2,1)(2,1){%
\begin{fmfchar*}(21,24)
\fmftop{v2}
\fmfbottom{v3}
\fmfforce{(w,h)}{v2}
\fmfforce{(w,0)}{v3}
\fmfpoly{phantom}{v1,v3,v2}
\fmf{phantom,tension=2}{v1,vg1}
\fmf{phantom}{v2,vg1}
\fmf{phantom}{v3,vg1}
\fmffreeze
\fmf{phantom,tension=1}{vg2,v2}
\fmf{phantom,tension=1}{vg3,v3}
\fmf{phantom}{vg1,v1}
\fmf{phantom,tension=0.5}{vg1,vg2}
\fmf{phantom,tension=0.5}{vg1,vg3}
\fmffreeze
\fmfposition
\fmf{phantom}{vg2,vg3}
\fmfipath{p[],pca}
\fmfipair{vm[],vo[],vi[]}
\fmfiset{p1}{vpath(__vg1,__v1)}
\fmfiset{p2}{vpath(__vg2,__v2)}
\fmfiset{p3}{vpath(__vg3,__v3)}
\fmfiset{p4}{vpath(__vg2,__vg3)}
\fmfiset{p6}{vpath(__vg3,__vg1)}
\fmfiset{p5}{vpath(__vg1,__vg2)}
{#1}
\fmfis{#2,ptext.clen=7,ptext.hin=3,ptext.hout=3,ptext.oin=6,ptext.oout=6,ptext.sep=;}{reverse(p1)}
\fmfis{#3,ptext.clen=7,ptext.hin=3,ptext.hout=3,ptext.oin=6,ptext.oout=6,ptext.sep=;}{p5}
\fmfis{#4,ptext.clen=7,ptext.hin=-9,ptext.hout=-9,ptext.oin=6,ptext.oout=6,ptext.sep=;}{p6}
\fmfis{#5,ptext.clen=7,ptext.hin=3,ptext.hout=3,ptext.oin=6,ptext.oout=6,ptext.sep=;}{p4}
\fmfis{#6,ptext.clen=7,ptext.hin=3,ptext.hout=3,ptext.oin=6,ptext.oout=6,ptext.sep=;}{p2}
\fmfis{#7,ptext.clen=7,ptext.hin=-9,ptext.hout=-9,ptext.oin=6,ptext.oout=6,ptext.sep=;}{p3}
\end{fmfchar*}}}
}

\newcommand{\cVatg}[6][]{%
\settoheight{\eqoff}{$\times$}%
\setlength{\eqoff}{0.5\eqoff}%
\addtolength{\eqoff}{-13\unitlength}%
\raisebox{\eqoff}{%
\fmfframe(2,1)(2,1){%
\begin{fmfchar*}(21,24)
\fmftop{v3}
\fmfbottom{v2}
\fmfforce{(w,h)}{v3}
\fmfforce{(w,0)}{v2}
\fmfpoly{phantom}{v1,v2,v3}
\fmf{phantom,tension=2}{v1,vg1}
\fmf{phantom}{v2,vg1}
\fmf{phantom}{v3,vg1}
\fmffreeze
\fmf{phantom,tension=1}{vgo2,v2}
\fmf{phantom,tension=1}{vgo3,v3}
\fmf{phantom}{vg1,vgm1}
\fmf{phantom}{vgm1,v1}
\fmf{#3,tension=0.5}{vg1,vgi2}
\fmf{#4,tension=0.5}{vg1,vgi3}
\fmf{#3,tension=1}{vgi2,vgm2}
\fmf{#3,tension=1}{vgm2,vgo2}
\fmf{#4,tension=1}{vgi3,vgm3}
\fmf{#4,tension=1}{vgm3,vgo3}
\fmffreeze
\fmfposition
\fmfipath{p[]}
\fmfipair{vm[],vo[],vi[]}
\fmfiset{p1}{vpath(__vg1,__v1)}
\fmfiset{p2}{vpath(__vgo2,__v2)}
\fmfiset{p3}{vpath(__vgo3,__v3)}
\fmfiset{vm1}{vloc(__vgm1)}
\fmfiset{vi2}{vloc(__vgi2)}
\fmfiset{vm2}{vloc(__vgm2)}
\fmfiset{vo2}{vloc(__vgo2)}
\fmfiset{vi3}{vloc(__vgi3)}
\fmfiset{vm3}{vloc(__vgm3)}
\fmfiset{vo3}{vloc(__vgo3)}
{#1}
\fmfis{#2,ptext.clen=7,ptext.hout=3,ptext.oin=0,ptext.oout=6,ptext.sep=;}{reverse(p1)}
\fmfis{#5,ptext.clen=7,ptext.hin=3,ptext.oin=6,ptext.oout=0,ptext.sep=;}{p2}
\fmfis{#6,ptext.clen=7,ptext.hin=-8,ptext.oin=6,ptext.oout=0,ptext.sep=;}{p3}
\end{fmfchar*}}}
}

\newcommand{\cVatt}[9][]{%
\settoheight{\eqoff}{$\times$}%
\setlength{\eqoff}{0.5\eqoff}%
\addtolength{\eqoff}{-14\unitlength}%
\raisebox{\eqoff}{%
\fmfframe(2,2)(2,2){%
\begin{fmfchar*}(21,24)
\fmftop{v3}
\fmfbottom{v2}
\fmfforce{(w,h)}{v2}
\fmfforce{(w,0)}{v3}
\fmfpoly{phantom}{v1,v3,v2}
\fmf{phantom,tension=2}{v1,vg1}
\fmf{phantom}{v2,vg1}
\fmf{phantom}{v3,vg1}
\fmffreeze
\fmf{phantom,tension=1}{vgo2,v2}
\fmf{phantom,tension=1}{vgo3,v3}
\fmf{phantom}{vg1,v1}
\fmf{#3,tension=0.5}{vg1,vgi2}
\fmf{#4,tension=0.5}{vg1,vgi3}
\fmf{#6,tension=0.5}{vgi2,vgo2}
\fmf{#7,tension=0.5}{vgi3,vgo3}
\fmffreeze
\fmfposition
\fmf{#5,label=$ $}{vgi2,vgi3}
\fmf{#5}{vgo2,vgo3}
\fmfipath{p[],pca}
\fmfipair{vm[],vo[],vi[]}
\fmfiset{p1}{vpath(__vg1,__v1)}
\fmfiset{p2}{vpath(__vgo2,__v2)}
\fmfiset{p3}{vpath(__vgo3,__v3)}
\fmfiset{p4}{vpath(__vgi2,__vgi3)}
\fmfiset{p5}{vpath(__vgi3,__vg1)}
\fmfiset{p6}{vpath(__vg1,__vgi2)}
\fmfiset{p7}{vpath(__vgo2,__vgo3)}
\fmfiset{p8}{vpath(__vgo3,__vgi3)}
\fmfiset{p9}{vpath(__vgi2,__vgo2)}
{#1}
\fmfis{#2,ptext.clen=7,ptext.hout=3,ptext.oin=0,ptext.oout=6,ptext.sep=;}{reverse(p1)}
\fmfis{#8,ptext.clen=7,ptext.hin=3,ptext.oin=6,ptext.oout=0,ptext.sep=;}{p2}
\fmfis{#9,ptext.clen=7,ptext.hin=-8,ptext.oin=6,ptext.oout=0,ptext.sep=;}{p3}
\end{fmfchar*}}}
}

\newcommand{\cVab}[7][plain]{%
\settoheight{\eqoff}{$\times$}%
\setlength{\eqoff}{0.5\eqoff}%
\addtolength{\eqoff}{-13\unitlength}%
\raisebox{\eqoff}{%
\fmfframe(3,1)(7,1){%
\begin{fmfchar*}(21,24)
\fmftop{v3}
\fmfbottom{v2}
\fmfforce{(w,h)}{v3}
\fmfforce{(w,0)}{v2}
\fmffixed{(0,0.75h)}{v2,vg3}
\fmffixed{(0,0.25h)}{v2,vg2}
\fmfpoly{phantom}{v1,v2,v3}
\fmf{#1}{vg2,vg3}
\fmf{#1}{vg2,v2}
\fmf{#1}{vg3,v3}
\fmf{photon}{vg #2,v1}
\fmf{photon,left=#3}{vg3,vg2}
\fmffreeze
\fmfposition
\fmfipath{p[],pca}
\fmfipair{vm[],vo[],vi[]}
\fmfiset{p1}{vpath(__vg #2,__v1)}
\fmfiset{p2}{vpath(__vg2,__v2)}
\fmfiset{p3}{vpath(__vg3,__v3)}
\fmfiset{p4}{vpath(__vg2,__vg3)}
\fmfiset{p5}{reverse(vpath(__vg3,__vg2))}
\fmfis{phantom,ptext.clen=6,ptext.hin=3,ptext.hout=3,ptext.oin=6,ptext.oout=6,ptext.out=#4,ptext.sep=;}{reverse(p1)}
\fmfis{phantom,ptext.clen=6,ptext.hin=3,ptext.hout=3,ptext.oin=6,ptext.oout=6,ptext.out=#5,ptext.sep=;}{p3}
\fmfis{phantom,ptext.clen=6,ptext.hin=-9,ptext.hout=-9,ptext.oin=6,ptext.oout=6,ptext.out=#6,ptext.sep=;}{p2}
\fmfis{phantom,ptext.clen=6,ptext.hin=-9,ptext.hout=-9,ptext.oin=8,ptext.oout=8,#7,ptext.sep=;}{p4}
\end{fmfchar*}}}
}

\newcommand{\seone}[9][phantom]{%
\settoheight{\eqoff}{$\times$}%
\setlength{\eqoff}{0.5\eqoff}%
\addtolength{\eqoff}{-7.5\unitlength}%
\raisebox{\eqoff}{%
\fmfframe(1,0)(1,0){%
\begin{fmfchar*}(20,15)
\fmfleft{v1}
\fmfright{v2}
\fmffixed{(0.66w,0)}{vc1,vc2}
\fmf{#4}{v1,vc1}
\fmf{#5}{vc2,v2}
\fmf{phantom,left=#2}{vc1,vc2}
\fmf{phantom,left=#3}{vc2,vc1}
\fmffreeze
\fmfposition
\fmfipath{p[]}
\fmfipair{vm[]}
\fmfiset{p1}{vpath(__vc1,__vc2)}
\fmfiset{p2}{vpath(__vc2,__vc1)}
\fmfi{#6}{subpath (0,length(p1)/2) of p1}
\fmfi{#7}{subpath (length(p1)/2,length(p1)) of p1}
\fmfi{#9}{subpath (0,length(p2)/2) of p2}
\fmfi{#8}{subpath (length(p2)/2,length(p2)) of p2}
\svertex{vm1}{p1}
\svertex{vm2}{p2}
\fmfi{#1}{vm1--vm2}
\end{fmfchar*}}}}

\newcommand{\swftwotwo}[3][plain]{%
\settoheight{\eqoff}{$\times$}%
\setlength{\eqoff}{0.5\eqoff}%
\addtolength{\eqoff}{-7.5\unitlength}%
\raisebox{\eqoff}{%
\fmfframe(1,0)(1,0){%
\begin{fmfchar*}(20,15)
\fmfleft{v1}
\fmfright{v2}
\fmffixed{(0.33w,0)}{vc1,vc2}
\fmffixed{(0.33w,0)}{vc2,vc3}
\fmf{#1}{v1,vc1}
\fmf{#1}{vc1,vc2}
\fmf{#1}{vc2,vc3}
\fmf{#1}{vc3,v2}
\fmf{photon,tension=0.5,left=#3}{vc1,vc2}
\fmf{photon,tension=0.5,left=#2}{vc1,vc3}
\fmffreeze
\fmfposition
\fmfipath{p[]}
\end{fmfchar*}}}}

\newcommand{\swftwotwor}[3][plain]{%
\settoheight{\eqoff}{$\times$}%
\setlength{\eqoff}{0.5\eqoff}%
\addtolength{\eqoff}{-7.5\unitlength}%
\raisebox{\eqoff}{%
\fmfframe(1,0)(1,0){%
\begin{fmfchar*}(20,15)
\fmfleft{v1}
\fmfright{v2}
\fmffixed{(0.33w,0)}{vc1,vc2}
\fmffixed{(0.33w,0)}{vc2,vc3}
\fmf{#1}{v1,vc1}
\fmf{#1}{vc1,vc2}
\fmf{#1}{vc2,vc3}
\fmf{#1}{vc3,v2}
\fmf{photon,tension=0.5,left=#3}{vc2,vc3}
\fmf{photon,tension=0.5,left=#2}{vc1,vc3}
\fmffreeze
\fmfposition
\fmfipath{p[]}
\end{fmfchar*}}}}

\newcommand{\swftwofour}[3][plain]{%
\settoheight{\eqoff}{$\times$}%
\setlength{\eqoff}{0.5\eqoff}%
\addtolength{\eqoff}{-7.5\unitlength}%
\raisebox{\eqoff}{%
\fmfframe(1,0)(1,0){%
\begin{fmfchar*}(20,15)
\fmfleft{v1}
\fmfright{v2}
\fmffixed{(0.33w,0)}{vc1,vc2}
\fmffixed{(0.33w,0)}{vc2,vc3}
\fmf{#1}{v1,vc1}
\fmf{#1}{vc1,vc2}
\fmf{#1}{vc2,vc3}
\fmf{#1}{vc3,v2}
\fmf{photon,tension=0.5,left=#2}{vc1,vc2}
\fmf{photon,tension=0.5,left=#3}{vc2,vc3}
\fmffreeze
\fmfposition
\fmfipath{p[]}
\end{fmfchar*}}}}

\newcommand{\swftwofive}[2][plain]{%
\settoheight{\eqoff}{$\times$}%
\setlength{\eqoff}{0.5\eqoff}%
\addtolength{\eqoff}{-7.5\unitlength}%
\raisebox{\eqoff}{%
\fmfframe(1,0)(1,0){%
\begin{fmfchar*}(20,15)
\fmfleft{v1}
\fmfright{v2}
\fmffixed{(0.22w,0)}{vc1,vc2}
\fmffixed{(0.22w,0)}{vc2,vc3}
\fmffixed{(0.22w,0)}{vc3,vc4}
\fmf{#1}{v1,vc1}
\fmf{#1}{vc1,vc2}
\fmf{#1}{vc2,vc3}
\fmf{#1}{vc3,vc4}
\fmf{#1}{vc4,v2}
\fmf{photon,tension=0.5,left=#2}{vc1,vc3}
\fmf{photon,tension=0.5,left=-(#2)}{vc2,vc4}
\fmffreeze
\fmfposition
\fmfipath{p[]}
\fmfiset{p1}{vpath(__v1,__vc1)}
\fmfiset{p2}{vpath(__vc1,__vc2)}
\fmfiset{p3}{vpath(__vc2,__vc3)}
\fmfiset{p4}{vpath(__vc3,__vc4)}
\fmfiset{p5}{vpath(__vc4,__v2)}
\end{fmfchar*}}}}

\newcommand{\nvml}[3][1]{%
\fmfcmd{%
begingroup;
save a, vp, tvp, nvp, tv, nv, ip, ts, tt, is, it, n, m, scale, t, r, s, ttpr, tnpr, ep, mm;
path lcirc;
pair vp[][], tvp[][], tv[][], nvp[][], nv[][], ip[][], ts[], is[], tt[], it[], ep[], mid;
n := #2;
m:=3;
for i=1 upto n:
for j=1 upto m:
a[i][j] := arctime ((j-1)/(m-1)*arclength pm[i]) of pm[i];
vp[i][j] := point a[i][j] of pm[i];
tvp[i][j] := unitvector direction a[i][j] of pm[i];
nvp[i][j] := tvp[i][j] rotated -90;
endfor;
endfor;
if(vp[1][1]=vp[n][m]):
vp[n+1][1] := vp[1][1];
tvp[0][m] := tvp[n][m];
nvp[0][m] := nvp[n][m];
tvp[n+1][1] :=tvp[1][1];
nvp[n+1][1] :=nvp[1][1];
else:
vp[n+1][1] := vp[n][m];
tvp[0][m] := (0,0);
nvp[0][m] := (0,0);
tvp[n+1][1] :=tvp[n][m];
nvp[n+1][1] :=nvp[n][m];
fi;
s := 1;
for i=1 upto n:
for j=1 upto m:
if (j=1):
tv[i][1] := (tvp[i-1][m]+tvp[i][1]);
nv[i][1] := (nvp[i-1][m]+nvp[i][1]);
if (not(tv[i][1]=(0,0))):
tv[i][1] := unitvector tv[i][1];
fi;
if (not(nv[i][1]=(0,0))):
nv[i][1] := unitvector nv[i][1];
fi;
ttpr := tvp[i][1] dotprod tvp[i-1][m];
tnpr := tvp[i][1] dotprod nvp[i-1][m];
elseif (j=m):
tv[i][m] := (tvp[i][m]+tvp[i+1][1]);
nv[i][m] := (nvp[i][m]+nvp[i+1][1]);
if (not(tv[i][m]=(0,0))):
tv[i][m] := unitvector tv[i][m];
fi;
if (not(nv[i][m]=(0,0))):
nv[i][m] := unitvector nv[i][m];
fi;
ttpr := tvp[i][m] dotprod tvp[i+1][1];
tnpr := -tvp[i][m] dotprod nvp[i+1][1];
else:
nv[i][j] :=nvp[i][j];
tv[i][j] :=tvp[i][j];
fi;
scale := 25;
if ((j=1) or (j=m)):
 if ((tnpr<=0) and not((tv[i][j]=(0,0)) or (nv[i][j]=(0,0)))):
  ip[i][j] := vp[i][j] shifted(0.15*scale*nvp[i][j]);
  ts[s] := tvp[i][j];
  is[s] := ip[i][j];
  s:=s+1;
 else:
  if ((j=1) and (ttpr>0)):
  fi;
 fi;
else:
 ip[i][j] := vp[i][j] shifted(0.15*scale*nv[i][j]);
 ts[s] := tv[i][j];
 is[s] := ip[i][j];
 s:=s+1;
fi;
endfor;
endfor;
if(vp[1][1]=vp[n][m]):
ts[s] := ts[1];
is[s] := is[1];
else:
tv[n+1][1] := unitvector (tvp[n][m]+tvp[n+1][1]);
nv[n+1][1] := unitvector (nvp[n][m]+nvp[n+1][1]);
ip[n+1][1] := vp[n+1][1] shifted(0.15*scale*nv[n+1][1]);
ts[s] := tv[n+1][1];
is[s] := ip[n+1][1];
fi;
t=#1;
lcirc:=is[1];
for k=2 upto s:
lcirc := lcirc{ts[k-1]}..tension t..{ts[k]}is[k];
endfor;
mm := arctime (0.5* arclength lcirc) of lcirc;
if(vp[1][1]=vp[n][m]):
ep1 := point arctime (0* arclength lcirc) of lcirc of lcirc;
ep2 := point mm of lcirc;
mid := 1/2[ep1,ep2];
else:
ep1 := point mm of lcirc;
ep2 :=unitvector direction mm of lcirc rotated -90;
mid:= ep1 shifted(0.2*scale*ep2);
fi;
draw(lcirc) withpen pencircle scaled 0.25;
drawarrow(subpath(mm*0.8,mm*1.1) of lcirc) withpen pencircle scaled 0.25;
endgroup;
}
\fmfiv{label=#3,l.dist=0}{mid}
}

\DeclareMathOperator{\tr}{tr}

\DeclareMathOperator{\re}{Re}
\DeclareMathOperator{\im}{Im}

\DeclareMathOperator{\perm}{P}
\DeclareMathOperator{\Lam}{\Lambda}
\DeclareMathOperator{\Top}{T}
\DeclareMathOperator{\Kop}{K}
\DeclareMathOperator{\Rop}{R}
\DeclareMathOperator{\chiop}{\mathbf{\chi}}
\DeclareMathOperator{\D}{D}
\DeclareMathOperator{\barD}{\vphantom{\D}\smash[t]{\bar{\mathrm{D}}}}
\DeclareMathOperator{\Ld}{L}

\newcommand{\chiimp}[4][]{%
\smash[t]{#1{\chiop}}_{(#2)}^{(#3)}(#4)
}
\newlength{\eqoff}

\newlength{\unit}
\psset{xunit=\unit,yunit=\unit,runit=\unit}
\newlength{\linew}
\psset{linewidth=\linew}

\usepackage[small,bf]{caption}

\numberwithin{equation}{section}

\newcommand{\mympostgrey}{0.75 white}

\begin{document}
\begin{fmffile}{graphs}

\fmfcmd{%
input Dalgebra
}

\fmfcmd{%
def getmid(suffix p) =
  pair p.mid[], p.off[], p.dir[];
  for i=0 upto 36:
    p.dir[i] = dir(5*i);
    p.mid[i]+p.off[i] = directionpoint p.dir[i] of p;
    p.mid[i]-p.off[i] = directionpoint -p.dir[i] of p;
  endfor
enddef;
}

\fmfcmd{%
marksize=2mm;
def draw_mark(expr p,a) =
  begingroup
    save t,tip,dma,dmb; pair tip,dma,dmb;
    t=arctime a of p;
    tip =marksize*unitvector direction t of p;
    dma =marksize*unitvector direction t of p rotated -45;
    dmb =marksize*unitvector direction t of p rotated 45;
    linejoin:=beveled;
    draw (-.5dma.. .5tip-- -.5dmb) shifted point t of p;
  endgroup
enddef;
style_def derplain expr p =
    save amid;
    amid=.5*arclength p;
    draw_mark(p, amid);
    draw p;
enddef;
style_def derphoton expr p =
    save amid;
    amid=.5*arclength p;
    draw_mark(p, amid);
    draw wiggly p;
enddef;
style_def derdots expr p =
    save amid;
    amid=.5*arclength p;
    draw_mark(p, amid);
    draw_dots p;
enddef;
def draw_marks(expr p,a) =
  begingroup
    save t,tip,dma,dmb,dmo; pair tip,dma,dmb,dmo;
    t=arctime a of p;
    tip =marksize*unitvector direction t of p;
    dma =marksize*unitvector direction t of p rotated -45;
    dmb =marksize*unitvector direction t of p rotated 45;
    dmo =marksize*unitvector direction t of p rotated 90;
  linejoin:=beveled;
    draw (-.5dma.. .5tip-- -.5dmb) shifted point t of p withcolor 0white;
    draw (-.5dmo.. .5dmo) shifted point t of p;
  endgroup
enddef;
style_def derplains expr p =
    save amid;
    amid=.5*arclength p;
    draw_marks(p, amid);
    draw p;
enddef;
def draw_markss(expr p,a) =
  begingroup
    save t,tip,dma,dmb,dmo; pair tip,dma,dmb,dmo;
    t=arctime a of p;
    tip =marksize*unitvector direction t of p;
    dma =marksize*unitvector direction t of p rotated -45;
    dmb =marksize*unitvector direction t of p rotated 45;
    dmo =marksize*unitvector direction t of p rotated 90;
    linejoin:=beveled;
    draw (-.5dma.. .5tip-- -.5dmb) shifted point t of p withcolor 0white;
    draw (-.5dmo.. .5dmo) shifted point arctime a+0.25 mm of p of p;
    draw (-.5dmo.. .5dmo) shifted point arctime a-0.25 mm of p of p;
  endgroup
enddef;
style_def derplainss expr p =
    save amid;
    amid=.5*arclength p;
    draw_markss(p, amid);
    draw p;
enddef;
style_def dblderplains expr p =
    save amidm;
    save amidp;
    amidm=.5*arclength p-0.75mm;
    amidp=.5*arclength p+0.75mm;
    draw_mark(p, amidm);
    draw_marks(p, amidp);
    draw p;
enddef;
style_def dblderplainss expr p =
    save amidm;
    save amidp;
    amidm=.5*arclength p-0.75mm;
    amidp=.5*arclength p+0.75mm;
    draw_mark(p, amidm);
    draw_markss(p, amidp);
    draw p;
enddef;
style_def dblderplainsss expr p =
    save amidm;
    save amidp;
    amidm=.5*arclength p-0.75mm;
    amidp=.5*arclength p+0.75mm;
    draw_marks(p, amidm);
    draw_markss(p, amidp);
    draw p;
enddef;
}

%
%

\fmfcmd{%
thin := 1pt; 
thick := 2thin;
arrow_len := 4mm;
arrow_ang := 15;
curly_len := 3mm;
dash_len := 1.5mm; 
dot_len := 1mm; 
wiggly_len := 2mm; 
wiggly_slope := 60;
zigzag_len := 2mm;
zigzag_width := 2thick;
decor_size := 5mm;
dot_size := 2thick;
}


\begingroup\parindent0pt
\vspace*{2em}
\begingroup\LARGE
From $\mathcal{N}=4$ gauge theory to $\mathcal{N}=2$ conformal QCD:
three-loop mixing of scalar composite operators
\par\endgroup
\vspace{1.5em}
\begingroup\large
Elli Pomoni, Christoph Sieg
\par\endgroup
\vspace{1em}
\begingroup\itshape
Institut f\"ur Mathematik und Institut f\"ur Physik, Humboldt-Universit\"at zu Berlin\\
Johann von Neumann Haus, Rudower Chaussee 25, 12489 Berlin, Germany
\par\endgroup
\vspace{1em}
\begingroup\ttfamily
pomoni@math.hu-berlin.de\\
csieg@math.hu-berlin.de
\par\endgroup
\vspace{1.5em}
\endgroup

\paragraph{Abstract.}
We derive the planar dilatation operator in the closed subsector 
of scalar composite operators of an $\mathcal{N}=2$ superconformal
quiver gauge theory to three loops. 
By tuning the ratio of its two gauge couplings we
interpolate between a $\mathds{Z}_2$ orbifold of 
$\mathcal{N}=4$ SYM theory and $\mathcal{N}=2$ superconformal QCD.
We find $\zeta(3)$ contributions at three loops that disappear when the 
theory is at the orbifold point. 
They are responsible for imaginary contributions to the dispersion relation 
of a single scalar excitation in the spin-chain picture. 
This points towards an interpretation of the individual scalar excitations as 
effective rather than as elementary magnons. We argue that the elementary
excitations should be associated with certain fermions and 
covariant derivatives, and that integrability in the respective subsectors 
should persist at least to two loops.



\newpage


\tableofcontents

\newpage

\section{Introduction}

The $\AdS/\CFT$ correspondence 
\cite{Maldacena:1997re,Gubser:1998bc,Witten:1998qj} in its original 
formulation relates type $\twob$ string 
theory in $\AdS_5\times\text{S}^5$ to $\mathcal{N}=4$ SYM theory.
Some closely related examples include 
$\mathds{Z}_k$ orbifolds \cite{Kachru:1998ys,Lawrence:1998ja}
and the $\beta$-deformation \cite{Lunin:2005jy} of the original correspondence.
The resulting gravity backgrounds are of the form 
$\AdS_5\times\mathcal{M}_5$ with a compact five-dimensional
manifold $\mathcal{M}_5$ that is given by $\text{S}^5/\mathds{Z}_k$
in case of the $\mathds{Z}_k$ orbifold or by a deformed $\text{S}^5$.
The dual field theory preserves conformal symmetry and respectively is some 
quiver gauge theory or the $\beta$-deformed $\mathcal{N}=4$ 
SYM theory with reduced
supersymmetry. All these examples have in common that their dual 
string backgrounds are critical, i.e.\ ten-dimensional, and
they share certain universal properties \cite{Gadde:2009dj}.
In the planar limit \cite{'tHooft:1973jz} the rank of the underlying gauge
group is taken to infinity, while the number of 
matter fields has to be kept finite. 
This is the quenched approximation, where
backreaction from matter fields is suppressed \cite{Karch:2002sh}. 
Furthermore, in the theories of the above type, the individual $a$ and $c$ 
anomaly coefficients \cite{Henningson:1998gx} become equal $a=c$ in the
limit. It is also important to note that
planar Feynman diagrams in $\mathcal{N}=4$ SYM theory and its 
$\mathds{Z}_k$ orbifold theories are identical 
\cite{Bershadsky:1998mb,Bershadsky:1998cb}.
The aforementioned four-dimensional superconformal theories should
be considered as being members of the $\mathcal{N}=4$ SYM 
universality class.

Lower-dimensional examples of the $\AdS/\CFT$ correspondence 
include the ABJM duality \cite{Aharony:2008ug} and its ABJ 
generalization \cite{Aharony:2008gk} that involve $\mathcal{N}=6$ 
supersymmetric Chern-Simons theory with levels $k$, $-k$ and 
respective product gauge groups $U(N)\times U(N)$ and $U(M)\times U(N)$ 
that are dual to M-theory on $\text{AdS}_4\times\text{S}^7/\mathds{Z}_k$
or to generalizations thereof involving flux. A gravity description
in terms of type II\,A string theory in the critical background 
$\text{AdS}_4\times\text{CP}^3$ is only possible in a certain regime
of the parameters.

Recently, it was argued \cite{Gadde:2009dj}
that $\mathcal{N}=2$ superconformal QCD (SCQCD) with gauge group 
$SU(N)$ and $N_\text{f}=2N$ fundamental hypermultiplets 
should have a dual gravity description. $\mathcal{N}=2$ SCQCD lies outside 
the $\mathcal{N}=4$ SYM universality class, and its gravity dual is 
non-critical with $\AdS_5$ and $\text{S}^1$ factors. It does
not admit a quenched approximation, and its planar limit is the 
leading contribution in the Veneziano expansion \cite{Veneziano:1976wm}.
The anomaly coefficients $a$ and $c$ differ even in the planar limit.

There is nevertheless a connection between $\mathcal{N}=2$ SCQCD and 
the $\mathcal{N}=4$ SYM universality class. 
The $\mathcal{N}=2$ quiver gauge theory with product gauge group 
$SU(N)\times SU(\hat N)$ and two coupling constants $g_\YM$ and $\hat g_\YM$
at its conformal point where $N=\hat N$, interpolates between 
$\mathcal{N}=2$ SCQCD and the $\mathds{Z}_2$ orbifold of $\mathcal{N}=4$ 
SYM theory \cite{Gadde:2009dj,Gadde:2010zi}.
We will refer to this superconformal theory as the \emph{interpolating theory}.
By considering this theory we can study how 
the transition from the $\mathcal{N}=4$ SYM universality
class to a theory outside this class is realized.


Of particular interest in a theory with conformal invariance is
the spectrum of conformal dimensions of the gauge invariant composite 
operators. 
In perturbation 
theory where the coupling $\bar\lambda$ is small, 
these operators 
mix under renormalization as
\begin{equation}\label{opren}
\mathcal{O}_{a,\text{ren}}
=\mathcal{Z}_{a}{}^b(\bar\lambda,\varepsilon)\mathcal{O}_{b,\text{bare}}
\col\qquad
\end{equation}
where $a$ labels the composite operators in 
an appropriate basis and $\varepsilon$ is a UV regulator.
The conformal dimensions of the composite operators are determined as
a sum of their common bare scaling dimension and of individual 
anomalous dimensions. The latter follow as eigenvalues of the 
dilatation operator that in terms of the renormalization constant 
$\mathcal{Z}$ is defined as 
\begin{equation}\label{DinZ}
\mathcal{D}
=\mu\frac{\de}{\de\mu}\ln\mathcal{Z}(\bar\lambda\mu^{2\varepsilon},\varepsilon)
=\lim_{\varepsilon\rightarrow0}\left[2\varepsilon\bar\lambda
\frac{\de}{\de\bar\lambda}\ln\mathcal{Z}(\bar\lambda,\varepsilon)\right]
\col\qquad
\end{equation}
where $\mu$ is the scale introduced by the regularization.
In the planar limit, mixing occurs within the subset of single-trace operators
in which the color contractions form a single cycle.
Due to the equivalence of $\mathcal{N}=4$ SYM theory and its orbifolds 
in the planar limit \cite{Bershadsky:1998mb,Bershadsky:1998cb},
the respective dilatation operators are identical.
In these theories, in the $\beta$-deformed theory and also in the
$\mathcal{N}=6$ CS theory, the operator mixing problem
appears to be integrable \cite{Minahan:2002ve,Ideguchi:2004wm,Roiban:2003dw,Berenstein:2004ys,Minahan:2008hf,
Gaiotto:2008cg,Grignani:2008is,Bak:2008cp,Bak:2008vd}. 
This can be seen by
mapping the problem to an interacting spin chain \cite{Minahan:2002ve}.
The composite single-trace operators are interpreted as 
spin chain states, and the dilatation operator \eqref{DinZ} is identified with 
the Hamiltonian acting on these chains. Integrability allows one to 
determine the eigenvalues from respective Bethe ans\"atze
\cite{Minahan:2002ve,Beisert:2004hm,Beisert:2005fw,Ideguchi:2004wm,Beisert:2005he,Gromov:2008qe,Grignani:2008is,Bak:2008cp}. The latter rely  
on the factorization of multiple-particle scattering processes
into products of two-particle scatterings captured in terms of a 
respective S-matrix with a dressing phase 
\cite{Arutyunov:2004vx,Beisert:2006ez}. 
The particles that are 
fundamental excitations of the spin chain are called magnons.
Further details can be found 
in the collection of reviews \cite{Beisert:2010jr}, see in particular
\cite{Minahan:2010js,Staudacher:2010jz,Ahn:2010ka,Klose:2010ki,Zoubos:2010kh}.


In $\mathcal{N}=4$ SYM theory, operator mixing can be studied in 
the closed $SU(2)$ subsector. This is the simplest setup 
since the respective operators contain only two different kinds of 
scalar field flavors. In the planar limit, mixing only occurs among operators 
with identical field content but different orderings within the single trace.
The dilatation operator can hence not alter the length $L$, i.e.\ the total 
number of fields within the gauge trace, and it can be expressed in terms of 
permutations that act on the individual field flavors. 
Based on the assumption of integrability and some further input, 
the dilatation operator of $\mathcal{N}=4$ SYM theory
has been constructed to three loops \cite{Beisert:2003tq} 
and then also to higher loops \cite{Beisert:2003jb,Fiamberti:2009jw}. 
The three-loop result was then confirmed by only using the underlying
symmetry, leaving undetermined some constants \cite{Beisert:2003ys}.
It was then also tested by field theory calculations that yield some
of its eigenvalues \cite{Moch:2004pa,Kotikov:2004er,Eden:2004ua}.
At higher loops some particular terms in the dilatation operator
were computed \cite{Gross:2002su,Beisert:2007hz,Fiamberti:2007rj,Fiamberti:2008sh,Fiamberti:2009jw}. More details can be found in the collection
of reviews \cite{Beisert:2010jr}, see in particular
\cite{Minahan:2010js,Sieg:2010jt,Rej:2010ju,Zoubos:2010kh}.
An explicit Feynman diagram calculation of the three-loop dilatation 
operator itself was accomplished recently \cite{Sieg:2010tz}. 
In the employed $\mathcal{N}=1$ superspace formalism the 
composite operators of the $SU(2)$ subsector appear as lowest 
components of superfields that are chiral. This chirality is crucial for 
the formulation of generalized finiteness conditions \cite{Sieg:2010tz}
that reduce significantly the calculational effort. 
The result is expressed in terms of the so-called chiral 
functions \cite{Fiamberti:2007rj,Fiamberti:2008sh} that naturally 
emerge in the $\mathcal{N}=1$ superspace formalism and are
linear combinations of multiple permutations. They 
are generated by the chiral structures, 
i.e.\ by the configurations of the elementary chiral and 
anti-chiral propagators and vertices of the underlying Feynman 
diagrams. 
In the $\beta$-deformed theory the chiral functions have to be
slightly modified \cite{Fiamberti:2008sm} since the permutations contain phase 
factors \cite{Berenstein:2004ys}.
However, the expression of the dilatation operator in terms of 
these chiral functions is not altered.
In the aforementioned theories of the $\mathcal{N}=4$ SYM universality 
class, the chiral functions are insensitive to potential further 
interactions involving gauge fields.

In the \emph{interpolating theory}, operator mixing can be studied
in a closed subsector that resembles the $SU(2)$ subsector of the 
$\mathcal{N}=4$ SYM theory.
Its composite operators consist only of certain scalar fields, and in 
an $\mathcal{N}=1$ superfield formulation they are chiral. 
However, unlike in the $SU(2)$ subsector, they are built by choosing 
out of more than just two different types of fields.
Moreover, their gauge traces involve the color indices from the 
two different $SU(N)$ factors in the product gauge group. 
Hence, the Feynman diagrams of operator renormalization are functions of 
the two gauge coupling constants $g_\YM$ and $\hat g_\YM$. 
They generate chiral functions that depend on
 the ratio of these couplings.
Similarly to the case of the $\beta$-deformation 
\cite{Berenstein:2004ys,Fiamberti:2008sm} part of this  
dependence is associated with the chiral structure itself.  
It is thus included in the natural definition of the flavor operations 
as defined below in \eqref{PLTdef}. But not all of the coupling 
dependence can be captured in this way. Gauge interactions 
that appear at higher loops introduce additional coupling dependence 
into the chiral functions \eqref{chifuncdef}.
The relative coefficients that combine the flavor operations to 
the chiral functions become functions of the coupling ratio,
yielding the deformed chiral functions \eqref{defchifunc}.
This kind of deformation should not occur in any 
theory of the $\mathcal{N}=4$ SYM universality class.
Yet, it should appear from six loop on
in the $\mathcal{N}=6$ CS theory of the ABJ generalization, involving the 
ratio of its two independent 't Hooft coupling constants.
With the \emph{interpolating theory} at hand we have a simpler setup than 
the CS theory that allows us to study the effects due to the 
presence of two gauge couplings.

The spin chain of the \emph{interpolating theory} was constructed in 
\cite{Gadde:2010zi,Liendo} with the purpose of investigating integrability
of $\mathcal{N}=2$ SCQCD. Already at one loop the 
scalar sector of the  \emph{interpolating theory} is not integrable, 
but the question is still open for its $\mathcal{N}=2$ SCQCD limit. 
The purpose of this paper is to provide the dilatation operator to 
a loop order where it is sensitive to deformations that are generated 
by the gauge interactions. This requires a calculation at three loops.
The insights obtained from our analysis allow us to predict that 
in other subsectors a first non-trivial test of integrability requires a three
loop calculation. In particular, in the $SU(1|1)$ and $SL(2)$ subsectors
integrability is inherited from the  $\mathcal{N}=4$ SYM theory to two loops, 
even for the \emph{interpolating theory}. Similar arguments should hold 
for the respective subsectors in $\mathcal{N}=1$ SQCD in the 
Veneziano limit with $N_\text{f}=3N$ where the theory becomes conformal. 
So far, in this theory, only a one-loop study of the non-dynamical scalar 
excitations is available \cite{Poland:2011kg}.
A recent review on integrability in pure QCD and in its supersymmetric
cousins can be found in \cite{Korchemsky:2010kj}.

The paper is organized as follows:\\
Firstly, in section \ref{sec:threeloops}, we explain why 
in the \emph{interpolating theory} the deformations of the chiral functions 
show up first at three loops. 
In section \ref{sec:N2SYM} we introduce the $\mathcal{N}=1$ superfield 
formulation of the theory and describe the closed chiral subsector 
which is the subject of our investigation.
Then, as a warm-up, 
in section \ref{sec:onetwoloopD} we present the calculation of the
one- and two-loop dilatation operator.
The three-loop calculation along the lines 
of \cite{Sieg:2010tz} is presented in 
section \ref{sec:threeloopD}.
In section \ref{sec:wrappingint} we extend the result
beyond the asymptotic limit by computing the leading wrapping 
correction.
Three-loop eigenvalues of the dilatation operator are summarized 
in section \ref{sec:evalues}. This includes the calculations 
of the dispersion relation \eqref{Ep}
for a single impurity and of the three-loop anomalous
dimensions of the composite single-trace operators that contain up to 
four elementary fields.
Based on our calculation and its spin chain interpretation, 
in section \ref{sec:elemex} we discuss subsectors with elementary 
excitations in which integrability should persist to two loops.
Finally, we conclude in
section \ref{sec:concl} and point towards 
interesting open problems that could be the subject of further studies.
Details of our conventions, the Feynman rules, the derivation of 
required one- and two-loop subdiagrams, a list of the relevant 
integrals and a discussion of similarity transformations
have been delegated to various appendices.

\section{Why three loops?}
\label{sec:threeloops}

As we have mentioned in the introduction, the presence of two gauge 
parameters in the \emph{interpolating theory}
leads to a dependence of the chiral functions on the ratio of these 
couplings. In particular, the chiral functions are deformed when additional 
gauge interactions are present in the 
underlying Feynman diagrams. We will now explain why 
in the closed chiral subsector these deformations
first show up at three loops and lead to new effects.
 
\begin{figure}[h]
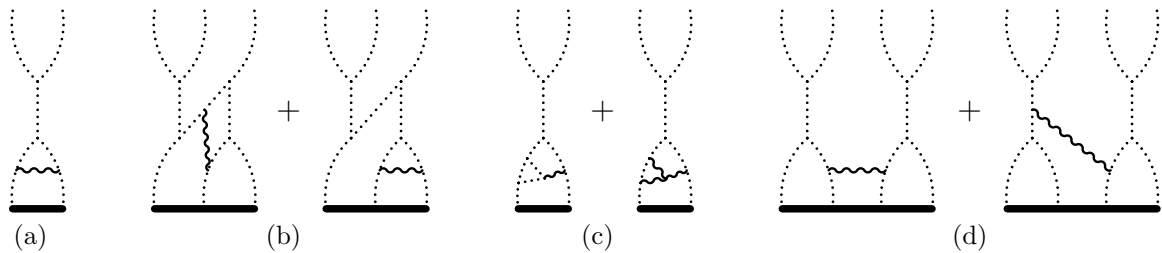

\begin{center}
\subfigure[\label{fig:twoloopdefchi}]{%
\settoheight{\eqoff}{$\times$}%
\setlength{\eqoff}{0.5\eqoff}%
\addtolength{\eqoff}{-17\unitlength}%
\raisebox{\eqoff}{%
\fmfframe(-1,2)(-16.5,2){%
\begin{fmfchar*}(30,30)
\chionei{dots}{dots}{dots}{dots}{dots}
\fmfi{photon}{vm4--vm5}
\end{fmfchar*}}}
}
\hspace{0.5cm}
\subfigure[\label{fig:threeloopdefchi1}]{%
$\settoheight{\eqoff}{$\times$}%
\setlength{\eqoff}{0.5\eqoff}%
\addtolength{\eqoff}{-17\unitlength}%
\raisebox{\eqoff}{%
\fmfframe(-1,2)(-9,2){%
\begin{fmfchar*}(30,30)
\chionetwoi{dots}{dots}{dots}{dots}{dots}{dots}{dots}{dots}{dots}
\fmfi{photon}{vm5{dir -90}..{dir -90}vm8}
\end{fmfchar*}}}
+
\settoheight{\eqoff}{$\times$}%
\setlength{\eqoff}{0.5\eqoff}%
\addtolength{\eqoff}{-17\unitlength}%
\raisebox{\eqoff}{%
\fmfframe(-1,2)(-9,2){%
\begin{fmfchar*}(30,30)
\chionetwoi{dots}{dots}{dots}{dots}{dots}{dots}{dots}{dots}{dots}
\fmfi{photon}{vm8--vm9}
\end{fmfchar*}}}
$}
\hspace{0.5cm}
\subfigure[\label{fig:threeloopdefchi2}]{%
$
\settoheight{\eqoff}{$\times$}%
\setlength{\eqoff}{0.5\eqoff}%
\addtolength{\eqoff}{-17\unitlength}%
\raisebox{\eqoff}{%
\fmfframe(-1,2)(-16.5,2){%
\begin{fmfchar*}(30,30)
\chionei{phantom}{dots}{dots}{dots}{dots}
\fmfiset{p11}{subpath (0,length(p4)/3) of p4}
\fmfiset{p12}{subpath (length(p4)/3,2*length(p4)/3) of p4}
\fmfiset{p13}{subpath (2*length(p4)/3,length(p4)) of p4}
\fmfipair{vg,tv,nv}
\fmfcmd{
as := arctime (0.5*arclength p4) of p4;
tv := unitvector direction as of p4;
nv := tv rotated -90;
vg := point as of p4 shifted(9*nv);}
\fmfi{dots}{vd4--vg}
\fmfi{dots}{vu4--vg}
\fmfi{photon}{vg--vm5}
\fmfi{dots}{p11}
\fmfi{dots}{p12}
\fmfi{dots}{p13}
\end{fmfchar*}}}
+
\settoheight{\eqoff}{$\times$}%
\setlength{\eqoff}{0.5\eqoff}%
\addtolength{\eqoff}{-17\unitlength}%
\raisebox{\eqoff}{%
\fmfframe(-1,2)(-16.5,2){%
\begin{fmfchar*}(30,30)
\chionei{phantom}{dots}{dots}{dots}{dots}
\fmfiset{p11}{subpath (0,length(p4)/3) of p4}
\fmfiset{p12}{subpath (length(p4)/3,2*length(p4)/3) of p4}
\fmfiset{p13}{subpath (2*length(p4)/3,length(p4)) of p4}
\fmfipair{vg,tv,nv}
\fmfcmd{
as := arctime (0.5*arclength p4) of p4;
tv := unitvector direction as of p4;
nv := tv rotated -90;
vg := point as of p4 shifted(9*nv);}
\fmfi{photon}{vd4--vg}
\fmfi{photon}{vu4--vg}
\fmfi{photon}{vg--vm5}
\fmfi{dots}{p11}
\fmfi{dots}{p12}
\fmfi{dots}{p13}
\end{fmfchar*}}}
$}
\hspace{0.5cm}
\subfigure[\label{fig:threeloopdefchi3}]{%
$
\settoheight{\eqoff}{$\times$}%
\setlength{\eqoff}{0.5\eqoff}%
\addtolength{\eqoff}{-17\unitlength}%
\raisebox{\eqoff}{%
\fmfframe(-1,2)(-1,2){%
\begin{fmfchar*}(30,30)
\chionethreei{dots}{dots}{dots}{dots}{dots}{dots}{dots}{dots}
\fmfi{photon}{vlm5--vrm4}
\end{fmfchar*}}}
{}+{}
\settoheight{\eqoff}{$\times$}%
\setlength{\eqoff}{0.5\eqoff}%
\addtolength{\eqoff}{-17\unitlength}%
\raisebox{\eqoff}{%
\fmfframe(-1,2)(-1,2){%
\begin{fmfchar*}(30,30)
\chionethreei{dots}{dots}{dots}{dots}{dots}{dots}{dots}{dots}
\fmfi{photon}{vlm3--vrm4}
\end{fmfchar*}}}
$
}
\caption{Feynman diagrams to three loops that generate deformed chiral 
functions. The dotted lines denote all possible arrangements of bifundamental 
and adjoint field flavors.
(a): The only two-loop diagram that generates a deformed $\chi(1)$ but is finite.
(b): Three-loop diagrams that generate a deformed $\chi(1,2)$. 
The second diagram contains the diagram (a) as subdiagram.
(c): Three-loop diagrams that generate a deformed $\chi(1)$.
(d): Three-loop diagrams that generate a deformed $\chi(1,3)$ and only contribute to scattering.
}
\end{center}
\end{figure}

In $\mathcal{N}=1$ superspace the Feynman diagrams of the closed chiral 
subsector have the same form as the respective diagrams in $\mathcal{N}=4$
SYM theory. As we conclude from the analysis in \cite{Sieg:2010tz},
in any diagram with an overall UV divergence at least one loop is
generated by its chiral structure. Thus, 
one has to work at least at two loops in order to find a UV divergent diagram 
that also involves a gauge field generating a deformation of 
its chiral function.
At two loops, the only diagram associated with such a deformation 
is depicted in figure \ref{fig:twoloopdefchi}.
This diagram is finite and hence does not contribute 
to the dilatation operator of the theory.
Deformations of the chiral functions can therefore
appear in the dilatation operator only from three loops on.
They originate from the diagrams with overall UV divergences that
are displayed 
in figure \ref{fig:threeloopdefchi1}, \ref{fig:threeloopdefchi2}
and \ref{fig:threeloopdefchi3} and 
from the reflected diagrams where applicable.
Note that the finite two-loop diagram of figure \ref{fig:twoloopdefchi} 
appears as subdiagram within the second diagram of
\ref{fig:threeloopdefchi1}. 

The above considerations imply that the one- and two-loop 
results for the dilatation operator depend only on chiral functions
\eqref{chifuncdef}
that are in one-to-one correspondence to the chiral functions
of $\mathcal{N}=4$ SYM theory.
This is not the case from three loops on. In the dilatation 
operator we find chiral functions that have their counterparts in
the $\mathcal{N}=4$ SYM case, but there are further contributions 
that depend on deformed chiral functions \eqref{defchifunc}. 
They originate from the
diagrams that are displayed in figures  
\ref{fig:threeloopdefchi1} and \ref{fig:threeloopdefchi2}.
These diagrams contribute with simple poles that 
have maximum transcendentality at this loop order, i.e.\ they 
are proportional to $\zeta(3)$. Some of these contributions add
anti-Hermitean terms to the dilatation operator. 
Besides these contributions, we find additional transcendentality-three
terms with undeformed chiral functions in the dilatation operator. 
All these terms disappear at the orbifold point, i.e.\ for equal couplings 
$g=\hat g$, such that one obtains the rational result of the 
$\mathcal{N}=4$ SYM theory. Note that in \cite{Andree:2010na}
the circular Wilson loop of $\mathcal{N}=2$ SCQCD was computed, and 
it deviates from its $\mathcal{N}=4$ SYM counterpart first at three loops
by $\zeta(3)$ terms.

The fact that three loops is the first loop order at which new
contributions can arise can also be understood from the basis 
of loop integrals in four dimensions. Only from three loops on
there appear several integrals with overall UV divergences that 
contain simple poles in $\varepsilon$. Among them one finds 
for the first time
an integral with a simple pole that is proportional to $\zeta(3)$. 
The occurring integrals are listed in appendix \ref{app:integrals}.

Three loops is also interesting for another reason. 
This is the lowest loop order at which the dilatation operator 
contains terms that do not contribute to the dispersion relation.
We refer to them as pure scattering terms, since they only contribute to the
scattering matrix. 
At three loops the only scattering term is
generated by Feynman diagrams in figure \ref{fig:threeloopdefchi3}, including
their reflections where applicable.
They contain a gauge field that deforms the chiral function and hence
also modifies the 
scattering matrix compared to the $\mathcal{N}=4$ SYM case.

\section{The interpolating theory}
\label{sec:N2SYM}

The \emph{interpolating theory}
is given by an $\mathcal{N}=2$ superconformal quiver gauge theory 
that depends on two couplings $g_\YM$ and $\hat g_\YM$, each of which 
is associated with one factor in the product gauge group 
$SU(N)\times SU(N)$. Since it has the same field content as a $\mathds{Z}_2$
orbifold of $\mathcal{N}=4$ SYM theory, its action is 
roughly given by two copies of the $\mathcal{N}=4$ SYM action, but with 
modified superpotentials that involve the bifundamental fields.

\subsection{Action and parameters}
\label{sec:action}

In terms of $\mathcal{N}=1$ superfields, 
with the conventions of \cite{Gates:1983nr}, the action reads
\begin{equation}\label{action}
\begin{aligned}
S
&=\frac{1}{2}\int\de^4x\de^2\theta\Big[\frac{1}{g_\YM^2}\tr\big(W^{\alpha}W_{\alpha}\big)
+\frac{1}{\hat g_\YM^2}\tr\big(\hat W^{\alpha}\hat W_{\alpha}\big)\Big]
\col\\
&\phantom{{}={}}+\int\de^4x\de^4\theta\big[\tr\big(\e^{-g_\YM V} \bar{\Phi}\e^{g_\YM V}\Phi\big)
+\tr\big(\e^{-\hat{g}_\YM \hat{V}} \hat{\bar{\Phi}}\e^{\hat{g}_\YM \hat{V}}\hat{\Phi}\big)\big]\\
&\phantom{{}={}}
+
\int\de^4x\de^4\theta\big[\tr\big(\bar{Q}^{\dot I}\e^{g_\YM V}Q_{\dot I}\e^{-\hat{g}_\YM \hat{V}}\big)
+\tr\big(\tilde{\bar{Q}}_{\dot I}\e^{\hat g_\YM \hat V} \tilde{Q}^{\dot I}\e^{-g_\YM V}\big)\big]\\
 &\phantom{{}={}}
+i\int\de^4x\de^2\theta\big[g_\YM\tr\big(\tilde{Q}^{\dot I} \Phi Q_{\dot I}\big)-\hat g_\YM\tr\big(Q_{\dot I} \hat{\Phi}\tilde{Q}^{\dot I}\big)\big]\\
&\phantom{{}={}}
-i\int\de^4x\de^2\bar\theta\big[g_\YM\tr\big(\bar Q^{\dot I}\bar\Phi\tilde{\bar{Q}}_{\dot I}\big)-\hat g_\YM\tr\big(\tilde{\bar{Q}}_{\dot I}\hat{\bar{\Phi}}\bar Q^{\dot I}\big)\big]
\col
\end{aligned}
\end{equation}
where $W_\alpha = i\barD^2 \left(\e^{-g_\YM V}\D_\alpha\,\e^{g_\YM V}\right)$
and $\hat W_\alpha = i\barD^2 \left(\e^{-\hat g_\YM\hat V}\D_\alpha\,\e^{\hat g_\YM\hat V}\right)$
are the chiral superfield strengths of the 
vector superfields $V$ and $\hat V$ that contain the gauge fields and
transform in the 
adjoint representation of respectively the 
first and second copy of the gauge group.
The field content  of the theory and its transformation properties 
under the $SU(N)\times SU(N)$ gauge and global $SU(2)_\text{L}$ and 
$SU(2)_\text{R}\times U(1)$ R-symmetry groups
is shown in table \ref{tab:fcont}.
\begin{table}[h]
\begin{center}
\setlength{\extrarowheight}{1.5pt}
\begin{tabular}{c|c|c|c|c}
$\text{field}$ & $SU(N)\times SU(N)$ & $SU(2)_\text{L}$ & $SU(2)_\text{R}$ & $U(1)$  \\
\hline
$V$ & $(\text{adj}.,1)$ & $1$ & $1$ & $0$ \\
$\Phi$ & $(\text{adj}.,1)$ & $1$ & $1$ & $1$ \\
$\bar\Phi$ & $(\text{adj}.,1)$ & $1$ & $1$ & $-1$ \\
\hline
$\hat V$ & $(1,\text{adj}.)$ & $1$ & $1$ & $0$\\
$\hat\Phi$ & $(1,\text{adj}.)$ & $1$ & $1$ & $1$\\
$\hat{\bar\Phi}$ & $(1,\text{adj}.)$ & $1$ & $1$ & $-1$\\
\hline
$Q_{\dot I}$ & $(\fun,\afun)$ & $\fun$ & \multirow{2}{*}{$\fun$} & $0$ \\
$\tilde{\bar Q}_{\dot I}$ & $(\fun,\afun)$ & $\fun$ & & $0$ \\
$\bar Q^{\dot I}$ & $(\afun,\fun)$ & $\afun$ & \multirow{2}{*}{$\afun$} & $0$\\
$\tilde Q^{\dot I}$ & $(\afun,\fun)$ & $\afun$ & & $0$ \\
\end{tabular}
\caption{The field content of the \emph{interpolating theory} in terms of $\mathcal{N}=1$ superfields. The fields are grouped according to their gauge group representations. $\fun$ and $\afun$ 
respectively denote fundamental and anti-fundamental representations. The global 
$SU(2)_{\text{R}}$ symmetry that transforms chiral into anti-chiral superfields
is not manifest in the $\mathcal{N}=1$ superspace formulation.\label{tab:fcont}}
\end{center}
\end{table}
We have grouped the fields according to their gauge group representations. 
As in the  $\mathcal{N}=4$ SYM theory the superpotential is 
a cubic interaction of three different types of chiral 
fields, but here it contains a contraction of the $SU(2)_\text{L}$ global 
symmetry index. The additional gauge fixing and ghost terms 
together with the Feynman 
rules required for a three-loop calculation in Fermi-Feynman gauge 
can be found in appendix \ref{app:actfrules}. 
Moreover, in order to regulate the UV divergences of the Feynman diagrams, we 
will use dimensional reduction \cite{Siegel:1979wq} in $D=4-2\varepsilon$
dimensions.

In the following we will only consider the planar limit. 
The respective 't Hooft couplings are given by 
\begin{equation}\label{lambdadef}
\lambda=g_\YM^2N\col\qquad
\hat\lambda=\hat g_\YM^2N\col\qquad
\bar\lambda=g_\YM\hat g_\YM N\col
\end{equation}
where we have also introduced their geometric mean $\bar\lambda$.
For later convenience we also define the rescaled coupling constants
\begin{equation}\label{gdef}
g=\frac{\sqrt{\lambda}}{4\pi}\col\qquad
\hat g=\frac{\sqrt{\hat\lambda}}{4\pi}\col\qquad
\bar g=\frac{\sqrt{\bar\lambda}}{4\pi}
\col\qquad
\end{equation}
and introduce the following ratios
\begin{equation}\label{rhodef}
\rho=\frac{\lambda}{\bar\lambda}
=\frac{g}{\hat g}
\col\qquad
\hat\rho=\rho^{-1}=\frac{\hat\lambda}{\bar\lambda}
=\frac{\hat g}{g}
\pnt
\end{equation}
Even if $\rho\hat\rho=1$, we will display the results in terms of both
parameters for convenience.
The renormalization constant and dilatation operator are then 
given to three loops as
\begin{equation}\label{ZDexp}
\begin{aligned}
\mathcal{Z}&=\unitmatrix+\bar\lambda\mathcal{Z}_1+\bar\lambda^2\mathcal{Z}_2
+\bar\lambda^3\mathcal{Z}_3
+\mathcal{O}(\bar\lambda^4)\\
\mathcal{D}
&=\bar g^2\mathcal{D}_1+\bar g^4\mathcal{D}_2+\bar g^6\mathcal{D}_3
+\mathcal{O}(\bar g^8)
\col
\end{aligned}
\end{equation}
where the $\ell$-loop expansion coefficients $\mathcal{Z}_\ell$ and
$\mathcal{D}_\ell$ are polynomial functions of degree $\ell$ in the
two coupling ratios $\rho$ and $\hat\rho$.

\subsection{Closed chiral subsector}
\label{sec:cop}

From the perspective of $\mathcal{N}=1$ superfields the 
flavor $SU(2)$ subsector of $\mathcal{N}=4$ SYM is chiral, since 
its operators are composed only out of the elementary chiral superfields.
Furthermore, it is closed, i.e.\ operator mixing due to renormalization 
only occurs among its members.
In the \emph{interpolating theory} there exists a closed 
chiral subsector that resembles the flavor $SU(2)$ subsector of 
$\mathcal{N}=4$ SYM theory.  Its composite  chiral single-trace operators 
have the form
\begin{equation}\label{cop}
\mathcal{O}=\tr\big(\Phi\dots\Phi Q_{\dot I}\hat\Phi\dots\hat\Phi\tilde Q^{\dot J}\dots\big)\col
\end{equation}
and they are symmetric traceless in their $SU(2)_\text{L}$ indices
$\dot I$, $\dot J$. These composite operators are highest weight states w.r.t.\ 
the $SU(2)_{\text{R}}$ symmetry, since they only include the 
chiral components of the $SU(2)_{\text{R}}$ doublets of table \ref{tab:fcont}.
The different types of chiral superfields within the operators \eqref{cop}
we call field flavors, including also 
their different $SU(2)_\text{L}$ components.
This is not to be confused with the $SU(N_\text{f})$ flavor of 
$\mathcal{N}=2$ SCQCD.\footnote{In the limit $\hat g_\YM\to0$, there is a symmetry enhancement. 
The respective gauge group becomes global and combines with the 
$SU(2)_\text{L}$ to the $SU(N_\text{f})$ flavor group of $\mathcal{N}=2$ SCQCD.}
The total number of elementary 
fields of the operator is denoted as its length $L$.
The fields $Q_{\dot I}$ and $\tilde Q^{\dot J}$ that transform in the 
bifundamental and anti-bifundamental representation of the gauge group 
$SU(N)\times SU(N)$ are regarded as impurities that have to appear 
pairwise in order to build a gauge invariant single-trace operator. 
Each impurity switches between the two different types of 
fields $\Phi$ and $\hat\Phi$ that transform in the adjoint representation
respectively of the first and second factor of the product gauge group.
The fields $\Phi$ and $\hat\Phi$ are vacuum fields, since
the operators
\begin{equation}\label{groundstates}
\tr(\Phi\dots\Phi)\col\qquad\tr(\hat\Phi\dots\hat\Phi)
\end{equation}
without impurities are protected from quantum corrections 
\cite{Dolan:2002zh,Gadde:2009dj,Gadde:2010zi} and 
hence yield two respective ground states. 

The operators that contain exclusively impurities 
and are symmetric traceless representations of $SU(2)_\text{L}$
read
\begin{equation}\label{impstate}
\tr\big(Q_{\dot I_1}\tilde Q^{\dot J_1}\dots Q_{\dot I_l}\tilde Q^{\dot J_l}\big)
\pnt
\end{equation}
They are also protected from quantum corrections 
\cite{Dolan:2002zh,Gadde:2009dj,Gadde:2010zi}. 
This can be easily understood 
in terms of the generalized finiteness conditions formulated in 
\cite{Sieg:2010tz}: for vanishing flavor subtraces the two fields cannot 
end up in a single anti-chiral vertex, i.e.\ the diagrams cannot contain 
the fundamental building block \eqref{buildingblock}. 
The remaining diagrams are all finite, since they either contain the finite 
self energies, or all their vertices appear in loops.

We close this section with a comparison of this closed chiral subsector with 
the $SU(2)$ subsector of $\mathcal{N}=4$ SYM theory.
The operators \eqref{cop} of the former are free of flavor traces, 
but still contain all different types of 
chiral fields. This is a difference compared to the  
$SU(2)$ subsector, where 
the operators are composed only out of two different kinds of 
chiral field flavors. In $\mathcal{N}=4$ SYM theory, an inclusion of all 
three chiral flavors extends the operator mixing 
at least to the bigger subsector $SU(2|3)$ \cite{Beisert:2003ys}. 
Furthermore, the $SU(2)$ subsector has only one type of vacuum and one 
maximally filled state. They are equivalent, and therefore it suffices 
to consider states in which the
number of one type of fields does not exceed the other type.
In contrast to this, 
the closed chiral subsector of the 
\emph{interpolating theory} has two types of vacua \eqref{cop}, 
and several maximally filled states \eqref{impstate} that are built as 
alternating products of $Q_{\dot I}$ and $\tilde Q^{\dot J}$. It is not 
obvious whether the information obtained from the states
above half-filling is redundant.
Bearing the above comments in mind, the 
closed chiral subsector of the \emph{interpolating theory}
has a richer structure than its $\mathcal{N}=4$ SYM counterpart. 


\subsection{Chiral functions}

As mentioned in the introduction, the chiral functions 
capture the structure of the chiral and anti-chiral field
lines in the Feynman diagrams of $\mathcal{N}=1$ superfields.
They are given as linear combinations of appropriately defined
flavor operations that act on the different field flavors within the 
composite operators. 
The elementary building block of the chiral 
structure of the Feynman diagrams is given by
a chiral and an anti-chiral vertex of the theory that are contracted with 
each other by a single chiral propagator. It includes the 
following combination of flavor operations
\begin{equation}\label{buildingblock}
\begin{aligned}
\settoheight{\eqoff}{$\times$}%
\setlength{\eqoff}{0.5\eqoff}%
\addtolength{\eqoff}{-12\unitlength}%
\raisebox{\eqoff}{%
\fmfframe(-1.5,2)(-12,2){%
\begin{fmfchar*}(20,20)
\chionei[phantom]{dots}{dots}{dots}{dots}{dots}
\fmfiv{label=$\scriptstyle i$,l.a=-90,l.dist=4}{vloc(__v3)}
\fmfiv{label=$\scriptstyle j$,l.a=-90,l.dist=4}{vloc(__v4)}
\end{fmfchar*}}}
=(\perm-\Lam_s-\Top)_{ij}
\col
\end{aligned}
\end{equation}
where the dotted lines denote any configuration of the chiral field flavors
that is admitted by the chiral and anti-chiral vertices given in 
\eqref{cvertices}.
The individual operations in the above expressions act on the 
different combinations of chiral field flavors at legs $(i,j)$ as
\begin{equation}\label{PLTdef}
\begin{aligned}
\perm_{ij}&=
\begin{cases}
\unitmatrix & (\Phi,\Phi)\cup (\hat\Phi,\hat\Phi)  \\
\mathcal{O}_{ij\to j\hat i} & (\Phi,Q_{\dot I})\cup(\hat\Phi,\tilde Q^{\dot I}) \\
\mathcal{O}_{ij\to\hat ji} & (Q_{\dot I},\hat\Phi)\cup(\tilde Q^{\dot I},\Phi) \\
\alpha\unitmatrix & (Q_{\dot I},\tilde Q^{\dot J}) \cup (\tilde Q^{\dot I},Q_{\dot J})
\end{cases}
\col\qquad
(\Lam_s)_{ij}=
\begin{cases}
\rho\unitmatrix & (\Phi,*)\cup(*,\Phi) \\
\hat\rho \unitmatrix & (\hat\Phi,*)\cup(*,\hat\Phi) \\
\alpha\unitmatrix & (Q_{\dot I},\tilde Q^{\dot J}) \cup (\tilde Q^{\dot I},Q_{\dot J})
\end{cases}
\col\\
\Top_{ij}&=
\begin{cases}
0 &  (\Phi,*)\cup(*,\Phi)\cup(\hat\Phi,*)\cup(*,\hat\Phi) \\
\rho\delta_{\dot I}^{\dot J}Q_{\dot K}\tilde Q^{\dot K} & (Q_{\dot I},\tilde Q^{\dot J})\\
\hat\rho\delta^{\dot I}_{\dot J}\tilde Q^{\dot K}Q_{\dot K} & (\tilde Q^{\dot I},Q_{\dot J})
\end{cases}
\col
\end{aligned}
\end{equation}
where we regard the `hat' as an involutive operator, 
i.e.\ $\hat{\hat\Phi}=\Phi$.
Furthermore, we have introduced a parameter $\alpha$ that drops out 
in the combination \eqref{buildingblock}, and hence can be fixed to a 
convenient value. Note that the subscript $s$ of $\Lambda_s$ 
refers to the fact that from the perspective of the fields at 
positions $(i,j)$ the chiral building block is an $s$-channel
diagram.

In terms of the building block \eqref{buildingblock}, the
chiral functions are defined as 
\begin{equation}\label{chifuncdef}
\chiop(a_1,\dots,a_n)
=\sum_{r=0}^{L-1}\prod_{i=1}^n(\perm-\Lam_s-\Top)_{r+a_i\,r+a_i+1}
\col
\end{equation}
where the summation considers the insertion of the interactions 
at all possible positions along the operators \eqref{cop} of length $L$.
The identity operation is given by $\chi()$. When acting on single-trace
operators the cyclic identification  $r+L\simeq r$ is understood.

As mentioned in section \ref{sec:threeloops}, in the 
\emph{interpolating theory} the chiral functions
are subject to deformations caused by additional interactions involving the
vector fields. We therefore have to decompose the chiral functions 
according to the different possible incoming and outgoing
arrangements of the field flavors. 
The individual contributions are abbreviated as
\begin{equation}\label{chifuncimpdef}
\chiimp{i_1,\dots,i_I}{o_1,\dots,o_I}{a_1,\dots,a_n}\col\qquad
\chiimp[\tilde]{i_1,\dots,i_I}{o_1,\dots,o_I}{a_1,\dots,a_n}\pnt
\end{equation}
The two lists $(i_1,\dots,i_I)$ and $(o_1,\dots,o_I)$ 
denote the positions of the total number of $I$ impurities $Q_{\dot I}$ and 
$\tilde Q^{\dot J}$ 
that enter and respectively leave the interaction region, counting 
from left to right the legs involved in the interaction. 
The `tilde' thereby indicates whether the first
encountered impurity respectively is $Q_{\dot I}$ or 
$\tilde Q^{\dot J}$ . Gauge invariance
and planarity then fix uniquely the types of the other chiral fields. 
In terms of the individual contributions \eqref{chifuncimpdef}
the deformed chiral functions have the following form
\begin{equation}\label{defchifunc}
\sum_{\vec i,\vec o}
\big(c_{\vec i}^{\vec o}(\rho,\hat\rho)
\chiimp{\vec i}{\vec o}{a_1,\dots,a_n}
+c_{\vec i}^{\vec o}(\hat\rho,\rho)
\chiimp[\tilde]{\vec i}{\vec o}{a_1,\dots,a_n}\big)\col
\end{equation}
where the sum runs over all possible configurations $\vec i$, $\vec o$
of the ingoing and outgoing impurities. Note that for both 
terms in the sum the coefficients just differ by an exchange 
$\rho\leftrightarrow\hat\rho$. The undeformed chiral functions
\eqref{chifuncdef} are recovered if all coefficients $c_{\vec i}^{\vec o}$ are
set to one.
\begin{figure}[h]
\begin{center}
\setlength{\unit}{0.025\textwidth}%
\psset{xunit=\unit,yunit=\unit,runit=\unit}%
\SpecialCoor%
\begin{pspicture}(0,0)(32,28)%
\scriptsize%
\rput(2,9.5){\rnode{x}{%
\unitlength=0.75mm%
\settoheight{\eqoff}{$\times$}%
\setlength{\eqoff}{0.5\eqoff}%
\addtolength{\eqoff}{-10\unitlength}%
$\raisebox{\eqoff}{%
\fmfframe(-1.5,0)(-12,0){%
\begin{fmfchar*}(20,20)
\chionei[phantom]{dots}{dots}{dots}{dots}{dots}
\end{fmfchar*}}}
=\chiop(1)$}
}%
\rput(9.5,17){\rnode{x*}{%
\unitlength=0.75mm%
\settoheight{\eqoff}{$\times$}%
\setlength{\eqoff}{0.5\eqoff}%
\addtolength{\eqoff}{-10\unitlength}%
$\raisebox{\eqoff}{%
\fmfframe(-1.5,0)(-12,0){%
\begin{fmfchar*}(20,20)
\chionei[phantom]{dots}{dots}{dashes}{dots}{dots}
\end{fmfchar*}}}
=\chiimp{*}{*}{1}+\chiimp[\tilde]{*}{*}{1}$}
}%
\rput(29,2){\rnode{xt}{%
\unitlength=0.75mm%
\settoheight{\eqoff}{$\times$}%
\setlength{\eqoff}{0.5\eqoff}%
\addtolength{\eqoff}{-10\unitlength}%
$\raisebox{\eqoff}{%
\fmfframe(-1.5,0)(-12,0){%
\begin{fmfchar*}(20,20)
\chionei[phantom]{dashes}{dashes}{plain}{dashes}{dashes}
\end{fmfchar*}}}
=\chiimp{1,2}{1,2}{1}+\chiimp[\tilde]{1,2}{1,2}{1}$}
}%
\rput(19,23){\rnode{x1}{%
\unitlength=0.75mm%
\settoheight{\eqoff}{$\times$}%
\setlength{\eqoff}{0.5\eqoff}%
\addtolength{\eqoff}{-10\unitlength}%
$\raisebox{\eqoff}{%
\fmfframe(-1.5,0)(-12,0){%
\begin{fmfchar*}(20,20)
\chionei[phantom]{dashes}{plain}{dashes}{dots}{dots}
\end{fmfchar*}}}
=\chiimp{1}{*}{1}+\chiimp[\tilde]{1}{*}{1}$}
}%
\rput(19,11){\rnode{x2}{%
\unitlength=0.75mm%
\settoheight{\eqoff}{$\times$}%
\setlength{\eqoff}{0.5\eqoff}%
\addtolength{\eqoff}{-10\unitlength}%
$\raisebox{\eqoff}{%
\fmfframe(-1.5,0)(-12,0){%
\begin{fmfchar*}(20,20)
\chionei[phantom]{plain}{dashes}{dashes}{dots}{dots}
\end{fmfchar*}}}
=\chiimp{2}{*}{1}+\chiimp[\tilde]{2}{*}{1}$}
}%
\rput(28.5,26){\rnode{x11}{%
\unitlength=0.75mm%
\settoheight{\eqoff}{$\times$}%
\setlength{\eqoff}{0.5\eqoff}%
\addtolength{\eqoff}{-10\unitlength}%
$\raisebox{\eqoff}{%
\fmfframe(-1.5,0)(-12,0){%
\begin{fmfchar*}(20,20)
\chionei[phantom]{dashes}{plain}{dashes}{dashes}{plain}
\end{fmfchar*}}}
=\chiimp{1}{1}{1}+\chiimp[\tilde]{1}{1}{1}$}
}%
\rput(28.5,20){\rnode{x12}{%
\unitlength=0.75mm%
\settoheight{\eqoff}{$\times$}%
\setlength{\eqoff}{0.5\eqoff}%
\addtolength{\eqoff}{-10\unitlength}%
$\raisebox{\eqoff}{%
\fmfframe(-1.5,0)(-12,0){%
\begin{fmfchar*}(20,20)
\chionei[phantom]{dashes}{plain}{dashes}{plain}{dashes}
\end{fmfchar*}}}
=\chiimp{1}{2}{1}+\chiimp[\tilde]{1}{2}{1}$}
}%
\rput(28.5,14){\rnode{x21}{%
\unitlength=0.75mm%
\settoheight{\eqoff}{$\times$}%
\setlength{\eqoff}{0.5\eqoff}%
\addtolength{\eqoff}{-10\unitlength}%
$\raisebox{\eqoff}{%
\fmfframe(-1.5,0)(-12,0){%
\begin{fmfchar*}(20,20)
\chionei[phantom]{plain}{dashes}{dashes}{dashes}{plain}
\end{fmfchar*}}}
=\chiimp{2}{1}{1}+\chiimp[\tilde]{2}{1}{1}$}
}%
\rput(28.5,8){\rnode{x22}{%
\unitlength=0.75mm%
\settoheight{\eqoff}{$\times$}%
\setlength{\eqoff}{0.5\eqoff}%
\addtolength{\eqoff}{-10\unitlength}%
$\raisebox{\eqoff}{%
\fmfframe(-1.5,0)(-12,0){%
\begin{fmfchar*}(20,20)
\chionei[phantom]{plain}{dashes}{dashes}{plain}{dashes}
\end{fmfchar*}}}
=\chiimp{2}{2}{1}+\chiimp[\tilde]{2}{2}{1}$}
}%
\ncangle[angleA=90,angleB=180,nodesepB=1]{x}{x*}
\ncangle[angleA=-90,angleB=180,nodesepB=1]{x}{xt}
\ncangle[angleA=90,angleB=180,nodesepB=1]{x*}{x1}
\ncangle[angleA=-90,angleB=180,nodesepB=1]{x*}{x2}
\ncangle[angleA=90,angleB=180,nodesepB=1]{x1}{x11}
\ncangle[angleA=-90,angleB=180,nodesepB=1]{x1}{x12}
\ncangle[angleA=90,angleB=180,nodesepB=1]{x2}{x21}
\ncangle[angleA=-90,angleB=180,nodesepB=1]{x2}{x22}
\end{pspicture}
\caption{%
Decomposition of the chiral function $\chi(1)$ into sums of chiral functions 
with fixed position and type of the impurities. The positions of the 
incoming and outgoing impurities are given by lists respectively at the 
lower and upper index positions. We use a $*$ to indicate that 
a single impurity can assume any of the two possible positions at the respective
field lines that form a fork. The type of all impurities is fixed by
indicating the type of the first impurity. We put a tilde on top of the chiral 
function if the first impurity is a $\tilde Q^{\dot I}$ and 
no tilde if it is $Q_{\dot I}$.
\label{fig:chionedec}}
\end{center}
\end{figure}
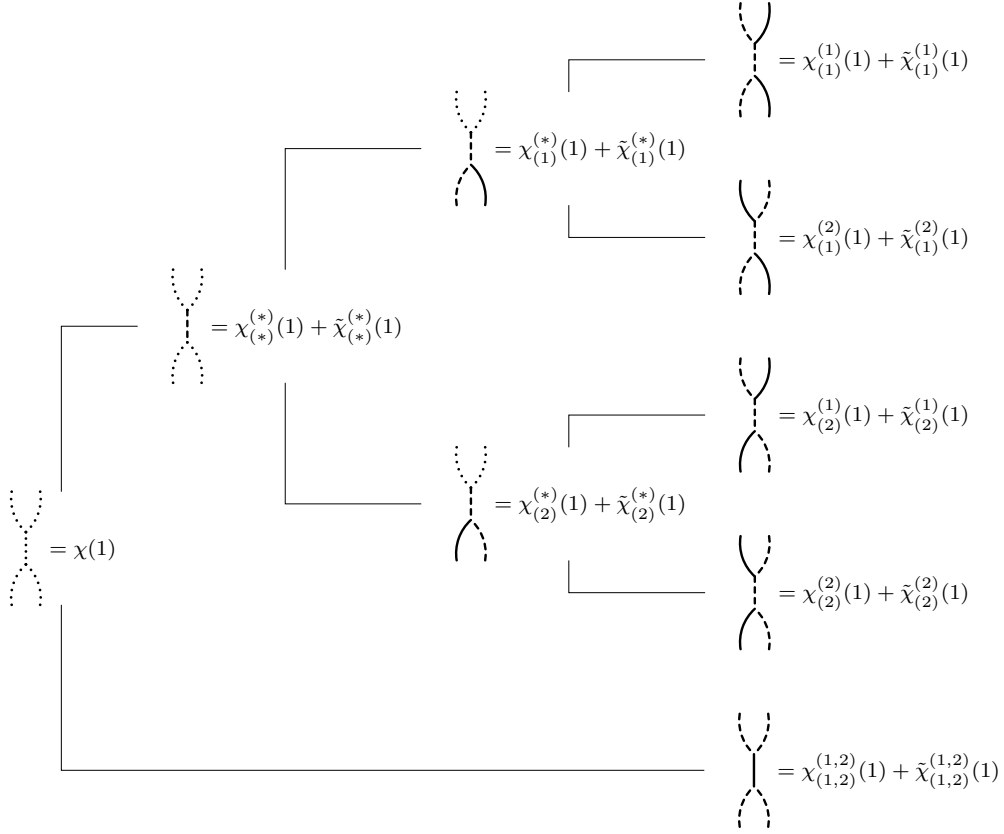
Apart from the summation in \eqref{chifuncdef} the
elementary building block \eqref{buildingblock} is the
simplest chiral function $\chi(1)$, and it is
decomposed into the individual contributions as shown in figure 
\ref{fig:chionedec}.
They read
\begin{equation}\label{chionedec}
\begin{aligned}
\chiimp{1}{1}{1}
&=
-\hat\rho\mathcal{O}_{Q_{\dot I}\hat\Phi\to Q_{\dot I}\hat\Phi}
\col\\
\chiimp{1}{2}{1}
&=
\mathcal{O}_{Q_{\dot I}\hat\Phi\to\Phi Q_{\dot I}}
\col\\
\chiimp{2}{1}{1}
&=
\mathcal{O}_{\Phi Q_{\dot I}\to Q_{\dot I}\hat\Phi}
\col\\
\chiimp{2}{2}{1}
&=
-\rho\mathcal{O}_{\Phi Q_{\dot I}\to\Phi Q_{\dot I}}
\col\\
\chiimp{1,2}{1,2}{1}
&=
-\rho\mathcal{O}_{Q_{\dot I}\tilde Q^{\dot J}\to\delta_{\dot I}^{\dot J}Q_{\dot K}\tilde Q^{\dot K}}\col\\
\end{aligned}
\qquad
\begin{aligned}
\chiimp[\tilde]{1}{1}{1}
&=
-\rho\mathcal{O}_{\tilde Q^{\dot I}\Phi\to\tilde Q^{\dot I}\Phi}
\col\\
\chiimp[\tilde]{1}{2}{1}
&=
\mathcal{O}_{\tilde Q^{\dot I}\Phi\to\hat\Phi\tilde Q^{\dot I}}
\col\\
\chiimp[\tilde]{2}{1}{1}
&=
\mathcal{O}_{\hat\Phi\tilde Q^{\dot I}\to\tilde Q^{\dot I}\Phi}
\col\\
\chiimp[\tilde]{2}{2}{1}
&=-\hat\rho\mathcal{O}_{\hat\Phi\tilde Q^{\dot I}\to\hat\Phi\tilde Q^{\dot I}}
\col\\
\chiimp[\tilde]{1,2}{1,2}{1}
&=-\hat\rho
\mathcal{O}_{\tilde Q^{\dot I}Q_{\dot J}\to\delta^{\dot I}_{\dot J}\tilde Q^{\dot K}Q_{\dot K}}
\col
\end{aligned}
\end{equation}
where the operators $\mathcal{O}_{A\to B}$ obey the Leibnitz rule when 
they replace the field configurations $A$ by the new configurations $B$.
For completeness, in the last line, we have included the contributions
that involve the trace operator
$\Top$ of \eqref{PLTdef}. They yield zero when applied to the 
composite operators \eqref{cop} of the closed chiral subsector.

We conclude this section by summarizing the action of Hermitean conjugation 
on the chiral functions.
The relations read
\begin{equation}\label{hcchi}
\begin{aligned}
\chiop(a_1,\dots,a_n)^\dagger&=\chi(a_n,\dots,a_1)\col\\
\chiimp{i_1,\dots,i_I}{o_1,\dots,o_I}{a_1,\dots,a_n}^\dagger
&=\chiimp{o_1,\dots,o_I}{i_1,\dots,i_I}{a_n,\dots,a_1}\col\\
\chiimp[\tilde]{i_1,\dots,i_I}{o_1,\dots,o_I}{a_1,\dots,a_n}^\dagger
&=\chiimp[\tilde]{o_1,\dots,o_I}{i_1,\dots,i_I}{a_n,\dots,a_1}\col\\
\end{aligned}
\end{equation}
i.e.\ under Hermitean conjugation the list of arguments is 
reversed and the lists of the 
incoming and outgoing impurities are interchanged.

\section{One- and two-loop dilatation operator}
\label{sec:onetwoloopD}

In this section we calculate the one- and two-loop contribution
to the dilatation operator of the closed chiral subsector introduced in 
section \ref{sec:cop}. 
At two loops the vanishing of flavor subtraces 
will allow us express the result exclusively in terms
of \eqref{chifuncdef}, even if the contributions involving flavor traces
undergo deformations.
In the $\mathcal{N}=1$ superfield formalism the one-loop self energies 
are identically zero, and the higher-loop chiral self energies are finite.
Due to the finiteness conditions of \cite{Sieg:2010tz},
the appearing diagrams can only have overall UV divergences if they contain
at least one chiral vertex that is not part of any loop. Therefore, all 
diagrams with only gauge-matter interactions or self-energy corrections 
are finite. 

For the one-loop dilatation operator in the closed subsector, 
we have to evaluate the following diagram
\begin{equation}
\begin{aligned}
\settoheight{\eqoff}{$\times$}%
\setlength{\eqoff}{0.5\eqoff}%
\addtolength{\eqoff}{-11\unitlength}%
\raisebox{\eqoff}{%
\fmfframe(-1,1)(-11,1){%
\begin{fmfchar*}(20,20)
\chionei{dots}{dots}{dots}{dots}{dots}
\end{fmfchar*}}}
&=\bar\lambda I_1\chiop(1)
\pnt
\end{aligned}
\end{equation}
The one-loop contribution to the renormalization constant is given as the 
pole part of the above result. It reads
\begin{equation}\label{Z1}
\begin{aligned}
\mathcal{Z}_{1}
&=-\mathcal{I}_1\chiop(1)
\col
\end{aligned}
\end{equation}
where $\mathcal{I}_1$=$\Kop(I_1)$ is the pole part extracted by the 
operation $\Kop$ from the integral $I_1$ in \eqref{Integralexpr}.

In order to determine the dilatation operator at two loops, 
we have to evaluate the diagrams that have interaction range $R=2,3$, 
i.e.\ in which two or three neighbouring field lines of the 
composite operator interact with each other. 
The maximum range $R=3$ diagrams are determined as
\begin{equation}\label{twoloopR3}
\begin{aligned}
\settoheight{\eqoff}{$\times$}%
\setlength{\eqoff}{0.5\eqoff}%
\addtolength{\eqoff}{-11\unitlength}%
\raisebox{\eqoff}{%
\fmfframe(-1,1)(-6,1){%
\begin{fmfchar*}(20,20)
\chionetwoi{dots}{dots}{dots}{dots}{dots}{dots}{dots}{dots}{dots}
\end{fmfchar*}}}
&=\bar\lambda^2I_2\chiop(1,2)
\col\qquad
&\settoheight{\eqoff}{$\times$}%
\setlength{\eqoff}{0.5\eqoff}%
\addtolength{\eqoff}{-11\unitlength}%
\raisebox{\eqoff}{%
\fmfframe(-1,1)(-6,1){%
\begin{fmfchar*}(20,20)
\chitwoonei{dots}{dots}{dots}{dots}{dots}{dots}{dots}{dots}{dots}
\end{fmfchar*}}}
&=\bar\lambda^2I_2\chiop(2,1)
\col
\end{aligned}
\end{equation}
The $R=2$ diagrams can be collectively written as a single diagram
that contains  as subdiagram the one-loop correction 
of the anti-chiral vertex given in \eqref{ccconeloop}.
We introduce generic relative couplings $\rho_i$,
$i=1,2,3$ for the individual faces of the diagram and obtain 
\begin{equation}\label{twoloopR2}
\begin{aligned}
\settoheight{\eqoff}{$\times$}%
\setlength{\eqoff}{0.5\eqoff}%
\addtolength{\eqoff}{-12\unitlength}%
\raisebox{\eqoff}{%
\fmfframe(-0.5,2)(-10.5,2){%
\begin{fmfchar*}(20,20)
\chionei{dots}{dots}{dots}{dots}{dots}
\fmfcmd{fill fullcircle scaled 8 shifted vloc(__vc2) withcolor black ;}
\fmfv{label=$\scriptstyle\textcolor{white}{1}$,l.dist=0}{vc2}
\fmfiv{label=$\scriptstyle\textcolor{black}{1}$,l.dist=0}{(0.25w,0.125h)}
\fmfiv{label=$\scriptstyle\textcolor{black}{2}$,l.dist=0}{(0.0625w,0.5h)}
\fmfiv{label=$\scriptstyle\textcolor{black}{3}$,l.dist=0}{(0.4375w,0.5h)}
\end{fmfchar*}}}
&=
-\bar\lambda^2I_2(\rho_2+\rho_3)\chiimp[\overset{\tau}]{\vec i}{\vec o}{1}
\pnt
\end{aligned}
\end{equation}
Each possible ingoing and outgoing flavor combination is encoded in terms of 
the 
respective positions $\vec i$, $\vec o$ of the impurities and by $\tau$,
the type of the first encountered impurity. Each such configuration
determines the $\rho_i$ in terms 
of $\rho$, $\hat\rho$.
With the explicit expressions for the chiral functions as given 
in \eqref{chionedec} it is easy to conclude from 
figure \ref{fig:chionedec} that in the closed chiral subsector 
the field separating faces $2$ and $3$ in the diagram 
has to be one of the bifundamentals. This means that the sum 
$\rho_2+\rho_3$ yields $\rho+\hat\rho$ for all contributions 
in this subsector. This is not 
true for the diagrams in which two impurities interact, but as
seen from \eqref{chionedec} such an interaction is possible only
for operators with non-vanishing flavor subtraces that are not members
of the closed chiral subsector.
We can therefore express the result in terms of $\chi(1)$, even if 
this does not hold for the trace terms. 
Summing up the contributions \eqref{twoloopR3} and \eqref{twoloopR2}, we 
obtain for the two-loop renormalization constant
\begin{equation}\label{Z2}
\begin{aligned}
\mathcal{Z}_2
&=-\mathcal{I}_2
\big[\chiop(1,2)+\chiop(2,1)-(\rho+\hat\rho)\chiop(1)\big]
\col
\end{aligned}
\end{equation}
where $\mathcal{I}_2=\Kop\Rop(I_2)$ is the overall UV divergence that
is extracted by $\Kop$ after the subdivergence has been removed by the 
operation $\Rop$. The expression of the two-loop integral is 
given in \eqref{Integralexpr}.

According to the definition \eqref{DinZ}, we
multiply the $\frac{1}{\varepsilon}$ pole of \eqref{Z1} by $2$, and
the one of \eqref{Z2} by $4$ and obtain for the one- and two-loop dilatation
operator the following expressions
\begin{equation}\label{D1D2}
\begin{aligned}
\mathcal{D}_1&=-2\chiop(1)\col\\
\mathcal{D}_2&=-2\big(\chiop(1,2)+\chiop(2,1)\big)+2(\rho+\hat\rho)\chiop(1)
\pnt
\end{aligned}
\end{equation}
At $\rho=\hat\rho=1$, where the \emph{interpolating theory} becomes the 
$\mathds{Z}_2$ orbifold
of the $\mathcal{N}=4$ SYM theory, 
this result is identical to the one in $\mathcal{N}=4$ SYM theory, apart from
a straightforward identification of the chiral functions in both theories.

\section{Three-loop dilatation operator}
\label{sec:threeloopD}

In this section we calculate the three-loop contribution to the 
renormalization constant and the dilatation operator, 
following closely the analysis of \cite{Sieg:2010tz}.
We classify the underlying Feynman diagrams according to their  
range $R$, defined by the number of interacting elementary fields, 
and according to their chiral structure as captured in terms of the chiral 
functions \eqref{chifuncdef} and \eqref{defchifunc}.

\subsection{Maximum range diagrams}

Following \cite{Sieg:2010tz}, the chiral maximum-range diagrams 
are evaluated as
\begin{equation}\label{threeloopR4}
\begin{aligned}
\settoheight{\eqoff}{$\times$}%
\setlength{\eqoff}{0.5\eqoff}%
\addtolength{\eqoff}{-12\unitlength}%
\raisebox{\eqoff}{%
\fmfframe(-1,2)(-1,2){%
\begin{fmfchar*}(20,20)
\chionetwothreei{dots}{dots}{dots}{dots}{dots}{dots}{dots}{dots}
\end{fmfchar*}}}
&=\bar\lambda^3I_3\chiop(1,2,3)
\col\quad
&\settoheight{\eqoff}{$\times$}%
\setlength{\eqoff}{0.5\eqoff}%
\addtolength{\eqoff}{-12\unitlength}%
\raisebox{\eqoff}{%
\fmfframe(-1,2)(-1,2){%
\begin{fmfchar*}(20,20)
\chithreetwoonei{dots}{dots}{dots}{dots}{dots}{dots}{dots}{dots}
\end{fmfchar*}}}
&=\bar\lambda^3I_3\chiop(3,2,1)
\col\qquad\\
\settoheight{\eqoff}{$\times$}%
\setlength{\eqoff}{0.5\eqoff}%
\addtolength{\eqoff}{-12\unitlength}%
\raisebox{\eqoff}{%
\fmfframe(-1,2)(-1,2){%
\begin{fmfchar*}(20,20)
\chitwoonethreei{dots}{dots}{dots}{dots}{dots}{dots}{dots}{dots}
\end{fmfchar*}}}
&=\bar\lambda^3I_{3\mathbf{bb}}\chiop(2,1,3)
\col\quad
&\settoheight{\eqoff}{$\times$}%
\setlength{\eqoff}{0.5\eqoff}%
\addtolength{\eqoff}{-12\unitlength}%
\raisebox{\eqoff}{%
\fmfframe(-1,2)(-1,2){%
\begin{fmfchar*}(20,20)
\chionethreetwoi{dots}{dots}{dots}{dots}{dots}{dots}{dots}{dots}
\end{fmfchar*}}}
&=\bar\lambda^3I_{3\mathbf{b}}\chiop(1,3,2)
\pnt
\end{aligned}
\end{equation}
The non-chiral maximum range diagrams involve one vector field that
leads to a deformation of the respective chiral function. For the 
contributions that have an overall UV divergence and 
that are free of flavor subtraces, one finds
\begin{equation}
\begin{aligned}\label{threeloopchi13}
&
\settoheight{\eqoff}{$\times$}%
\setlength{\eqoff}{0.5\eqoff}%
\addtolength{\eqoff}{-12\unitlength}%
\raisebox{\eqoff}{%
\fmfframe(-0.5,2)(-0.5,2){%
\begin{fmfchar*}(20,20)
\chionethreei{dots}{dots}{dots}{dots}{dots}{dots}{dots}{dots}
\fmfi{photon}{vlm3--vrm4}
\end{fmfchar*}}}
=
\settoheight{\eqoff}{$\times$}%
\setlength{\eqoff}{0.5\eqoff}%
\addtolength{\eqoff}{-12\unitlength}%
\raisebox{\eqoff}{%
\fmfframe(-0.5,2)(-0.5,2){%
\begin{fmfchar*}(20,20)
\chionethreei{dots}{dots}{dots}{dots}{dots}{dots}{dots}{dots}
\fmfi{photon}{vlm5--vrm3}
\end{fmfchar*}}}
=\bar\lambda^3I_3\big[
\hat\rho\chiimp{*,*}{*,*}{1,3}
+\rho\chiimp[\tilde]{*,*}{*,*}{1,3}\big]
\col\\
&
\settoheight{\eqoff}{$\times$}%
\setlength{\eqoff}{0.5\eqoff}%
\addtolength{\eqoff}{-12\unitlength}%
\raisebox{\eqoff}{%
\fmfframe(-0.5,2)(-0.5,2){%
\begin{fmfchar*}(20,20)
\chionethreei{dots}{dots}{dots}{dots}{dots}{dots}{dots}{dots}
\fmfi{photon}{vlm5--vrm4}
\end{fmfchar*}}}
=-2\bar\lambda^3(I_3+I_{32\mathbf{t}})\big[
\hat\rho\chiimp{*,*}{*,*}{1,3}
+\rho\chiimp[\tilde]{*,*}{*,*}{1,3}\big]
\col\\
&
\settoheight{\eqoff}{$\times$}%
\setlength{\eqoff}{0.5\eqoff}%
\addtolength{\eqoff}{-12\unitlength}%
\raisebox{\eqoff}{%
\fmfframe(-0.5,2)(-0.5,2){%
\begin{fmfchar*}(20,20)
\chionethreei{dots}{dots}{dots}{dots}{dots}{dots}{dots}{dots}
\fmfcmd{fill fullcircle scaled 8 shifted vloc(__vc2) withcolor black ;}
\fmfiv{plain,label=$\scriptstyle\textcolor{white}{1}$,l.dist=0}{vloc(__vc2)}
\end{fmfchar*}}}
=
\settoheight{\eqoff}{$\times$}%
\setlength{\eqoff}{0.5\eqoff}%
\addtolength{\eqoff}{-12\unitlength}%
\raisebox{\eqoff}{%
\fmfframe(-0.5,2)(-0.5,2){%
\begin{fmfchar*}(20,20)
\chionethreei{dots}{dots}{dots}{dots}{dots}{dots}{dots}{dots}
\fmfcmd{fill fullcircle scaled 8 shifted vloc(__vc4) withcolor black ;}
\fmfiv{plain,label=$\scriptstyle\textcolor{white}{1}$,l.dist=0}{vloc(__vc4)}
\end{fmfchar*}}}
=
-(\rho+\hat\rho)\lambda^3I_1I_2\chi(1,3)
\pnt
\end{aligned}
\end{equation}

The sum of the maximum range diagrams then contributes to the 
renormalization constant as
\begin{equation}
\begin{aligned}\label{Z3R4}
\mathcal{Z}_{3,R=4}
&=-\mathcal{I}_3\big(\chiop(1,2,3)+\chiop(3,2,1)\big)
-\mathcal{I}_{3\mathbf{bb}}\chiop(2,1,3)-\mathcal{I}_{3\mathbf{b}}\chiop(1,3,2)\\
&\phantom{{}={}}
+2\mathcal{I}_{32\mathbf{t}}\big(\hat\rho\chiimp{*,*}{*,*}{1,3}
+\rho\chiimp[\tilde]{*,*}{*,*}{1,3}\big)
-2\mathcal{I}_1\mathcal{I}_2(\rho+\hat\rho)\chiop(1,3)
\col
\end{aligned}
\end{equation}
where as in \cite{Sieg:2010tz} we have also added the respective contribution
with chiral function $\chi(1,3)$ that only involves higher order poles
in $\varepsilon$ and hence does not contribute to the dilatation operator. 
It is required for the check that these poles indeed cancel 
in the logarithm of the renormalization constant.
The first contribution in the second line contains a 
deformation of the chiral function $\chi(1,3)$ that has the form 
\eqref{defchifunc}.
It only contributes to the scattering of impurities.

\subsection{Next-to-maximum range diagrams}

The contributions from chiral next-to-maximum range diagrams are given by
\begin{equation}\label{threeloopR3k3chiral}
\begin{aligned}
\settoheight{\eqoff}{$\times$}%
\setlength{\eqoff}{0.5\eqoff}%
\addtolength{\eqoff}{-11\unitlength}%
\raisebox{\eqoff}{%
\fmfframe(-1,1)(-6,1){%
\begin{fmfchar*}(20,20)
\chionetwoonei
\end{fmfchar*}}}
&=\bar\lambda^3I_3\chiop(1,2,1)
\col\qquad
\settoheight{\eqoff}{$\times$}%
\setlength{\eqoff}{0.5\eqoff}%
\addtolength{\eqoff}{-11\unitlength}%
\raisebox{\eqoff}{%
\fmfframe(-1,1)(-6,1){%
\begin{fmfchar*}(20,20)
\chitwoonetwoi
\end{fmfchar*}}}
=\bar\lambda^3I_3\chiop(2,1,2)
\pnt
\end{aligned}
\end{equation}
In the closed chiral subsector, i.e.\ for the terms not involving 
the flavor trace operator, the sum of their chiral functions can be replaced
as
\begin{equation}
\begin{aligned}
\chiop(1,2,1)+\chiop(2,1,2)
&\to(\rho^2+\hat\rho^2)\chiop(1)
\pnt
\end{aligned}
\end{equation}

The results for the only next-to-maximum range diagrams that involve a 
single vector field line are summarized by introducing 
generic coupling ratios $\rho_i$, $i=1,\dots,5$,
for the different faces of the diagrams. We find
\begin{equation}\label{threeloopR3k3}
\begin{aligned}
\settoheight{\eqoff}{$\times$}%
\setlength{\eqoff}{0.5\eqoff}%
\addtolength{\eqoff}{-12\unitlength}%
\raisebox{\eqoff}{%
\fmfframe(-0.5,2)(-5.5,2){%
\begin{fmfchar*}(20,20)
\chionetwoi{dots}{dots}{dots}{dots}{dots}{dots}{dots}{dots}{dots}
\fmfiv{label=$\scriptstyle\textcolor{black}{1}$,l.dist=0}{(0.125w,0.5h)}
\fmfiv{label=$\scriptstyle\textcolor{black}{2}$,l.dist=0}{(0.25w,0.175h)}
\fmfiv{label=$\scriptstyle\textcolor{black}{3}$,l.dist=0}{(0.5w,0.175h)}
\fmfiv{label=$\scriptstyle\textcolor{black}{4}$,l.dist=0}{(0.25w,0.825h)}
\fmfiv{label=$\scriptstyle\textcolor{black}{5}$,l.dist=0}{(0.5w,0.825h)}
\fmfiv{label=$\scriptstyle\textcolor{black}{6}$,l.dist=0}{(0.625w,0.5h)}
\end{fmfchar*}}}
+
\settoheight{\eqoff}{$\times$}%
\setlength{\eqoff}{0.5\eqoff}%
\addtolength{\eqoff}{-3.75\unitlength}%
\raisebox{\eqoff}{%
\fmfframe(1,0)(1,0){%
\begin{fmfchar*}(10,7.5)
\fmfleft{v1}
\fmfright{v2}
\fmfforce{0.0625w,0.5h}{v1}
\fmfforce{0.9375w,0.5h}{v2}
\fmf{photon}{v1,v2}
\end{fmfchar*}}}
&=\bar\lambda^3\big(-I_3(\rho_1+\rho_2+\rho_5+\rho_6)
+I_{3\mathbf{t}}(\rho_2-\rho_3)\big)\chiimp[\overset{\tau}]{\vec i}{\vec o}{1,2}
\col\\
\settoheight{\eqoff}{$\times$}%
\setlength{\eqoff}{0.5\eqoff}%
\addtolength{\eqoff}{-12\unitlength}%
\raisebox{\eqoff}{%
\fmfframe(-0.5,2)(-5.5,2){%
\begin{fmfchar*}(20,20)
\chitwoonei{dots}{dots}{dots}{dots}{dots}{dots}{dots}{dots}{dots}
\fmfiv{label=$\scriptstyle\textcolor{black}{1}$,l.dist=0}{(0.125w,0.5h)}
\fmfiv{label=$\scriptstyle\textcolor{black}{2}$,l.dist=0}{(0.25w,0.175h)}
\fmfiv{label=$\scriptstyle\textcolor{black}{3}$,l.dist=0}{(0.5w,0.175h)}
\fmfiv{label=$\scriptstyle\textcolor{black}{4}$,l.dist=0}{(0.25w,0.825h)}
\fmfiv{label=$\scriptstyle\textcolor{black}{5}$,l.dist=0}{(0.5w,0.825h)}
\fmfiv{label=$\scriptstyle\textcolor{black}{6}$,l.dist=0}{(0.625w,0.5h)}
\end{fmfchar*}}}
+
\settoheight{\eqoff}{$\times$}%
\setlength{\eqoff}{0.5\eqoff}%
\addtolength{\eqoff}{-3.75\unitlength}%
\raisebox{\eqoff}{%
\fmfframe(1,0)(1,0){%
\begin{fmfchar*}(10,7.5)
\fmfleft{v1}
\fmfright{v2}
\fmfforce{0.0625w,0.5h}{v1}
\fmfforce{0.9375w,0.5h}{v2}
\fmf{photon}{v1,v2}
\end{fmfchar*}}}
&=\bar\lambda^3\big(-I_3(\rho_1+\rho_3+\rho_4+\rho_6)
+I_{3\mathbf{t}}(\rho_3-\rho_2)\big)\chiimp[\overset{\tau}]{\vec i}{\vec o}{2,1}
\col
\end{aligned}
\end{equation}
where the contributions with the integral $I_{3\mathbf{t}}$ come from 
the diagrams displayed in figure \ref{fig:threeloopdefchi1}. The individual 
diagrams and their expressions in the $\mathcal{N}=4$ SYM theory can be found 
in \cite{Sieg:2010tz}.
We have to evaluate the above expression for all possible combinations of 
ingoing and outgoing field flavors that are encoded in terms of the positions
$\vec i$, $\vec o$ and type $\tau$ of the respective ingoing and outgoing 
impurities. Thereby, configurations that lead to flavor traces are neglected. 
The gauge structure of a given combination fixes each $\rho_i$
to either $\rho$ or $\hat\rho$.

There are further contributions from next-to-maximum range diagrams that 
involve two vector field lines. Following again closely the analysis in 
\cite{Sieg:2010tz}, the remaining contributions with an overall 
UV divergence are given by
\begin{equation}
\begin{aligned}\label{threeloopR3k2}
\settoheight{\eqoff}{$\times$}%
\setlength{\eqoff}{0.5\eqoff}%
\addtolength{\eqoff}{-12\unitlength}%
\raisebox{\eqoff}{%
\fmfframe(-0.5,2)(-5.5,2){%
\begin{fmfchar*}(20,20)
\chionegi{dots}{dots}{dots}{dots}{dots}{dots}
\fmfi{photon}{vu3--vgm3}
\fmfi{photon}{vd3--vgm3}
\end{fmfchar*}}}
&=-\frac{1}{2}\bar\lambda^3I_3\big(
\hat\rho^2\chiimp{*}{*}{1}+\rho^2\chiimp[\tilde]{*}{*}{1}
\big)
\col\qquad
\settoheight{\eqoff}{$\times$}%
\setlength{\eqoff}{0.5\eqoff}%
\addtolength{\eqoff}{-12\unitlength}%
\raisebox{\eqoff}{%
\fmfframe(-0.5,2)(-5.5,2){%
\begin{fmfchar*}(20,20)
\chionegi{dots}{dots}{dots}{dots}{dots}{dots}
\fmfi{photon}{vm3--vg2}
\fmfi{photon}{vm5--vg2}
\end{fmfchar*}}}
=
\frac{1}{2}\bar\lambda^3I_3\big(
\hat\rho^2\chiimp{*}{*}{1}+\rho^2\chiimp[\tilde]{*}{*}{1}
\big)
\col\\
\settoheight{\eqoff}{$\times$}%
\setlength{\eqoff}{0.5\eqoff}%
\addtolength{\eqoff}{-12\unitlength}%
\raisebox{\eqoff}{%
\fmfframe(-0.5,2)(-5.5,2){%
\begin{fmfchar*}(20,20)
\chionegi{dots}{dots}{dots}{dots}{dots}{dots}
\fmfi{photon}{vu5--vgm5}
\fmfi{photon}{vd5--vgm5}
\end{fmfchar*}}}
&=-\frac{1}{2}\bar\lambda^3I_{3\mathbf{t}}\big(
\hat\rho^2\chiimp{*}{*}{1}+\rho^2\chiimp[\tilde]{*}{*}{1}\big)
\col\\
\settoheight{\eqoff}{$\times$}%
\setlength{\eqoff}{0.5\eqoff}%
\addtolength{\eqoff}{-12\unitlength}%
\raisebox{\eqoff}{%
\fmfframe(-0.5,2)(-5.5,2){%
\begin{fmfchar*}(20,20)
\gchionei{dots}{dots}{dots}{dots}{dots}{dots}
\fmfi{photon}{vu3--vgm3}
\fmfi{photon}{vd3--vgm3}
\end{fmfchar*}}}
&=-\frac{1}{2}\bar\lambda^3I_3\big(
\rho^2\chiimp{*}{*}{1}+\hat\rho^2\chiimp[\tilde]{*}{*}{1}
\big)
\col\qquad
\settoheight{\eqoff}{$\times$}%
\setlength{\eqoff}{0.5\eqoff}%
\addtolength{\eqoff}{-12\unitlength}%
\raisebox{\eqoff}{%
\fmfframe(-0.5,2)(-5.5,2){%
\begin{fmfchar*}(20,20)
\gchionei{dots}{dots}{dots}{dots}{dots}{dots}
\fmfi{photon}{vm3--vg2}
\fmfi{photon}{vm4--vg2}
\end{fmfchar*}}}
=
\frac{1}{2}\bar\lambda^3I_3\big(
\rho^2\chiimp{*}{*}{1}+\hat\rho^2\chiimp[\tilde]{*}{*}{1}
\big)
\col\\
\settoheight{\eqoff}{$\times$}%
\setlength{\eqoff}{0.5\eqoff}%
\addtolength{\eqoff}{-12\unitlength}%
\raisebox{\eqoff}{%
\fmfframe(-0.5,2)(-5.5,2){%
\begin{fmfchar*}(20,20)
\gchionei{dots}{dots}{dots}{dots}{dots}{dots}
\fmfi{photon}{vu4--vgm5}
\fmfi{photon}{vd4--vgm5}
\end{fmfchar*}}}
&=-\frac{1}{2}\bar\lambda^3I_{3\mathbf{t}}\big(
\rho^2\chiimp{*}{*}{1}+\hat\rho^2\chiimp[\tilde]{*}{*}{1}\big)
\pnt
\end{aligned}
\end{equation}
The results from the individual diagrams contain 
different relative coefficients and hence seem to combine into 
a deformed chiral function. However, they in fact sum up to the 
undeformed chiral function $\chi(1)$ with a common global factor
as long as contributions from flavor subtraces are of no concern.

From \eqref{threeloopR3k3chiral}, \eqref{threeloopR3k3} and 
\eqref{threeloopR3k2} we determine the contribution of all range $R=3$ diagrams
to the renormalization constant as
\begin{equation}
\begin{aligned}\label{Z3R3}
\mathcal{Z}_{3,R=3}
&=
-\mathcal{I}_3\big(\chiop(1,2,1)+\chiop(2,1,2)\big)
+2(\rho+\hat\rho)\mathcal{I}_3\big(\chiop(1,2)+\chiop(2,1)\big)
\\
&\phantom{{}={}}
-(\rho-\hat\rho)\mathcal{I}_{3\mathbf{t}}\big(
\chiimp{2}{*}{1,2}
-\chiimp{1,2}{*,3}{1,2}
-\chiimp{2}{*}{2,1}
-\chiimp{2,3}{1,*}{2,1}
\\
&\phantom{{}={}-(\rho-\hat\rho)\mathcal{I}_{3\mathbf{t}}\big(}
-\chiimp[\tilde]{2}{*}{1,2}
+\chiimp[\tilde]{1,2}{*,3}{1,2}
+\chiimp[\tilde]{2}{*}{2,1}
+\chiimp[\tilde]{2,3}{1,*}{2,1}
\big)\\
&\phantom{{}={}}
+\frac{1}{2}\mathcal{I}_{3\mathbf{t}}(\rho^2+\hat\rho^2)\chiop(1)
\pnt
\end{aligned}
\end{equation}
The contributions with individual configurations of the impurities 
in the second and third line have the form \eqref{defchifunc} of
a deformed chiral function. They are due to the second terms
in \eqref{threeloopR3k3} that originate from the diagrams displayed 
in figure \ref{fig:threeloopdefchi1}.

\subsection{Nearest-neighbour interactions}

The nearest-neighbour interactions either involve the finite two-loop
corrections of the anti-chiral vertex  
or the finite two-loop self energies of the chiral fields.
Introducing again generic coupling ratios $\rho_i$, $i=1,2,3$ for the 
different faces in the diagram, we obtain from \eqref{ccctwoloop}
the result
\begin{equation}\label{threeloopR2ccc}
\begin{aligned}
\settoheight{\eqoff}{$\times$}%
\setlength{\eqoff}{0.5\eqoff}%
\addtolength{\eqoff}{-12\unitlength}%
\raisebox{\eqoff}{%
\fmfframe(-0.5,2)(-10.5,2){%
\begin{fmfchar*}(20,20)
\chionei{dots}{dots}{dots}{dots}{dots}
\fmfcmd{fill fullcircle scaled 8 shifted vloc(__vc2) withcolor black ;}
\fmfv{label=$\scriptstyle\textcolor{white}{2}$,l.dist=0}{vc2}
\fmfiv{label=$\scriptstyle\textcolor{black}{1}$,l.dist=0}{(0.25w,0.125h)}
\fmfiv{label=$\scriptstyle\textcolor{black}{2}$,l.dist=0}{(0.0625w,0.5h)}
\fmfiv{label=$\scriptstyle\textcolor{black}{3}$,l.dist=0}{(0.4375w,0.5h)}
\end{fmfchar*}}}
&=\bar\lambda^3\Big(
I_3(\rho_2+\rho_3)^2
-\frac{1}{2}I_{3\mathbf{t}}(
2\rho_1(\rho_2+\rho_3)
+\rho_2^2
+\rho_3^2)\Big)\chiimp[\overset{\tau}]{\vec i}{\vec o}{1}
\pnt
\end{aligned}
\end{equation}
Like the respective two-loop expression \eqref{twoloopR2} also this
result has to be evaluated for all possible field flavor arrangements, 
i.e.\ positions $\vec i$, $\vec o$ and type $\tau$ of the respective 
in- and outgoing impurities in the closed chiral subsector. 
The second type of diagram involves the finite two-loop self energies of the 
different chiral field flavors. Since according to \eqref{sigmac} they are 
the same for all flavors, we find immediately
\begin{equation}\label{threeloopR2se}
\begin{aligned}
\settoheight{\eqoff}{$\times$}%
\setlength{\eqoff}{0.5\eqoff}%
\addtolength{\eqoff}{-12\unitlength}%
\raisebox{\eqoff}{%
\fmfframe(-0.5,2)(-10.5,2){%
\begin{fmfchar*}(20,20)
\chionei{dots}{dots}{dots}{dots}{dots}
\fmfcmd{fill fullcircle scaled 8 shifted vm4 withcolor black ;}
\fmfiv{label=$\scriptstyle\textcolor{white}{2}$,l.dist=0}{vm4}
\end{fmfchar*}}}
&=
\settoheight{\eqoff}{$\times$}%
\setlength{\eqoff}{0.5\eqoff}%
\addtolength{\eqoff}{-12\unitlength}%
\raisebox{\eqoff}{%
\fmfframe(-0.5,2)(-10.5,2){%
\begin{fmfchar*}(20,20)
\chionei{dots}{dots}{dots}{dots}{dots}
\fmfcmd{fill fullcircle scaled 8 shifted vm5 withcolor black ;}
\fmfiv{label=$\scriptstyle\textcolor{white}{2}$,l.dist=0}{vm5}
\end{fmfchar*}}}
=2\bar\lambda^3I_{3\mathbf{t}}\chiop(1)
\col\\
\end{aligned}
\end{equation}
The sum of all $R=2$ diagrams yields the following contribution to the
renormalization constant
\begin{equation}
\begin{aligned}\label{Z3R2}
\mathcal{Z}_{3,R=2}
&=\Big(-\mathcal{I}_3(\rho+\hat\rho)^2
+\frac{1}{2}\mathcal{I}_{3\mathbf{t}}(\rho^2+\hat\rho^2-4)\Big)
\chiop(1)\\
&\phantom{{}={}}
+\mathcal{I}_{3\mathbf{t}}
\big((\hat\rho^2-1)\big(\chiimp{1}{*}{1}+\chiimp[\tilde]{2}{*}{1}\big)
+(\rho^2-1)\big(\chiimp{2}{*}{1}+\chiimp[\tilde]{1}{*}{1}\big)\big)
\pnt
\end{aligned}
\end{equation}
By examining the individual diagrams of the two-loop chiral vertex 
correction in \eqref{threeloopR2ccc}, we find that the diagrams 
displayed in figure \ref{fig:threeloopdefchi2} generate the 
contributions with deformed chiral functions \eqref{defchifunc} in 
\eqref{Z3R2}.

\subsection{Result}
\label{sec:result}

The renormalization constant is a sum of \eqref{Z3R4}, 
\eqref{Z3R3} and \eqref{Z3R2}, 
and 
it is given by
\begin{equation}
\begin{aligned}
\mathcal{Z}_3
&={}-{}\mathcal{I}_3\big(\chiop(1,2,3)+\chiop(3,2,1)-2(\rho+\hat\rho)(\chiop(1,2)+\chiop(2,1))\\
&\phantom{{}={}-\mathcal{I}_3\big(}
+\chiop(1,2,1)+\chiop(2,1,2)+(\rho+\hat\rho)^2\chiop(1)\big)\\
&\phantom{{}={}}
-\mathcal{I}_{3\mathbf{bb}}\chiop(2,1,3)-\mathcal{I}_{3\mathbf{b}}\chiop(1,3,2)\\
&\phantom{{}={}}
+2\mathcal{I}_{32\mathbf{t}}\big(\hat\rho\chiimp{*,*}{*,*}{1,3}
+\rho\chiimp[\tilde]{*,*}{*,*}{1,3}\big)
-2\mathcal{I}_1\mathcal{I}_2(\rho+\hat\rho)\chiop(1,3)\\
&\phantom{{}={}}
+\mathcal{I}_{3\mathbf{t}}(\rho-\hat\rho)\big(
(\rho-\hat\rho)
\chiop(1)
-\hat\rho\big(\chiimp{1}{*}{1}+\chiimp[\tilde]{2}{*}{1}\big)
+\rho\big(\chiimp{2}{*}{1}+\chiimp[\tilde]{1}{*}{1}\big)\big)\\
&\phantom{{}={}+I_{3\mathbf{t}}(\rho-\hat\rho)\big(}
-\chiimp{2}{*}{1,2}
+\chiimp{1,2}{*,3}{1,2}
+\chiimp{2}{*}{2,1}
+\chiimp{2,3}{1,*}{2,1}
\\
&\phantom{{}={}+I_{3\mathbf{t}}(\rho-\hat\rho)\big(}
+\chiimp[\tilde]{2}{*}{1,2}
-\chiimp[\tilde]{1,2}{*,3}{1,2}
-\chiimp[\tilde]{2}{*}{2,1}
-\chiimp[\tilde]{2,3}{1,*}{2,1}
\big)
\pnt
\end{aligned}
\end{equation}
According to the definition \eqref{DinZ}, we have to 
extract the $\frac{1}{\varepsilon}$ pole and multiply the result by $6$.
The three-loop contribution to the dilatation operator then reads
\begin{equation}\label{D3}
\begin{aligned}
\mathcal{D}_3
&={}-{}4\big(\chiop(1,2,3)+\chiop(3,2,1)-2(\rho+\hat\rho)(\chiop(1,2)+\chiop(2,1))\\
&\phantom{{}={}-4\big(}
+\chiop(1,2,1)+\chiop(2,1,2)+(\rho+\hat\rho)^2\chiop(1)\big)\\
&\phantom{{}={}}
+2\big(\chiop(2,1,3)-\chiop(1,3,2)\big)
-4\big(\hat\rho\chiimp{*,*}{*,*}{1,3}
+\rho\chiimp[\tilde]{*,*}{*,*}{1,3}\big)\\
&\phantom{{}={}}
+2\zeta(3)(\rho-\hat\rho)\big(
(\rho-\hat\rho)
\chiop(1)
-\hat\rho\big(\chiimp{1}{*}{1}+\chiimp[\tilde]{2}{*}{1}\big)
+\rho\big(\chiimp{2}{*}{1}+\chiimp[\tilde]{1}{*}{1}\big)\\
&\phantom{{}={}+2\zeta(3)(\rho-\hat\rho)\big(}
-\chiimp{2}{*}{1,2}
+\chiimp{1,2}{*,3}{1,2}
+\chiimp{2}{*}{2,1}
+\chiimp{2,3}{1,*}{2,1}
\\
&\phantom{{}={}+2(\rho-\hat\rho)\zeta(3)\big(}
+\chiimp[\tilde]{2}{*}{1,2}
-\chiimp[\tilde]{1,2}{*,3}{1,2}
-\chiimp[\tilde]{2}{*}{2,1}
-\chiimp[\tilde]{2,3}{1,*}{2,1}
\big)
\pnt
\end{aligned}
\end{equation}
At the orbifold point, where $\rho=\hat\rho=1$, the result
is identical to the one of the $\mathcal{N}=4$ SYM theory after 
a straightforward identification of the chiral functions that is required
due to the differing chiral field content.
Note that apart from the deformed pure scattering term in the third line, 
all the other deformations are homogeneous maximal transcendental due to
$\zeta(3)$.

Using the relations \eqref{hcchi} for the conjugation of the chiral functions,
we see that the above result \eqref{D3} of the Feynman diagram 
calculation is not 
Hermitean. A similar phenomenon has already been observed in the context of 
QCD \cite{Braun:2001qx}. While there the mixing matrix is non-Hermitean at
leading order, here the one- and two-loop results \eqref{D1D2} are Hermitean.
As in \cite{Braun:2001qx} also in our case the eigenvalues of gauge invariant 
operator are real. Therefore, there should exist a non-unitary similarity
transformation that transforms the dilatation operator to an Hermitean form
in a new basis.
In appendix \ref{app:strafo}, we construct general similarity transformations
and determine the one that casts \eqref{D3} into the simplest Hermitean 
form given by
\begin{equation}\label{D3red}
\begin{aligned}
\mathcal{D}_3
&={}-{}4\big(\chiop(1,2,3)+\chiop(3,2,1)-2(\rho+\hat\rho)(\chiop(1,2)+\chiop(2,1))\\
&\phantom{{}={}-4\big(}
+\chiop(1,2,1)+\chiop(2,1,2)+(\rho+\hat\rho)^2\chiop(1)\big)\\
&\phantom{{}={}}
-4\big(\hat\rho\chiimp{*,*}{*,*}{1,3}
+\rho\chiimp[\tilde]{*,*}{*,*}{1,3}\big)\\
&\phantom{{}={}}
+2\zeta(3)(\rho-\hat\rho)\big(
(\rho-\hat\rho)
\chiop(1)
-(\rho+\hat\rho)\big(\chiimp{1}{1}{1}-\chiimp{2}{2}{1}
-\chiimp[\tilde]{1}{1}{1}+\chiimp[\tilde]{2}{2}{1}\big)
\big)
\pnt
\end{aligned}
\end{equation}

\section{Wrapping interactions}
\label{sec:wrappingint}

The mixing of operators of length $L\le3$
cannot be studied by using the dilatation operator \eqref{D3}, since 
it contains contributions from Feynman diagrams with interaction range 
$R\ge L$. 
In case of shorter operators, these contributions have to be replaced
\cite{Fiamberti:2007rj,Fiamberti:2008sh} by the so-called
wrapping interactions \cite{Serban:2004jf,Beisert:2004hm}, 
that arise due to the truncation of the genus expansion beyond
the planar contributions \cite{Sieg:2005kd}. 
For operators of length $L=3$ in the closed chiral subsector, 
the respective analysis is very similar to the one in the 
$\beta$-deformed case \cite{Fiamberti:2008sm,Fiamberti:2008sn}. 
Finite size corrections in the $SL(2)$ subsectors of $\mathds{Z}_k$ orbifolds 
have recently been studied by means of the TBA and $Y$-system in 
\cite{Arutyunov:2010gu,Beccaria:2011qd}.

Following \cite{Fiamberti:2008sm,Fiamberti:2008sn}, the chiral wrapping diagrams
\begin{equation}
\begin{aligned}
\settoheight{\eqoff}{$\times$}%
\setlength{\eqoff}{0.5\eqoff}%
\addtolength{\eqoff}{-12.5\unitlength}%
\raisebox{\eqoff}{%
\fmfframe(-1,1)(1,4){%
\begin{fmfchar*}(20,20)
\chionetwothreei[phantom]{phantom}{dots}{dots}{dots}{dots}{dots}{dots}{phantom}
\fmf{plain,tension=0.5,right=0,width=1mm}{v6,v8}
\wline[0.125w]{dots}{vloc(__vc4)}{v6}{v8}{vloc(__vc3)}
\end{fmfchar*}}}
&=\bar\lambda^3I_3\chiop(1,2,3)
\col\qquad
\settoheight{\eqoff}{$\times$}%
\setlength{\eqoff}{0.5\eqoff}%
\addtolength{\eqoff}{-12.5\unitlength}%
\raisebox{\eqoff}{%
\fmfframe(-1,1)(1,4){%
\begin{fmfchar*}(20,20)
\chithreetwoonei[phantom]{dots}{dots}{dots}{phantom}{phantom}{dots}{dots}{dots}
\fmf{plain,tension=0.5,right=0,width=1mm}{v5,v7}
\wline[0.125w]{dots}{vloc(__vc1)}{v5}{v7}{vloc(__vc6)}
\end{fmfchar*}}}
&=\bar\lambda^3I_3\chiop(3,2,1)
\end{aligned}
\end{equation}
replace the chiral maximum range diagrams in \eqref{threeloopR4}. 
Thereby, a cyclic
identification $(\perm-\Lam_s-\Top)_{3\,4}\simeq(\perm-\Lam_s-\Top)_{3\,1}$
in the definition of the chiral functions \eqref{chifuncdef} is understood.
The above diagrams yield the same expressions as the diagrams
in the first line of \eqref{threeloopR4}. Therefore, these terms 
persist in the dilatation operator. However,
the contributions from the diagrams in the second line of 
\eqref{threeloopR4} and the ones from \eqref{threeloopchi13}
are removed by the subtraction procedure.

It turns out that there are no further contributions from wrapping 
diagrams. At three loops their overall UV divergences cancel with each other 
as in the case of the $\beta$-deformed theory \cite{Fiamberti:2008sn}.
The respective cancellations read
\begin{equation}
\begin{aligned}
\settoheight{\eqoff}{$\times$}%
\setlength{\eqoff}{0.5\eqoff}%
\addtolength{\eqoff}{-12.5\unitlength}%
\raisebox{\eqoff}{%
\fmfframe(2,1)(-3,4){%
\begin{fmfchar*}(20,20)
\chionetwoi{dots}{dots}{dots}{dots}{dots}{dots}{dots}{dots}{dots}
\wigglywrap{vm4}{v4}{v6}{vm9}
\end{fmfchar*}}}
+
\settoheight{\eqoff}{$\times$}%
\setlength{\eqoff}{0.5\eqoff}%
\addtolength{\eqoff}{-12.5\unitlength}%
\raisebox{\eqoff}{%
\fmfframe(2,1)(-3,4){%
\begin{fmfchar*}(20,20)
\chionetwoi{dots}{dots}{dots}{dots}{dots}{dots}{dots}{dots}{dots}
\wigglywrap{vm4}{v4}{v6}{vm7}
\end{fmfchar*}}}
+
\settoheight{\eqoff}{$\times$}%
\setlength{\eqoff}{0.5\eqoff}%
\addtolength{\eqoff}{-12.5\unitlength}%
\raisebox{\eqoff}{%
\fmfframe(2,1)(-3,4){%
\begin{fmfchar*}(20,20)
\chionetwoi{dots}{dots}{dots}{dots}{dots}{dots}{dots}{dots}{dots}
\wigglywrap{vm3}{v4}{v6}{vm9}
\end{fmfchar*}}}
+
\settoheight{\eqoff}{$\times$}%
\setlength{\eqoff}{0.5\eqoff}%
\addtolength{\eqoff}{-12.5\unitlength}%
\raisebox{\eqoff}{%
\fmfframe(2,1)(-3,4){%
\begin{fmfchar*}(20,20)
\chionetwoi{dots}{dots}{dots}{dots}{dots}{dots}{dots}{dots}{dots}
\wigglywrap{vm3}{v4}{v6}{vm7}
\end{fmfchar*}}}
&=0\col\\
\settoheight{\eqoff}{$\times$}%
\setlength{\eqoff}{0.5\eqoff}%
\addtolength{\eqoff}{-12.5\unitlength}%
\raisebox{\eqoff}{%
\fmfframe(2,1)(-3,4){%
\begin{fmfchar*}(20,20)
\chionegi{dots}{dots}{dots}{dots}{dots}{dots}
\fmfi{photon}{vu3--vgm3}
\wigglywrap{vd3}{v4}{v6}{vgm3}
\end{fmfchar*}}}
+
\settoheight{\eqoff}{$\times$}%
\setlength{\eqoff}{0.5\eqoff}%
\addtolength{\eqoff}{-12.5\unitlength}%
\raisebox{\eqoff}{%
\fmfframe(2,1)(-3,4){%
\begin{fmfchar*}(20,20)
\chionegi{dots}{dots}{dots}{dots}{dots}{dots}
\fmfi{photon}{vm3--vgm3}
\wigglywrap{vd4}{v4}{v6}{vgm3}
\end{fmfchar*}}}
&=0
\col
\end{aligned}
\end{equation}
and they also hold for the respective reflected diagrams. 

With the above described modifications, the dilatation operator 
that considers the leading wrapping correction for length $L=3$ operators
then reads
\begin{equation}\label{D3w}
\begin{aligned}
\mathcal{D}_{3,\text{w}}
&=-4\big(\chiop(1,2,3)+\chiop(3,2,1)-2(\rho+\hat\rho)(\chiop(1,2)+\chiop(2,1))\\
&\phantom{{}={}-4\big(}
+\chiop(1,2,1)+\chiop(2,1,2)+(\rho+\hat\rho)^2\chiop(1)\big)\\
&\phantom{{}={}}
+2\zeta(3)(\rho-\hat\rho)\big(
(\rho-\hat\rho)
\chiop(1)
-\hat\rho\big(\chiimp{1}{*}{1}+\chiimp[\tilde]{2}{*}{1}\big)
+\rho\big(\chiimp{2}{*}{1}+\chiimp[\tilde]{1}{*}{1}\big)\\
&\phantom{{}={}+2\zeta(3)(\rho-\hat\rho)\big(}
-\chiimp{2}{*}{1,2}
+\chiimp{1,2}{*,3}{1,2}
+\chiimp{2}{*}{2,1}
+\chiimp{2,3}{1,*}{2,1}
\\
&\phantom{{}={}+2(\rho-\hat\rho)\zeta(3)\big(}
+\chiimp[\tilde]{2}{*}{1,2}
-\chiimp[\tilde]{1,2}{*,3}{1,2}
-\chiimp[\tilde]{2}{*}{2,1}
-\chiimp[\tilde]{2,3}{1,*}{2,1}
\big)
\pnt
\end{aligned}
\end{equation}
Since all $L=2$ operators are protected (they correspond to the types 
of states given in \eqref{groundstates}, \eqref{impstate}), there is no 
need to calculate the next wrapping correction explicitly. 

\section{Eigenvalues}
\label{sec:evalues}

In this section we calculate some eigenvalues of the three-loop dilatation 
operator. First, we determine the 
dispersion relations of the scalar impurities.
Then, we derive the anomalous dimensions for the shortest 
non-protected operators of length $L=3$ and $L=4$.

\subsection{Dispersion relation}
\label{sec:disprel}

The momentum eigenstates of 
a single impurity are given by
\begin{equation}
\begin{aligned}\label{momeigenstates}
\psi(p)
&=\sum_{m}\e^{imp}
\underbrace{\Phi\dots\Phi}_{m-1}Q_{\dot I}\hat\Phi\dots\hat\Phi
\col\qquad
\tilde\psi(p)
=\sum_{m}\e^{imp}
\underbrace{\hat\Phi\dots\hat\Phi}_{m-1}\tilde Q^{\dot I}\Phi\dots\Phi
\pnt
\end{aligned}
\end{equation}
Note that these states are bifundamental in contrast to the ones in
$\mathcal{N}=4$ SYM theory that are in the adjoint representation.
Due to this difference, already at one-loop the local action of the dilatation 
operator depends on the gauge fixing parameter \cite{Gadde:2010ku,Liendo}.
Since we only work in Fermi-Feynman gauge, we can make no prediction 
of how a different gauge choice affects the result.
Of course, gauge independence is guaranteed whenever the dilatation operator
acts on a gauge invariant composite operator.

When the chiral functions \eqref{chifuncimpdef}
with a specific position of the impurity within 
the ingoing and outgoing interacting field lines
are applied to the states in \eqref{momeigenstates}, 
they yield the phase shifts
\begin{equation}\label{chiimpphaseshifts}
\begin{aligned}
\chiimp{n}{*}{1,\dots,n}\psi&=\e^{i(n-1)p}\big(\e^{-ip}-\hat\rho\big)\psi\col\\
\chiimp{n+1}{*}{1,\dots,n}\psi&=\e^{i(n-1)p}\big(\e^{ip}-\rho\big)\psi\col\\
\chiimp{1}{*}{n,\dots,1}\psi&=\e^{-i(n-1)p}\big(\e^{-ip}-\hat\rho\big)\psi\col\\
\chiimp{2}{*}{n,\dots,1}\psi&=\e^{-i(n-1)p}\big(\e^{ip}-\rho\big)\psi\col
\end{aligned}
\qquad
\begin{aligned}
\chiimp[\tilde]{n}{*}{1,\dots,n}\tilde\psi&=\e^{i(n-1)p}\big(\e^{-ip}-\rho\big)\tilde\psi\col\\
\chiimp[\tilde]{n+1}{*}{1,\dots,n}\tilde\psi&=\e^{i(n-1)p}\big(\e^{ip}-\hat\rho\big)\tilde\psi\col\\
\chiimp[\tilde]{1}{*}{n,\dots,1}\tilde\psi&=\e^{-i(n-1)p}\big(\e^{-ip}-\rho\big)\tilde\psi\col\\
\chiimp[\tilde]{2}{*}{n,\dots,1}\tilde\psi&=\e^{-i(n-1)p}\big(\e^{ip}-\hat\rho\big)\tilde\psi\pnt\\
\end{aligned}
\end{equation}
The above results for either $\psi$ or $\tilde\psi$ at fixed $n$ 
combine to the phase shift for the respective impurity generated by the sum 
\begin{equation}\label{chisumphaseshifts}
\begin{aligned}
&\frac{1}{2}\big[\chiop(1,2,\dots,n)+\chiop(n,\dots,2,1)\big]
\to
-\cos(n-1)p\,\big[4\sin^2\tfrac{p}{2}+(\rho^{\frac{1}{2}}-\hat\rho^{\frac{1}{2}})^2\big]
\pnt
\end{aligned}
\end{equation}
We then find that the phase shifts generated by an application 
of the dilatation operator \eqref{D1D2}, \eqref{D3} as obtained from 
Feynman diagrams to the states in \eqref{momeigenstates}
are given by 
\begin{equation}\label{phaseshift}
\begin{aligned}
E(p)
&=
2\bar g^2\big[4\sin^2\tfrac{p}{2}+(\rho^{\frac{1}{2}}-\hat\rho^{\frac{1}{2}})^2\big]
-2\bar g^4\big[4\sin^2\tfrac{p}{2}+(\rho^{\frac{1}{2}}-\hat\rho^{\frac{1}{2}})^2\big]^2
\\
&\phantom{{}={}}
+\bar g^6\Big(4\big[4\sin^2\tfrac{p}{2}+(\rho^{\frac{1}{2}}-\hat\rho^{\frac{1}{2}})^2\big]^3
-2\zeta(3)(\rho-\hat\rho)^2\big[4\sin^2\tfrac{p}{2}+(\rho^{\frac{1}{2}}-\hat\rho^{\frac{1}{2}})^2\big]\\
&\hphantom{={}+{}\bar g^4\Big(}
-2\zeta(3)(\rho^2-\hat\rho^2)(\rho-\hat\rho\mp2i\sin p)\Big)
+\mathcal{O}(\bar g^8)
\col\\
\end{aligned}
\end{equation}
where the upper and lower sign is assumed respectively 
for the momentum eigenstate 
$\psi(p)$ and $\tilde\psi(p)$ in \eqref{momeigenstates}.
The above expression determines the first three orders in the ansatz of
the dispersion relation 
\begin{equation}\label{Ep}
\begin{aligned}
E(p)=\sqrt{1+h^2(g,\hat g)
\big[4\sin^2\tfrac{p}{2}+(\rho^{\frac{1}{2}}-\hat\rho^{\frac{1}{2}})^2\big]
}-1
+f(g,\hat g)(\rho-\hat\rho\mp2i\sin p)
\col
\end{aligned}
\end{equation}
where the functions $h^2(g,\hat g)$ and $f(g,\hat g)$
are given by
\begin{equation}
h^2(g,\hat g)=4\bar g^2-4\bar g^6(\rho-\hat\rho)^2\zeta(3)
+\mathcal{O}(\bar g^8)\col\qquad
f(g,\hat g)=-2\bar g^6(\rho^2-\hat\rho^2)\zeta(3)
+\mathcal{O}(\bar g^8)
\pnt
\end{equation}
The function $h^2(g,\hat g)$ begins with a one-loop contribution 
and is corrected at three loops, where $f(g,\hat g)$ appears for the 
first time. 
Both three-loop contributions are proportional to $\zeta(3)$, 
i.e.\ they are homogeneous maximal transcendental at that loop order
and they vanish at the orbifold point where $g=\hat g$. 
Note also that 
the function $h^2(g,\hat g)$ is strikingly similar to 
its counterpart in
the ABJM and ABJ setups \cite{Minahan:2009aq,Minahan:2009wg,Leoni:2010tb}.
There, its four-loop contribution 
is proportional to $\zeta(2)$ and hence homogeneous maximal 
transcendental in three dimensions.
The three-loop term in $h^2(g,\hat g)$ is due to an undeformed chiral function
and thus fits into the all-order expression conjectured in 
\cite{Gadde:2010ku}. This is not the case for
the complex contribution with coefficient $f(g,\hat g)$
that is generated by deformed chiral functions. Its imaginary part
is generated by an anti-Hermitean combination in the dilatation operator.

Remarkably, the dispersion relation \eqref{Ep} is not real for real 
momentum $p$, but its three-loop term contains an imaginary contribution.
This is due to anti-Hermitean terms in the three loop dilatation 
operator \eqref{D3}.
As explained in appendix \ref{app:strafo}, the dilatation operator 
and the basis of operators can be altered by similarity transformations 
without affecting the eigenvalues of gauge invariant operators.
We find that it can be transformed into the Hermitean expression \eqref{D3red}
that leads to a dispersion relation without an 
imaginary term which read
\begin{equation}\label{Epreal}
\begin{aligned}
E'(p')=\sqrt{1+h^2(g,\hat g)
\big[4\sin^2\tfrac{p'}{2}+(\rho^{\frac{1}{2}}-\hat\rho^{\frac{1}{2}})^2\big]
}-1+f(g,\hat g)(\rho-\hat\rho)
\pnt
\end{aligned}
\end{equation}
The transformed momentum eigenstates $\psi'(p')$ and $\tilde\psi'(p')$
given in \eqref{strafomomeigenstates}
have the same form as $\psi(p)$ and $\tilde\psi(p)$ in \eqref{momeigenstates}
apart from a normalization factor and an imaginary shift of the original  
momentum $p$. It is given by
\begin{equation}
p'=p\pm i\bar g^4(\rho^2-\hat\rho^2)\zeta(3)+\mathcal{O}(\bar g^6)
\col
\end{equation}
such that the energies obey $E(p)=E'(p')$.

We conclude this section with a discussion of the limiting cases.
When the interpolating theory becomes the orbifold theory at 
$\rho=\hat\rho=1$ or $\mathcal{N}=2$ SCQCD, the imaginary part in
\eqref{phaseshift} drops out and the dispersion relations become identical 
for both states $\psi(p)$ and $\tilde\psi(p)$.
The respective limits read to three-loop order
\begin{equation}
E(p)=
\begin{cases}
8g^2\sin^2\tfrac{p}{2}-32g^4\sin^4\tfrac{p}{2}+256g^6\sin^6\tfrac{p}{2} 
+\mathcal{O}(g^8)& \text{orbifold} \\
2g^2-2g^4+4g^6(1-\zeta(3))+\mathcal{O}(g^8)& \text{$\mathcal{N}=2$ SCQCD}
\end{cases}
\col
\end{equation}
where at the orbifold point one recovers the dispersion relation of 
$\mathcal{N}=4$ SYM theory \cite{Beisert:2004hm}, while in 
$\mathcal{N}=2$ SCQCD it is independent of $p$
as observed before at one loop \cite{Gadde:2010zi}. 


\subsection{Eigenvalues of some short operators}

In this section we explicitly give the results for the anomalous dimensions of $L=3,4$ operators. For later convenience we define the following parameters
\begin{equation}\label{Gsigma}
G=\frac{g+\hat g}{2}
\col\qquad
\sigma
=\frac{g-\hat g}{g+\hat g}
\pnt
\end{equation}

The chiral composite operators of length $L=2$ are all protected, since 
they correspond to the types of states given in \eqref{groundstates}
and \eqref{impstate}.

The chiral composite operators of length $L=3$ are the two 
groundstates of the form given in \eqref{groundstates} and the two states
\begin{equation}
\mathcal{O}_0
=\hat\rho\tr\big(Q_{\dot I}\tilde Q^{\dot J}\Phi\big)
+\tr\big(Q_{\dot I}\hat\Phi\tilde Q^{\dot J}\big)
\col\qquad
\mathcal{O}_1
=-\rho\tr\big(Q_{\dot I}\tilde Q^{\dot J}\Phi\big)
+\tr\big(Q_{\dot I}\hat\Phi\tilde Q^{\dot J}\big)
\col
\end{equation}
that are eigenstates of the dilatation operator given by  
\eqref{D1D2} and by the wrapping corrected three-loop contribution 
\eqref{D3w}. Their eigenvalues read 
\begin{equation}
\gamma_0=0\col\qquad
\gamma_1=8G^2(1+\sigma^2)-32G^4(1+\sigma^2)^2
+256G^6(1+\sigma^2)((1+\sigma^2)^2-\sigma^2\zeta(3)\big)
\pnt
\end{equation}

The chiral composite operators of length $L=4$ are the three protected 
states of the form given in \eqref{groundstates} and \eqref{impstate}
and the three states
\begin{equation}
\mathcal{O}_1=\tr\big(Q_{\dot I}\tilde Q^{\dot J}\Phi\Phi\big)
\col\qquad
\mathcal{O}_2=\tr\big(Q_{\dot I}\hat\Phi\tilde Q^{\dot J}\Phi\big)
\col\qquad
\mathcal{O}_3=\tr\big(Q_{\dot I}\hat\Phi\hat\Phi\tilde Q^{\dot J}\big)
\col
\end{equation}
that mix under renormalization.
At the orbifold point, where the couplings are equal and hence 
$G=g=\hat g$, $\sigma=0$, the eigenvectors
do not depend on the coupling, since all chiral functions become proportional 
to a single mixing matrix. This simplification does not hold in the generic case.
In terms of the parameters \eqref{Gsigma} the
three eigenvalues are given by
\begin{equation}
\begin{aligned}
\gamma_0&=0\col\\
\gamma_{\pm}&=
8G^2(1+\sigma^2)
-4G^4(7(1+\sigma^4)+10\sigma^2)
+8G^6(1+\sigma^2)(23(1+\sigma^4)+2\sigma^2(13-16\zeta(3)))\\
&\phantom{{}={}}
\pm 4G^2\sqrt{X}\col\\
X&=
(1-\sigma^2)^2(1-10G^2(1+\sigma^2))\\
&\phantom{{}={}}
+G^4(101(1+\sigma^8)+12\sigma^2(1+\sigma^4)-162\sigma^4
-32\sigma^2(1-\sigma^2)^2\zeta(3))\\
&\phantom{{}={}}
-4G^6((1+\sigma^2)(95(1+\sigma^8)+84\sigma^2(1+\sigma^4)+90\sigma^4)
-48\sigma^2(1-\sigma^2)^2(1+\sigma^2)\zeta(3))\\
&\phantom{{}={}}
+4G^8(361(1+\sigma^{12})+1298\sigma^2(1+\sigma^8)+2759\sigma^4(1+\sigma^4)
+3708\sigma^6\\
&\phantom{{}={}+4G^8(}
-16\sigma^2(1-\sigma^2)^2(27(1+\sigma^4)+58\sigma^2)\zeta(3))
\pnt
\end{aligned}
\end{equation}
At the orbifold point $G=g=\hat g$, $\sigma=0$, the anomalous dimensions 
simplify to 
\begin{equation}
\begin{aligned}
\gamma_0&=0\col\qquad
\gamma_{\pm}=8g^2-28g^4+184g^6
\pm 4g^2(1-5g^2+38g^4)\pnt\\
\end{aligned}
\end{equation}
For $\mathcal{N}=2$ supersymmetric QCD, where $G=\frac{g}{2}$, $\hat g=0$, 
$\sigma=1$, the anomalous dimensions simplify to 
\begin{equation}
\begin{aligned}
\gamma_0&=0\col\qquad
\gamma_{\pm}=
4g^2-6g^4+g^6(18-8\zeta(3))\pm 2g^4(1-7g^2)
\pnt
\end{aligned}
\end{equation}
In both cases, the square-root could be taken exactly without employing further 
series expansions.

\section{Where are the elementary excitations?}
\label{sec:elemex}

The calculation presented in this paper determines the three-loop mixing in
the closed chiral subsector that contains the operators \eqref{cop}.
The excitations in this sector are the bifundamental fields 
$Q_{\dot I}$ and $\tilde Q^{\dot J}$. As we have found they have complex 
dispersion relations \eqref{phaseshift}, and this makes it difficult 
to interpret them 
as elementary excitations. In this section, we discuss the excitations 
of other closed subsectors that might be regarded as fundamental ones.

The fields of the theory have a definite scaling dimension $\Delta$ and
$U(1)_\text{R}$ charge $J$. The excitations above the vacuum fields 
are classified according to the
difference of $\Delta-J$. While the vacuum fields
have $\Delta=1$, $J=1$ and hence $\Delta-J=0$, the excitations have 
$\Delta-J\ge1$.  In $\mathcal{N}=4$ SYM theory, all excitations
with $\Delta-J=1$ are elementary, i.e.\ they are magnons, while 
the ones with $\Delta-J\ge2$ are their bound states.
In contrast to this, in the \emph{interpolating theory} 
the $\Delta-J=1$ excitations $Q_{\dot I}$ and 
$\tilde Q^{\dot I}$ of the closed chiral subsector should not be thought 
of as being elementary. Therefore, we should search in other closed 
subsectors of the theory for $\Delta-J=1$ excitations with real 
dispersion relations.

From our analysis we conclude that the complex terms in the dispersion 
relations of $Q_{\dot I}$ and $\tilde Q^{\dot J}$ appear due to the
bifundamental nature of these excitations, resulting in a dependence on two
gauge couplings. In order to avoid such terms we should focus on 
sectors that involve one gauge group only. We should look for $\Delta-J=1$ 
excitations that transform in the same adjoint representation as the 
vacuum fields. They are either given by two of the adjoint fermions 
or by a lightcone component of covariant spacetime derivatives involving 
the respective gauge field. In a vacuum given by the first
of the groundstates in \eqref{groundstates}, the respective fermions 
are the ones included in the superfields $\Phi$ and $V$.
The composite operators that contain only one
type of the fermionic or derivative excitations respectively 
correspond to the closed $SU(1|1)$ or $SL(2)$ subsector of 
$\mathcal{N}=4$ SYM theory. 

We will now argue why 
in these subsectors non-trivial deviations from the 
case of $\mathcal{N}=4$ SYM theory should only show up from three loops on.
Firstly, since the vacuum fields and excitations in these sectors 
transform in the adjoint representation of only one factor of
the gauge group, a dependence on the coupling constant of the other gauge 
group factor can only arise when bifundamental 
matter fields form a loop in which at least one further interaction 
occurs. This requires at least two loops. For example,
a diagram of this type is given by the first one in \eqref{sedia}. It 
contributes to the two-loop self energies of the adjoint chiral matter 
fields. Due to its presence, the final result \eqref{ctwoloopse}
for the self energy is the same for all chiral matter fields, and it 
depends on the product of both gauge couplings.
This dependence then occurs in the overall UV divergence
of diagrams at three and higher loops, in which the finite two-loop 
chiral self energy appears as subdiagram.
Such diagrams contribute in the orbifold theory, where the two couplings
are equal, while they are absent in $\mathcal{N}=2$ SCQCD, where one of the 
couplings is zero. If one does not find other diagrams that compensate 
this behavior, the dilatation operator of $\mathcal{N}=2$ SCQCD 
would start to deviate from the $\mathcal{N}=4$ SYM result in the 
respective subsector at three loops. 
The one- and two-loop dilatation operators of the 
$SU(1|1)$ and $SL(2)$ subsectors is hence identical to their 
counterparts in $\mathcal{N}=4$ SYM theory up to a trivial identification of the coupling constants. Deviations can
first appear at three loops, where integrability can then be 
non-trivially tested for the first time.


\section{Conclusions}
\label{sec:concl}

The central result of this paper is the expression for the dilatation operator
as given in \eqref{D3} at three loops. For 
the discovery of new effects in the \emph{interpolating theory} and their 
investigation it was necessary to work at least at this order.
Three loops is the first order at which the chiral functions of the theory are 
deformed by gauge interactions. There occur
homogeneous transcendental contributions that involve $\zeta(3)$.
They vanish at the point of equal couplings, where the theory becomes
a $\mathds{Z}_2$ orbifold of $\mathcal{N}=4$ SYM theory, and the dilatation 
operator reduces to the $\mathcal{N}=4$ SYM result.

We have found that the dispersion relation \eqref{Ep} contains a function 
$h^2(g,\hat g)$ with non-vanishing three-loop contribution. 
Furthermore, for real momentum $p$ it develops an imaginary part away 
from the orbifold and $\mathcal{N}=2$ SCQCD points.
Both of these deviations from the $\mathcal{N}=4$ result 
are homogeneous maximal transcendental, i.e.\
proportional to $\zeta(3)$ at three loops.
The imaginary contribution is due to anti-Hermitean terms in the three-loop 
dilatation operator. They can be removed by a non-unitary similarity 
transformation that also transforms the basis of eigenstates. In particular, 
the momenta of eigenstates of single excitations acquire imaginary two-loop 
shifts.
The dispersion relation \eqref{Ep} contains extra terms involving 
$f(g,\hat g)$ that are missing in a symmetry-based
all-order conjecture in \cite{Gadde:2010ku}.
 Concerning a spin chain interpretation of operator mixing in the
closed chiral subsector of the theory, the excitations 
$Q_{\dot I}$ and $\tilde Q^{\dot I}$ should be regarded as effective 
rather than as elementary magnons. In addition, compared to $\mathcal{N}=4$ SYM theory, the two-body S-matrix is further deformed by
pure scattering terms that first show up at three loops 
and come with a deformed chiral function.

Based on our findings we have argued that 
one should search for integrability in other subsectors of the 
\emph{interpolating theory} that 
contain elementary excitations. 
They should correspond to the $SU(1|1)$ and $SL(2)$ subsectors
of $\mathcal{N}=4$ SYM theory and involve the adjoint fields 
associated with only one factor of the 
product gauge group. The respective excitations should have simpler 
dispersion relations than \eqref{Ep}.
On the basis of the finite two-loop self energy
corrections we have predicted that
non-trivial deviations from the respective sectors of 
$\mathcal{N}=4$ SYM theory can only show up from three loops on.

\section*{Acknowledgements}

We are very grateful to Matthias Staudacher for discussions
concerning the interpretation of our result and for useful comments 
on the manuscript. We also thank Jan Plefka for useful discussions and
Vladimir Mitev for comments on the manuscript.
E.\ P.\ thanks IHES and C.\ S.\ thanks University of Milan,  
INFN Milan and Jagellonian University 
for kind hospitality during the course of this work.
The work of E.\ P.\ is supported by Humboldt 
Foundation. The work of C.\ S.\ is supported by DFG, SFB 647
\emph{Raum -- Zeit -- Materie. Analytische und Geometrische Strukturen}.

\appendix

\section{Gauge fixed action and Feynman rules in $\mathcal{N}=1$ superspace}
\label{app:actfrules}

The \emph{interpolating theory} in an $\mathcal{N}=1$ superspace formulation
is presented in section \ref{sec:N2SYM}. It
contains two real vector superfields $V$, $\hat V$ and 
two chiral superfields $\Phi$, $\hat\Phi$ that all transform in the 
adjoint representation of respectively the first and second factor 
of the product gauge group $SU(N)\times SU(N)$. 
Furthermore, it contains chiral superfields
$Q_{\dot I}$, $\tilde Q^{\dot I}$ that respectively transform in the bifundamental
and anti-bifundamental representation of $SU(N)\times SU(N)$.
The gauge fixing proceeds independently for each gauge field $V$, $\hat V$
as in the case of $\mathcal{N}=4$ SYM theory \cite{Gates:1983nr}, and we obtain 
the gauge fixed action
\begin{equation}
\begin{aligned}
S_{\text{gf}}
&=S-\frac{1}{2}\int\de^4x\de^4\theta\Big[\frac{1}{\alpha}\tr\big((\D^2V)(\barD^2V)\big)+\frac{1}{\hat\alpha}\tr\big((\D^2\hat V)(\barD^2\hat V)\big)\Big]\\
&\phantom{{}={}}+\int\de^4x\de^4\theta\big[\tr\big((c'+\bar c')\Ld_{\frac{1}{2}g_\YM V}(c+\bar c+\coth\Ld_{\frac{1}{2}g_\YM V}(c-\bar c))\big)\\
&\hphantom{{}={}+\int\de^4x\de^4\theta\big[}
+\tr\big((\hat c'+\hat{\bar c}')\Ld_{\frac{1}{2}\hat g_\YM\hat V}(\hat c+\hat{\bar c}+\coth\Ld_{\frac{1}{2}\hat g_\YM\hat V}(\hat c-\hat{\bar c}))\big)\big]
\col\\
\end{aligned}
\end{equation}
where $S$ is the action given in \eqref{action}, and
$\Ld_VX=\comm{V}{X}$.
The fields are decomposed as
\begin{equation}
\begin{aligned}
V&=V_aT^a\col\qquad
&\hat V&=\hat V_{\hat a}\hat T^{\hat a}\col\\
c&=c_aT^a\col\qquad
&\hat c&=\hat c_a\hat T^a\col\\
c'&=c'_aT^a\col\qquad
&\hat c'&=\hat c'_a\hat T^a\col\\
\Phi&=\Phi_aT^a\col\qquad
&\hat\Phi&=\hat \Phi_{\hat a}\hat T^{\hat a}\col\\
\end{aligned}
\qquad
\begin{aligned}
Q_{\dot I}&=Q_{\dot I}^{\underline{a}}B_{\underline{a}}\col\qquad
&\bar{Q}^{\dot I}&=\bar{Q}^{\dot I}_{\underline{a}}B^{\underline{a}}\col\\
\tilde{\bar Q}_{\dot I}&=\tilde{\bar Q}_{\dot I}^{\underline{a}}B_{\underline{a}}\col\qquad
&\tilde Q^{\dot I}&=\tilde Q^{\dot I}_{\underline{a}}B^{\underline{a}}
\end{aligned}
\end{equation}
in terms of representation matrices for the product gauge group 
$SU(N)\times SU(N)$
\begin{equation}
\begin{aligned}
(T^a)^i{}_j\col\qquad
(\hat T^{\hat a})^{\hat i}{}_{\hat j}\col\qquad
(B_{\underline a})^{i}{}_{\hat j}\col\qquad
(B^{\underline a})^{\hat i}{}_{j}\col\qquad
\end{aligned}
\end{equation}
that transform the fundamental indices $i$ and $\hat i$ of the 
respective $SU(N)$ factor into 
adjoint $a$ and $\hat a$, 
(anti-)bifundamental $\underline a$ indices.
The matrices of the adjoint representations fulfill the commutation relations
\begin{equation}
\comm{T^a}{T^b}=if_{abc}T^c\col\qquad\comm{\smash{\hat T^a}}{\smash{\hat T^b}}=if_{abc}\hat T^c
\end{equation}
of the respective Lie-algebra. Furthermore, the matrices obey
\begin{equation}
\begin{aligned}
(T^a)^i{}_j(T^a)^k{}_l
=\delta^i_l\delta^k_j-\frac{1}{N}\delta^i_j\delta^k_l
\col\quad
(\hat T^{\hat a})^{\hat i}{}_{\hat j}(\hat T^{\hat a})^{\hat k}{}_{\hat l}
=\delta^{\hat i}_{\hat l}\delta^{\hat k}_{\hat j}
-\frac{1}{N}\delta^{\hat i}_{\hat j}\delta^{\hat k}_{\hat l}
\col\quad
(B_{\underline a})^{i}{}_{\hat j}(B^{\underline a})^{\hat k}{}_{l}
=\delta^i_l\delta^{\hat k}_{\hat j}
\col
\end{aligned}
\end{equation}
where summations over $a$, $\hat a$, $\underline{a}$ are understood.

\section{Feynman rules}

In this appendix we present the Feynman rules that are required for a
three-loop calculation.
We use the Wick rotated rules, i.e.\ we have transformed
$\e^{-iS}\to\e^S$ in the path integral.
In supersymmetric Fermi-Feynman gauge where $\alpha=1+\mathcal{O}(g_\YM^2)$, 
the vector, chiral and ghost propagators are given 
by\footnote{The corrections from the gauge parameter $\alpha$ do not 
appear in the diagrams explicitly considered in this paper.}
\begin{equation}\label{propagators}
\begin{aligned}
\langle V_aV_b\rangle
&=
\settoheight{\eqoff}{$\times$}%
\setlength{\eqoff}{0.5\eqoff}%
\addtolength{\eqoff}{-3.75\unitlength}%
\raisebox{\eqoff}{%
\fmfframe(2,0)(2,0){%
\begin{fmfchar*}(15,7.5)
\fmfleft{v1}
\fmfright{v2}
\fmfforce{0.0625w,0.5h}{v1}
\fmfforce{0.9375w,0.5h}{v2}
\fmf{photon}{v1,v2}
\fmffreeze
\fmfposition
\fmfipath{pm[]}
\fmfiset{pm1}{vpath(__v1,__v2)}
\nvml{1}{$\scriptstyle p$}
\end{fmfchar*}}}
=
-\frac{\delta_{ab}}{p^2}\delta^4(\theta_1-\theta_2)
\col\\
\langle\Phi_a\bar\Phi_b\rangle
&=
\settoheight{\eqoff}{$\times$}%
\setlength{\eqoff}{0.5\eqoff}%
\addtolength{\eqoff}{-3.75\unitlength}%
\raisebox{\eqoff}{%
\fmfframe(2,0)(2,0){%
\begin{fmfchar*}(15,7.5)
\fmfleft{v1}
\fmfright{v2}
\fmfforce{0.0625w,0.5h}{v1}
\fmfforce{0.9375w,0.5h}{v2}
\fmf{plain}{v1,v2}
\fmffreeze
\fmfposition
\fmfipath{pm[]}
\fmfiset{pm1}{vpath(__v1,__v2)}
\nvml{1}{$\scriptstyle p$}
\end{fmfchar*}}}
=
\frac{\delta_{ab}}{p^2}\delta^4(\theta_1-\theta_2)
\col\\
\langle Q_{\dot I}^{\underline{a}}\bar Q^{\dot J}_{\underline{b}}\rangle
&=
\settoheight{\eqoff}{$\times$}%
\setlength{\eqoff}{0.5\eqoff}%
\addtolength{\eqoff}{-3.75\unitlength}%
\raisebox{\eqoff}{%
\fmfframe(2,0)(2,0){%
\begin{fmfchar*}(15,7.5)
\fmfleft{v1}
\fmfright{v2}
\fmfforce{0.0625w,0.5h}{v1}
\fmfforce{0.9375w,0.5h}{v2}
\fmf{dashes}{v1,v2}
\fmffreeze
\fmfposition
\fmfipath{pm[]}
\fmfiset{pm1}{vpath(__v1,__v2)}
\nvml{1}{$\scriptstyle p$}
\end{fmfchar*}}}
=
\frac{\delta_{\dot I}^{\dot J}\delta^{\underline{a}}_{\underline{b}}}{p^2}\delta^4(\theta_1-\theta_2)
\col\\
\langle\tilde Q^{\dot I}_{\underline{a}}\tilde{\bar Q}_{\dot J}^{\underline{b}}\rangle
&=
\settoheight{\eqoff}{$\times$}%
\setlength{\eqoff}{0.5\eqoff}%
\addtolength{\eqoff}{-3.75\unitlength}%
\raisebox{\eqoff}{%
\fmfframe(2,0)(2,0){%
\begin{fmfchar*}(15,7.5)
\fmfleft{v1}
\fmfright{v2}
\fmfforce{0.0625w,0.5h}{v1}
\fmfforce{0.9375w,0.5h}{v2}
\fmf{dashes}{v1,v2}
\fmffreeze
\fmfposition
\fmfipath{pm[]}
\fmfiset{pm1}{vpath(__v1,__v2)}
\nvml{1}{$\scriptstyle p$}
\end{fmfchar*}}}
=
\frac{\delta^{\dot I}_{\dot J}\delta_{\underline{a}}^{\underline{b}}}{p^2}\delta^4(\theta_1-\theta_2)
\col\\
\langle \bar c'_ac_b\rangle
=-\langle c'_a\bar c_b\rangle
&=
\settoheight{\eqoff}{$\times$}%
\setlength{\eqoff}{0.5\eqoff}%
\addtolength{\eqoff}{-3.75\unitlength}%
\raisebox{\eqoff}{%
\fmfframe(2,0)(2,0){%
\begin{fmfchar*}(15,7.5)
\fmfleft{v1}
\fmfright{v2}
\fmfforce{0.0625w,0.5h}{v1}
\fmfforce{0.9375w,0.5h}{v2}
\fmf{dots}{v1,v2}
\fmffreeze
\fmfposition
\fmfipath{pm[]}
\fmfiset{pm1}{vpath(__v1,__v2)}
\nvml{1}{$\scriptstyle p$}
\end{fmfchar*}}}
=\frac{\delta_{ab}}{p^2}\delta^4(\theta_1-\theta_2)
\pnt
\end{aligned}
\end{equation}
The expressions for the propagators of $\hat V$, $\hat\Phi$ and $\hat c$, $\hat c'$ are identical up to the replacement $\delta_{ab}\to\delta_{\hat a\hat b}$.

The cubic gauge vertex for the vector field $V$ is given by
\begin{equation}
\begin{aligned}\label{V3vertex}
V_{V^3}
&=
\Bigg(
\cvert{photon}{photon}{photon}{}{$\scriptstyle\D^\alpha$}{$\scriptstyle\barD^2\!\scriptstyle\D_{\!\alpha}$}
-
\cvert{photon}{photon}{photon}{}{$\scriptstyle\barD^2\!\scriptstyle\D_{\!\alpha}$}{$\scriptstyle\D^\alpha$}
+\cvert{photon}{photon}{photon}{$\scriptstyle\D_{\!\alpha}\!\scriptstyle\barD^2$}{}{$\scriptstyle\D^\alpha$}
-\cvert{photon}{photon}{photon}{$\scriptstyle\D_{\!\alpha}\!\scriptstyle\barD^2$}{$\scriptstyle\D^\alpha$}{}
+\cvert{photon}{photon}{photon}{$\scriptstyle\D^\alpha$}{$\scriptstyle\barD^2\!\scriptstyle\D_{\!\alpha}$}{}
-\cvert{photon}{photon}{photon}{$\scriptstyle\D^\alpha$}{}{$\scriptstyle\barD^2\!\scriptstyle\D_{\!\alpha}$}
\Bigg)
\frac{g_\YM }{2}\tr\big(T^a\comm{T^b}{T^c}\big)
\col
\end{aligned}
\end{equation}
where the color indices are labeled $(a,b,c)$ clockwise, 
starting with the leg to the left. 
The respective vertex for the field $\hat V$ is identical up to a
replacement of the gauge coupling and the color trace.
The $\D$-algebra has to be performed for all six permutations of the 
structure of the covariant derivatives at its legs.
The only purpose of the vertices that appear on the r.h.s.\ of the equation 
is to display this structure. They do not contain any other non-trivial 
factors. 

The cubic gauge-matter vertices 
are given by
\begin{equation}\label{cvertices}
\begin{gathered}
\begin{aligned}
V_{\bar\Phi V\Phi}
&=
\cvert{photon}{plain}{plain}{}{$\scriptstyle\barD^2$}{$\scriptstyle\D^2$}
g_\YM\tr\big(T^a\comm{T^b}{T^c}\big)
\col\quad
&V_{\hat{\bar\Phi}V\hat{\Phi}}
&=
\cvert{photon}{plain}{plain}{}{$\scriptstyle\barD^2$}{$\scriptstyle\D^2$}
\hat g_\YM\tr\big(\hat T^a\comm{\smash{\hat T^b}}{\smash{\hat T^c}}\big)
\col\\
V_{\bar Q^{\dot J} VQ_{\dot I}}
&=
\cvert{photon}{dashes}{dashes}{}{$\scriptstyle\barD^2$}{$\scriptstyle\D^2$}
g_\YM\delta^{\dot I}_{\dot J}\tr\big(T^aB_{\underline{b}}B^{\underline{c}}\big)
\col\quad
&V_{Q_{\dot J}\hat V\bar Q^{\dot I}}
&=
\cvert{photon}{dashes}{dashes}{}{$\scriptstyle\D^2$}{$\scriptstyle\barD^2$}
(-\hat g_\YM)\delta_{\dot I}^{\dot J}\tr\big(\hat T^{\hat a}B^{\underline{b}}B_{\underline{c}}\big)
\col\\
V_{\tilde Q^{\dot J} V\tilde{\bar Q}_{\dot I}}
&=
\cvert{photon}{dashes}{dashes}{}{$\scriptstyle\D^2$}{$\scriptstyle\barD^2$}
(-g_\YM)\delta^{\dot I}_{\dot J}\tr\big(T^aB_{\underline{b}}B^{\underline{c}}\big)
\col\quad
&V_{\tilde{\bar Q}_{\dot J}\hat V\tilde Q^{\dot I}}
&=
\cvert{photon}{dashes}{dashes}{}{$\scriptstyle\barD^2$}{$\scriptstyle\D^2$}
\hat g_\YM\delta_{\dot I}^{\dot J}\tr\big(\hat T^{\hat a}B^{\underline{b}}B_{\underline{c}}\big)
\col\\
V_{\tilde Q^{\dot J}\Phi Q_{\dot I}}
&=\cvert{plain}{dashes}{dashes}{}{$\scriptstyle\barD^2$}{$\scriptstyle\barD^2$}
ig_\YM\delta^{\dot I}_{\dot J}\tr\big(T^aB_{\underline{b}}B^{\underline{c}}\big)
\col\quad
&V_{Q_{\dot J}\hat\Phi\tilde Q^{\dot I}}
&=\cvert{plain}{dashes}{dashes}{}{$\scriptstyle\barD^2$}{$\scriptstyle\barD^2$}
(-i\hat g_\YM)\delta_{\dot I}^{\dot J}\tr\big(\hat T^{\hat a}B^{\underline{b}}B_{\underline{c}}\big)
\col\\
V_{\bar Q^{\dot J}\bar\Phi \tilde{\bar Q}_{\dot I}}
&=\cvert{plain}{dashes}{dashes}{}{$\scriptstyle\D^2$}{$\scriptstyle\D^2$}
(-ig_\YM)\delta^{\dot I}_{\dot J}\tr\big(T^aB_{\underline{b}}B^{\underline{c}}\big)
\col\quad
&V_{\tilde{\bar Q}_{\dot J}\hat{\bar\Phi}\bar Q_{\dot I}}
&=\cvert{plain}{dashes}{dashes}{}{$\scriptstyle\D^2$}{$\scriptstyle\D^2$}
i\hat g_\YM\delta^{\dot I}_{\dot J}\tr\big(\hat T^{\hat a}B^{\underline{b}}B_{\underline{c}}\big)
\col\\
\end{aligned}\\
\begin{aligned}
V_{Vcc'}
&=
\cvert{photon}{dots}{dots}{}{$\scriptstyle\barD^2$}{$\scriptstyle\barD^2$}\frac{g_\YM}{2}\tr\big(T^a\comm{T^b}{T^c}\big)
\col\qquad
&V_{V\bar c\bar c'}
&=
\cvert{photon}{dots}{dots}{}{$\scriptstyle\D^2$}{$\scriptstyle\D^2$}\frac{g_\YM}{2}\tr\big(T^a\comm{T^b}{T^c}\big)
\col\\
V_{Vc\bar c'}
&=
\cvert{photon}{dots}{dots}{}{$\scriptstyle\barD^2$}{$\scriptstyle\D^2$}\frac{g_\YM}{2}\tr\big(T^a\comm{T^b}{T^c}\big)
\col\qquad
&V_{V\bar cc'}
&=
\cvert{photon}{dots}{dots}{}{$\scriptstyle\D^2$}{$\scriptstyle\barD^2$}\frac{g_\YM}{2}\tr\big(T^a\comm{T^b}{T^c}\big)
\pnt
\end{aligned}
\end{gathered}
\end{equation}
The color indices are labeled $(a,b,c)$ clockwise, 
starting with the leg to the left.

For the three-loop calculation we only need some of the quartic gauge-matter 
vertices. They read
\begin{equation}\label{qvertices}
\begin{aligned}
V_{\bar\Phi V^2\Phi}&=
\qvert{photon}{photon}{plain}{plain}{}{}{$\scriptstyle\barD^2$}{$\scriptstyle\D^2$}\Big(-\frac{g_\YM^2}{2}\Big)
\big(\tr\big(\comm{T^a}{T^d}\comm{T^b}{T^c}\big)+\tr\big(\comm{T^a}{T^c}\comm{T^b}{T^d}\big)
\big)
\col\\
V_{V^2Q_{\dot I}\bar Q^{\dot J}}
&=
\qvert{photon}{photon}{plain}{plain}{}{}{$\scriptstyle\barD^2$}{$\scriptstyle\D^2$}\frac{g_\YM^2}{2}\delta^{\dot I}_{\dot J}\tr\big(\acomm{T^a}{T^b}B_{\underline{c}}B^{\underline{d}}\big)
\col\\
V_{\hat V^2\bar Q_{\dot I}Q^{\dot J}}
&=
\qvert{photon}{photon}{plain}{plain}{}{}{$\scriptstyle\D^2$}{$\scriptstyle\barD^2$}\frac{\hat g_\YM^2}{2}\delta^{\dot I}_{\dot J}\tr\big(\acomm{\vphantom{T^b}\smash[t]{\hat T^{\hat a}}}{\smash[t]{\hat T^{\hat b}}}B^{\underline{c}}B_{\underline{d}}\big)
\col\\
V_{VQ_{\dot I}\hat V\bar Q^{\dot J}}
&=\qvert{photon}{plain}{photon}{plain}{}{$\scriptstyle\barD^2$}{}{$\scriptstyle\D^2$}
(-g_\YM\hat g_\YM)\delta^{\dot I}_{\dot J}\tr\big(T^aB_{\underline{b}}\hat T^{\hat c}B^{\underline{d}}\big)
\col\\
V_{V^2\tilde{\bar Q}_{\dot I}\tilde Q^{\dot J}}
&=
\qvert{photon}{photon}{plain}{plain}{}{}{$\scriptstyle\D^2$}{$\scriptstyle\barD^2$}\frac{g_\YM^2}{2}\delta_{\dot I}^{\dot J}\tr\big(\acomm{T^a}{T^b}B_{\underline{c}}B^{\underline{d}}\big)
\col\\
V_{\hat V^2\tilde Q^{\dot I}\tilde{\bar Q}_{\dot J}}
&=
\qvert{photon}{photon}{plain}{plain}{}{}{$\scriptstyle\barD^2$}{$\scriptstyle\D^2$}\frac{\hat g_\YM^2}{2}\delta_{\dot I}^{\dot J}\tr\big(\acomm{\vphantom{T^b}\smash[t]{\hat T^{\hat a}}}{\smash[t]{\hat T^{\hat b}}}B^{\underline{c}}B_{\underline{d}}\big)
\col\\
V_{V\tilde{\bar Q}_{\dot I}\hat V\tilde Q^{\dot J}}
&=\qvert{photon}{plain}{photon}{plain}{}{$\scriptstyle\D^2$}{}{$\scriptstyle\barD^2$}
(-g_\YM\hat g_\YM)\tr\big(T^aB_{\underline{b}}\hat T^{\hat c}B^{\underline{d}}\big)
\col\\
\end{aligned}
\end{equation}
where the color indices are labeled $(a,b,c,d)$ clockwise 
starting with the leg in the upper left corner.

\section{One- and two-loop subdiagrams}

In this appendix we present the non-vanishing one- and two-loop subdiagrams that
appear in our calculation.
In order to collectively represent the subdiagrams with all possible 
combinations of adjoint and bifundamental fields, we introduce
generic 't Hooft couplings $\lambda_1$, $\lambda_2$, $\lambda_3$ and
dotted lines as chiral field lines that denote the adjoint and
(anti)-bifundamental fields. 
Each of the couplings $\lambda_i$ is then built from the Yang-Mills 
gauge coupling and rank that is associated with a respective face of the subdiagram that carries label $i$. In order to
obtain the subdiagram for a specific field configuration, one 
just has to replace the dotted field lines by the matter fields of the theory 
and then determine the gauge factor that is associated with 
each face.

\subsection{One-loop chiral vertex correction}

Omitting all factors from the respective tree-level vertex
of \eqref{cvertices} the
one-loop corrections to the chiral vertices are easily summarized
as
\begin{equation}
\begin{aligned}\label{ccconeloop}
\cVat[\fmfcmd{fill fullcircle scaled 10 shifted vloc(__vg1) withcolor black ;}
\fmfiv{label=\small$\textcolor{white}{1}$,l.dist=0}{vloc(__vg1)}
\fmfiv{label=$\scriptstyle\textcolor{black}{1}$,l.dist=0}{(0.875w,0.5h)}
\fmfiv{label=$\scriptstyle\textcolor{black}{2}$,l.dist=0}{(0.3125w,0.25h)}
\fmfiv{label=$\scriptstyle\textcolor{black}{3}$,l.dist=0}{(0.3125w,0.75h)}
]
{dots}{dots}{dots}{phantom}{dots,ptext.in=$\scriptstyle\barD^2$}{dots,ptext.in=$\scriptstyle\barD^2$}
&=
\cVat
{dots,ptext.in=$\scriptstyle\barD^2$}{dots,ptext.in=$\scriptstyle\barD^2$,ptext.out=$\scriptstyle\D^2$}{dots,ptext.out=$\scriptstyle\D^2$}{photon}{dots,ptext.in=$\scriptstyle\barD^2$}{dots,ptext.in=$\scriptstyle\barD^2$}
+\dots
=
\left(\cVat[\fmfiset{p10}{p4--p6--p5--cycle}\fmfcmd{fill p10 withcolor \mympostgrey ;}]
{dots,l.side=left,l.dist=3,label=$\scriptstyle\Box$}{plain}{plain}{plain}{dots,ptext.in=$\scriptstyle\barD^2$}{dots,ptext.in=$\scriptstyle\barD^2$}
\lambda_1 
+\dots\right)
\col
\end{aligned}
\end{equation}
where the ellipsis denote the remaining two diagrams obtained by 
clockwise cyclic permutations of the interactions of one sector to the next sector. Under each such 
permutation the coupling constants associated with the individual sectors 
also have to be replaced as $\lambda_1\to\lambda_2$, $\lambda_2\to\lambda_3$, 
$\lambda_3\to\lambda_1$. The d'Alembertian
$\Box$ cancels the respective
propagator and thereby produces a minus.

The one-loop correction to the cubic gauge-matter vertex is
given by 
\begin{equation}
\begin{aligned}
\cVat[\fmfcmd{fill fullcircle scaled 10 shifted vloc(__vg1) withcolor black ;}
\fmfiv{label=\small$\textcolor{white}{1}$,l.dist=0}{vloc(__vg1)}]%
{photon}{dots}{dots}{phantom}%
{dots}
{dots}
&=
\cVat
{photon}{photon}{photon}{dots}%
{dots}
{dots}
+
\cVat{photon}%
{dots}
{dots}
{photon}%
{dots}
{dots}
+
\cVat{photon}%
{dots}
{dots}
{dots}%
{dots}
{dots}
\\
&\phantom{{}={}}
+
\cVab[dots]{2}{-0.75}{}{$ $}{$ $}{ }
+
\cVab[dots]{2}{0.75}{}{$ $}{$ $}{ }
+
\cVab[dots]{3}{-0.75}{}{$ $}{$ $}{ }
+
\cVab[dots]{3}{0.75}{}{$ $}{$ $}{ }
\col
\end{aligned}
\end{equation}
where we have omitted the covariant derivatives. 
The result for the first term containing the cubic gauge vertex \eqref{V3vertex}
can be found e.g.\ in \cite{Sieg:2010tz}. We just have 
to generalize it including the different 't Hooft couplings, and then we obtain 
\begin{equation}
\begin{aligned}
\cVat
[
\fmfiv{label=$\scriptstyle\textcolor{black}{1}$,l.dist=0}{(1w,0.5h)}
\fmfiv{label=$\scriptstyle\textcolor{black}{2}$,l.dist=0}{(0.3125w,0.25h)}
\fmfiv{label=$\scriptstyle\textcolor{black}{2}$,l.dist=0}{(0.3125w,0.75h)}
]
{photon}{photon}{photon}{dots}{dots,ptext.in=$\scriptstyle\barD^2$}{dots,ptext.in=$\scriptstyle\D^2$}
&=\left(
\cVat[\fmfiset{p10}{p4--p6--p5--cycle}\fmfcmd{fill p10 withcolor \mympostgrey ;}]
{photon,ptext.in=$\hspace{0.4cm}\scriptstyle\D^\alpha\!\scriptstyle\barD^2\!\scriptstyle\D_{\!\alpha}$}{plain}{plain}{plain}{dots,ptext.in=$\scriptstyle\barD^2$}{dots,ptext.in=$\scriptstyle\D^2$}
+
\cVat[\fmfiset{p10}{p4--p6--p5--cycle}\fmfcmd{fill p10 withcolor \mympostgrey ;}]
{photon,ptext.out=$\scriptstyle\comm{\smash{\D_{\!\alpha}}}{\smash{\barD_{\!\dot\beta}}}\hspace{0.5cm}$}{plain}{plain}{derplain,label=$\scriptstyle l^{\alpha\dot\beta}$,l.side=left,l.dist=3}{dots,ptext.in=$\scriptstyle\barD^2$}{dots,ptext.in=$\scriptstyle\D^2$}
\right)\frac{\lambda_2}{2}
\pnt
\end{aligned}
\end{equation}
The covariant derivatives and also momenta are read-off 
when leaving the vertices. 
The other two contributions involving only cubic vertices 
are also easily adopted from the expressions in \cite{Sieg:2010tz}.
They read
\begin{equation}
\begin{aligned}
\cVat
{photon}{dots,ptext.in=$\scriptstyle\barD^2$,ptext.out=$\scriptstyle\D^2$}{dots,ptext.in=$\scriptstyle\D^2$,ptext.out=$\scriptstyle\barD^2$}{photon}{dots,ptext.in=$\scriptstyle\barD^2$}{dots,ptext.in=$\scriptstyle\D^2$}
&=
\left(
\cVat[\fmfiset{p10}{p4--p6--p5--cycle}\fmfcmd{fill p10 withcolor \mympostgrey ;}]
{photon,ptext.in=$\hspace{0.4cm}\scriptstyle\barD^2\!\scriptstyle\D^2$}{plain}{plain}{plain}{dots,ptext.in=$\scriptstyle\barD^2$}{dots,ptext.in=$\scriptstyle\D^2$}
-
\cVat[\fmfiset{p10}{p4--p6--p5--cycle}\fmfcmd{fill p10 withcolor \mympostgrey ;}]
{photon,ptext.in=$\hspace{0.4cm}\scriptstyle\barD_{\!\dot\beta}\D_{\!\alpha}$}{plain}{derplain,label=$\scriptstyle (p_3-l)^{\alpha\dot\beta}$,l.side=right,l.dist=3}{plain}{dots,ptext.in=$\scriptstyle\barD^2$}{dots,ptext.in=$\scriptstyle\D^2$}
+
\cVat[\fmfiset{p10}{p4--p6--p5--cycle}\fmfcmd{fill p10 withcolor \mympostgrey ;}]
{photon}{plain}{plain,label=$\scriptstyle\Box$,l.side=right,l.dist=3}{plain}{dots,ptext.in=$\scriptstyle\barD^2$}{dots,ptext.in=$\scriptstyle\D^2$}
\right)(-\lambda_1) 
\col\\
\cVat
{photon}{dashes,ptext.in=$\scriptstyle\D^2$,ptext.out=$\scriptstyle\barD^2$}{dashes,ptext.in=$\scriptstyle\barD^2$,ptext.out=$\scriptstyle\D^2$}{plain}{dashes,ptext.in=$\scriptstyle\barD^2$}{dashes,ptext.in=$\scriptstyle\D^2$}
&=
\left(\cVat[\fmfiset{p10}{p4--p6--p5--cycle}\fmfcmd{fill p10 withcolor \mympostgrey ;}]
{photon,ptext.in=$\hspace{0.4cm}\scriptstyle\barD^2\!\scriptstyle\D^2$}{plain}{plain}{plain}{dashes,ptext.in=$\scriptstyle\barD^2$}{dashes,ptext.in=$\scriptstyle\D^2$}
-
\cVat[\fmfiset{p10}{p4--p6--p5--cycle}\fmfcmd{fill p10 withcolor \mympostgrey ;}]
{photon,ptext.in=$\hspace{0.4cm}\scriptstyle\barD_{\!\dot\beta}\D_{\!\alpha}$}{derplain,label=$\scriptstyle (l+p_2)^{\alpha\dot\beta}$,l.side=left,l.dist=3}{plain}{plain}{dashes,ptext.in=$\scriptstyle\barD^2$}{dashes,ptext.in=$\scriptstyle\D^2$}
+
\cVat[\fmfiset{p10}{p4--p6--p5--cycle}\fmfcmd{fill p10 withcolor \mympostgrey ;}]
{photon}{plain,label=$\scriptstyle\Box$,l.side=left,l.dist=3}{plain}{plain}{dashes,ptext.in=$\scriptstyle\barD^2$}{dashes,ptext.in=$\scriptstyle\D^2$}
\right)(-\lambda_1)
\col\\
\cVat
{photon}{plain,ptext.in=$\scriptstyle\D^2$,ptext.out=$\scriptstyle\barD^2$}{plain,ptext.in=$\scriptstyle\barD^2$,ptext.out=$\scriptstyle\D^2$}{dashes}{dashes,ptext.in=$\scriptstyle\barD^2$}{dashes,ptext.in=$\scriptstyle\D^2$}
&=
\left(\cVat[\fmfiset{p10}{p4--p6--p5--cycle}\fmfcmd{fill p10 withcolor \mympostgrey ;}]
{photon,ptext.in=$\hspace{0.4cm}\scriptstyle\barD^2\!\scriptstyle\D^2$}{plain}{plain}{plain}{dashes,ptext.in=$\scriptstyle\barD^2$}{dashes,ptext.in=$\scriptstyle\D^2$}
-
\cVat[\fmfiset{p10}{p4--p6--p5--cycle}\fmfcmd{fill p10 withcolor \mympostgrey ;}]
{photon,ptext.in=$\hspace{0.4cm}\scriptstyle\barD_{\!\dot\beta}\D_{\!\alpha}$}{derplain,label=$\scriptstyle (l+p_2)^{\alpha\dot\beta}$,l.side=left,l.dist=3}{plain}{plain}{dashes,ptext.in=$\scriptstyle\barD^2$}{dashes,ptext.in=$\scriptstyle\D^2$}
+
\cVat[\fmfiset{p10}{p4--p6--p5--cycle}\fmfcmd{fill p10 withcolor \mympostgrey ;}]
{photon}{plain,label=$\scriptstyle\Box$,l.side=left,l.dist=3}{plain}{plain}{dashes,ptext.in=$\scriptstyle\barD^2$}{dashes,ptext.in=$\scriptstyle\D^2$}
\right)(-\lambda_2)
\col\\
\cVab[dots]{2}{-0.75}{}{$\scriptstyle\barD^2$}{$\scriptstyle\D^2$}{ptext.in=$\scriptstyle\barD^2$,ptext.out=$\scriptstyle\D^2$}
\!\!&=
\cVat[\fmfiset{p10}{p4--p6--p5--cycle}\fmfcmd{fill p10 withcolor \mympostgrey ;}]
{photon}{plain}{plain,label=$\scriptstyle\Box$,l.side=right,l.dist=3}{plain}{dots,ptext.in=$\scriptstyle\barD^2$}{dots,ptext.in=$\scriptstyle\D^2$}
\frac{\lambda_2}{2}
\col\qquad
\cVab[dots]{2}{0.75}{}{$\scriptstyle\barD^2$}{$\scriptstyle\D^2$}{ptext.hin=2,ptext.hout=2,ptext.in=$\scriptstyle\barD^2$,ptext.out=$\scriptstyle\D^2$}
=
\cVat[\fmfiset{p10}{p4--p6--p5--cycle}\fmfcmd{fill p10 withcolor \mympostgrey ;}]
{photon}{plain}{plain,label=$\scriptstyle\Box$,l.side=right,l.dist=3}{plain}{dots,ptext.in=$\scriptstyle\barD^2$}{dots,ptext.in=$\scriptstyle\D^2$}
\lambda_1
\col\\
\cVab[dots]{3}{-0.75}{}{$\scriptstyle\barD^2$}{$\scriptstyle\D^2$}{ptext.in=$\scriptstyle\barD^2$,ptext.out=$\scriptstyle\D^2$}
\!\!&=
\cVat[\fmfiset{p10}{p4--p6--p5--cycle}\fmfcmd{fill p10 withcolor \mympostgrey ;}]
{photon}{plain,label=$\scriptstyle\Box$,l.side=left,l.dist=3}{plain}{plain}{dots,ptext.in=$\scriptstyle\barD^2$}{dots,ptext.in=$\scriptstyle\D^2$}
\frac{\lambda_2}{2}
\col\qquad
\cVab[dots]{3}{0.75}{}{$\scriptstyle\barD^2$}{$\scriptstyle\D^2$}{ptext.hin=2,ptext.hout=2,ptext.in=$\scriptstyle\barD^2$,ptext.out=$\scriptstyle\D^2$}
=
\cVat[\fmfiset{p10}{p4--p6--p5--cycle}\fmfcmd{fill p10 withcolor \mympostgrey ;}]
{photon}{plain,label=$\scriptstyle\Box$,l.side=left,l.dist=3}{plain}{plain}{dots,ptext.in=$\scriptstyle\barD^2$}{dots,ptext.in=$\scriptstyle\D^2$}
\lambda_1
\col
\end{aligned}
\end{equation}
where we have inserted $-\frac{\Box}{p^2}=1$ in order to obtain 
triangle integrals.
We sum up the above expressions and simplify the result as explained in 
\cite{Sieg:2010tz}.
The expression for the one-loop corrections of the cubic gauge-matter vertices
is then given by
\begin{equation}
\begin{aligned}\label{cvaoneloop}
\cVat[\fmfcmd{fill fullcircle scaled 10 shifted vloc(__vg1) withcolor black ;}
\fmfiv{label=$\scriptstyle\textcolor{white}{1}$,l.dist=0}{vloc(__vg1)}
\fmfiv{label=$\scriptstyle\textcolor{black}{1}$,l.dist=0}{(0.875w,0.5h)}
\fmfiv{label=$\scriptstyle\textcolor{black}{2}$,l.dist=0}{(0.3125w,0.25h)}
\fmfiv{label=$\scriptstyle\textcolor{black}{2}$,l.dist=0}{(0.3125w,0.75h)}
]
{photon}{dots}{dots}{phantom}{dots,ptext.in=$\scriptstyle\barD^2$}{dots,ptext.in=$\scriptstyle\D^2$}
&=-
\cVat[\fmfiset{p10}{p4--p6--p5--cycle}\fmfcmd{fill p10 withcolor \mympostgrey ;}]
{photon,ptext.in=$\hspace{0.4cm}\scriptstyle\D^\alpha\!\scriptstyle\barD^2\!\scriptstyle\D_{\!\alpha}$}{plain}{plain}{plain}{dots,ptext.in=$\scriptstyle\barD^2$}{dots,ptext.in=$\scriptstyle\D^2$}\lambda_1
-\frac{1}{4}\left(
\cVat[\fmfiset{p10}{p4--p6--p5--cycle}\fmfcmd{fill p10 withcolor \mympostgrey ;}]
{photon,ptext.out=$\scriptstyle\comm{\smash{\D_{\!\alpha}}}{\smash{\barD_{\!\dot\beta}}}\hspace{0.5cm}$}{plain}{plain}{plain}{derdots,ptext.in=$\scriptstyle\barD^2$,label=$\scriptstyle p_2^{\alpha\dot\beta}$,l.side=right,l.dist=0.5}{dots,ptext.in=$\scriptstyle\D^2$}
-
\cVat[\fmfiset{p10}{p4--p6--p5--cycle}\fmfcmd{fill p10 withcolor \mympostgrey ;}]
{photon,ptext.out=$\scriptstyle\comm{\smash{\D_{\!\alpha}}}{\smash{\barD_{\!\dot\beta}}}\hspace{0.5cm}$}{plain}{plain}{plain}{dots,ptext.in=$\scriptstyle\barD^2$}{derdots,ptext.in=$\scriptstyle\D^2$,label=$\scriptstyle p_3^{\alpha\dot\beta}$,l.side=left,l.dist=2}\right)\lambda_2
\pnt
\end{aligned}
\end{equation}

\subsection{Two-loop chiral self energies}

All one-loop self energies are identically zero at the conformal 
point. The two-loop chiral self energies are finite and do not
contribute to the two-loop $\beta$-function. However, at higher  
loops overall UV divergences may be generated by diagrams that 
contain them as subdiagrams, and we therefore have to calculate them.
The two-loop self energies of the chiral adjoint and bifundamental
matter fields are determined by the following Feynman diagrams
\begin{equation}\label{sedia}
\begin{aligned}
S_{\Phi}&=\seone[photon]{1}{1}{plain}{plain}{dashes}{dashes}{dashes}{dashes}
=-2\lambda\hat\lambda I_{2\mathbf{t}}
\col\qquad
S_{Q,\tilde Q}=\seone[photon]{1}{1}{dashes}{dashes}{dots}{dots}{dots}{dots}
=-(\lambda^2+\hat\lambda^2)I_{2\mathbf{t}}
\col\\
S_{2a}&=\swftwotwo[dots]{1}{1}+\swftwotwor[dots]{1}{1}
+\swftwotwo[dots]{-1}{-1}+\swftwotwor[dots]{-1}{-1}
=(\lambda_1^2+\lambda_2^2)I_2
\col\\
S_{2b}&=\swftwotwo[dots]{1}{-1}+\swftwotwor[dots]{1}{-1}
+\swftwotwo[dots]{-1}{1}+\swftwotwor[dots]{-1}{1}
=4\lambda_1\lambda_2I_2
\col\\
S_{3a}&=\swftwofour[dots]{1}{1}+\swftwofour[dots]{-1}{-1}
=\frac{\lambda_1^2+\lambda_2^2}{2}I_1^2
\col\\
S_{3b}&=\swftwofour[dots]{1}{-1}+\swftwofour[dots]{-1}{1}
=2\lambda_1\lambda_2I_1^2
\col\\
S_{4}&=\swftwofive[dots]{1}+\swftwofive[dots]{-1}
=-2\lambda_1\lambda_2(I_1^2+I_{2\mathbf{t}})
\col\\
S_{5}&=\seone[dots]{1}{0}{dots}{dots}{dots}{photon}{dots}{dots}
+\seone[dots]{1}{0}{dots}{dots}{photon}{dots}{dots}{dots}
+\seone[dots]{-1}{0}{dots}{dots}{dots}{photon}{dots}{dots}
+\seone[dots]{-1}{0}{dots}{dots}{photon}{dots}{dots}{dots}
=-2(\lambda_1+\lambda_2)^2I_2
\col\\
S_{6}&=\seone[photon]{1}{0}{dots}{dots}{photon}{photon}{dots}{dots}
+\seone[photon]{-1}{0}{dots}{dots}{photon}{photon}{dots}{dots}
=\frac{\lambda_1^2+\lambda_2^2}{2}(-I_1^2+2I_2+2I_{2\mathbf{t}})
\col\\
\end{aligned}
\end{equation}
where $\lambda_1=\lambda_2=\lambda,\hat\lambda$ for $\Phi$, $\hat\Phi$
and $(\lambda_1,\lambda_2)=(\lambda,\hat\lambda),(\hat\lambda,\lambda)$ 
respectively for $Q_{\dot I}$, $\tilde Q^{\dot I}$.
These configuration yield a common result for the finite two-loop self energies
of the chiral matter fields that is given by
\begin{equation}\label{sigmac}
\begin{aligned}
\Sigma_{\Phi,Q,\tilde Q}&=S_{\Phi,Q,\tilde Q}+S_{2a}+S_{2b}+S_{3a}+S_{3b}+S_{4}+S_{5a}+S_{5b}+S_{6}
=-2\bar\lambda^2 I_{2\mathbf{t}}
\pnt
\end{aligned}
\end{equation}
Restoring the dependence on the spinor derivative and the correct 
proportionality to the external momentum $p$, 
the two-loop chiral self energy can be written as
\begin{equation}\label{ctwoloopse}
\settoheight{\eqoff}{$\times$}%
\setlength{\eqoff}{0.5\eqoff}%
\addtolength{\eqoff}{-7.5\unitlength}%
\raisebox{\eqoff}{%
\fmfframe(1,0)(1,0){%
\begin{fmfchar*}(20,15)
\fmfleft{v1}
\fmfright{v2}
\fmf{dots}{v1,v2}
\fmffreeze
\fmfposition
\fmfipair{vm[]}
\fmfipath{p[]}
\fmfiset{p1}{vpath(__v1,__v2)}
\svertex{vm1}{p1}
\fmfcmd{fill fullcircle scaled 10 shifted vm1 withcolor black ;}
\fmfiv{label=\small$\textcolor{white}{2}$,l.dist=0}{vm1}
\end{fmfchar*}}}
=
-2\bar\lambda^2p^{2(D-3)}
\settoheight{\eqoff}{$\times$}%
\setlength{\eqoff}{0.5\eqoff}%
\addtolength{\eqoff}{-7.5\unitlength}%
\raisebox{\eqoff}{%
\fmfframe(1,0)(1,0){%
\begin{fmfchar*}(20,15)
\fmfleft{v1}
\fmfright{v2}
\fmffixed{(0.66w,0)}{vc1,vc2}
\fmf{dots}{v1,vc1}
\fmf{dots}{vc2,v2}
\fmf{plain,left=0.5}{vc1,vc2}
\fmf{plain,left=0.5}{vc2,vc1}
\fmffreeze
\fmfposition
\fmfipair{vm[]}
\fmfipath{p[]}
\fmfiset{p1}{vpath(__v1,__vc1)}
\fmfiset{p2}{vpath(__v2,__vc2)}
\fmfiset{p3}{vpath(__vc1,__vc2)}
\fmfiset{p4}{vpath(__vc2,__vc1)}
\fmfcmd{fill p3--p4--cycle withcolor \mympostgrey;}
\svertex{vm1}{p3}
\svertex{vm2}{p4}
\fmfi{plain}{vm1--vm2}
\fmfis{phantom,ptext.in=$\scriptstyle\D^2$,ptext.clen=7,ptext.hin=3,ptext.hout=3,ptext.oin=5,ptext.oout=3,ptext.sep=;}{p1}
\fmfis{phantom
,ptext.out=$\scriptstyle\barD^2$,ptext.clen=7,ptext.hin=3,ptext.hout=3,ptext.oin=5,ptext.oout=3,ptext.sep=;}{p2}
\end{fmfchar*}}}
\pnt
\end{equation}
The gray scaled part of the graph is identified as the integral 
$I_{2\mathbf{t}}$ given in \eqref{I2t}.

\subsection{Two-loop chiral vertex corrections}

The two-loop correction of the chiral vertex is given as a sum of the 
following non-vanishing contributions
\begin{equation}
\begin{aligned}
\cVat[\fmfcmd{fill fullcircle scaled 10 shifted vloc(__vg1) withcolor black ;}
\fmfiv{label=\small$\textcolor{white}{2}$,l.dist=0}{vloc(__vg1)}]
{dots}{dots}{dots}{phantom}{dots}{dots}
&=
\cVatg[\fmfcmd{fill fullcircle scaled 8 shifted vloc(__vg1) withcolor black ;}\fmfiv{label=$\scriptstyle\textcolor{white}{1}$,l.dist=0}{vloc(__vg1)}\fmfi{photon}{vm2--vm3}]{dots}{dots}{dots}{dots}{dots}
+
\cVatg[\fmfi{photon}{vm1{dir 90}..{dir -45}vi3}\fmfi{phantom}{vo2--vo3}\fmfcmd{fill fullcircle scaled 8 shifted vm1 withcolor black ;}\fmfiv{label=$\scriptstyle\textcolor{white}{1}$,l.dist=0}{vm1}]{dots}{dots}{dots}{dots}{dots}
+
\cVatg[\fmfi{photon}{vm1{dir -90}..{dir 45}vi2}\fmfi{phantom}{vo2--vo3}\fmfcmd{fill fullcircle scaled 8 shifted vm1 withcolor black ;}\fmfiv{label=$\scriptstyle\textcolor{white}{1}$,l.dist=0}{vm1}]{dots}{dots}{dots}{dots}{dots}
+\cVatg[\fmfi{photon}{vm2--vi3}\fmfi{photon}{vm2--vo3}]{dots}{dots}{dots}{dots}{dots}
+\cVatg[\fmfi{photon}{vi2--vm3}\fmfi{photon}{vo2--vm3}]{dots}{dots}{dots}{dots}{dots}
\\
&\phantom{{}={}}
+\dots\col
\end{aligned}
\end{equation}
where the ellipsis denote again the omitted diagrams that are obtained by 
 cyclic rotations of the interactions of the displayed diagrams to 
the other sectors.
After $\D$-algebra, the final result can be cast into the form
\begin{equation}\label{ccctwoloop}
\begin{aligned}
\cVat[\fmfcmd{fill fullcircle scaled 10 shifted vloc(__vg1) withcolor black ;}
\fmfiv{label=$\scriptstyle\textcolor{white}{2}$,l.dist=0}{vloc(__vg1)}
\fmfiv{label=$\scriptstyle\textcolor{black}{1}$,l.dist=0}{(0.875w,0.5h)}
\fmfiv{label=$\scriptstyle\textcolor{black}{2}$,l.dist=0}{(0.3125w,0.25h)}
\fmfiv{label=$\scriptstyle\textcolor{black}{3}$,l.dist=0}{(0.3125w,0.75h)}
]
{dots}{dots}{dots}{phantom}{dots,ptext.in=$\scriptstyle\barD^2$}{dots,ptext.in=$\scriptstyle\barD^2$}
&=\frac{1}{2}\left(
2\lambda_1^2
\cVatt[\fmfcmd{fill p7--p8--p5--p6--p9--cycle withcolor \mympostgrey ;}]{dots,label=$\scriptstyle\Box^2$,l.side=left,l.dist=3}{plain}{plain}{plain}{plain}{plain}{dots,ptext.in=$\scriptstyle\barD^2$}{dots,ptext.in=$\scriptstyle\barD^2$}
-(2\lambda_2\lambda_3-(\lambda_2-\lambda_3)^2)
\cVatt[\fmfcmd{fill p7--p8--p5--p6--p9--cycle withcolor \mympostgrey ;}]{dots,label=$\scriptstyle\Box $,l.side=left,l.dist=3}{plain}{plain}{plain,label=$\scriptstyle\Box$,l.dist=3,l.side=left}{plain}{plain}{dots,ptext.in=$\scriptstyle\barD^2$}{dots,ptext.in=$\scriptstyle\barD^2$}
\right).
\\
&\phantom{{}={}\frac{1}{2}\Bigg({}\,}\left.
+(\lambda_2^2-\lambda_3^2)\left(
\cVatt[\fmfcmd{fill p7--p8--p5--p6--p9--cycle withcolor \mympostgrey ;}]{dots}{plain}{plain}{plain}{plain,label=$\scriptstyle\Box$,l.side=left,l.dist=3}{plain}{dots,ptext.in=$\scriptstyle\barD^2$}{dots,label=$\scriptstyle\Box$,l.side=left,l.dist=3,ptext.in=$\scriptstyle\barD^2$}
-
\cVatt[\fmfcmd{fill p7--p8--p5--p6--p9--cycle withcolor \mympostgrey ;}]{dots}{plain}{plain}{plain}{plain}{plain,label=$\scriptstyle\Box$,l.dist=3}{dots,label=$\scriptstyle\Box$,l.dist=3,ptext.in=$\scriptstyle\barD^2$}{dots,ptext.in=$\scriptstyle\barD^2$}
\right)
\right.\\
&\phantom{{}={}\frac{1}{2}\Bigg({}\,}\left.
-(\lambda_2^2+(\lambda_1-\lambda_2)^2)
\cVatt[\fmfcmd{fill p7--p8--p5--p6--p9--cycle withcolor \mympostgrey ;}]{dots,label=$\scriptstyle\Box$,l.side=left,l.dist=3}{plain}{plain}{plain}{plain,label=$\scriptstyle\Box$,l.side=left,l.dist=3}{plain}{dots,ptext.in=$\scriptstyle\barD^2$}{dots,ptext.in=$\scriptstyle\barD^2$}
-(\lambda_3^2+(\lambda_1-\lambda_3)^2)
\cVatt[\fmfcmd{fill p7--p8--p5--p6--p9--cycle withcolor \mympostgrey ;}]{dots,label=$\scriptstyle\Box$,l.side=left,l.dist=3}{plain}{plain}{plain}{plain}{plain,label=$\scriptstyle\Box$,l.side=right,l.dist=3}{dots,ptext.in=$\scriptstyle\barD^2$}{dots,ptext.in=$\scriptstyle\barD^2$}
\right.
\\
&\phantom{{}={}\frac{1}{2}\Bigg({}\,}\left.
+
\lambda_2^2\left(
\cVatt[\fmfcmd{fill p7--p8--p5--p6--p9--cycle withcolor \mympostgrey ;}]{dots,label=$\scriptstyle\Box$,l.side=left,l.dist=3}{plain}{plain}{plain}{plain}{plain}{dots,ptext.in=$\scriptstyle\barD^2$,label.side=right,label=$\scriptstyle\Box$,l.dist=3}{dots,ptext.in=$\scriptstyle\barD^2$}
+
\cVatt[\fmfcmd{fill p7--p8--p5--p6--p9--cycle withcolor \mympostgrey ;}]{dots}{plain}{plain}{plain}{plain,l.side=left,label=$\scriptstyle\Box$,l.dist=3}{plain}{dots,ptext.in=$\scriptstyle\barD^2$,l.side=right,label=$\scriptstyle\Box$,l.dist=3}{dots,ptext.in=$\scriptstyle\barD^2$}
\right)
\right.
\\
&\phantom{{}={}\frac{1}{2}\Bigg({}\,}\left.
+
\lambda_3^2\left(
\cVatt[\fmfcmd{fill p7--p8--p5--p6--p9--cycle withcolor \mympostgrey ;}]{dots,label=$\scriptstyle\Box$,l.side=left,l.dist=3}{plain}{plain}{plain}{plain}{plain}{dots,ptext.in=$\scriptstyle\barD^2$}{dots,ptext.in=$\scriptstyle\barD^2$,label.side=left,label=$\scriptstyle\Box$,l.dist=3}
+
\cVatt[\fmfcmd{fill p7--p8--p5--p6--p9--cycle withcolor \mympostgrey ;}]{dots}{plain}{plain}{plain}{plain}{plain,l.side=right,label=$\scriptstyle\Box$,l.dist=3}{dots,ptext.in=$\scriptstyle\barD^2$}{dots,ptext.in=$\scriptstyle\barD^2$,label.side=left,label=$\scriptstyle\Box$,l.dist=3}
\right)
+\dots
\right)
\col
\end{aligned}
\end{equation}
where the ellipsis denotes the clockwise cyclic permutations of the above structures, thereby replacing $\lambda_1\to\lambda_2$, $\lambda_2\to\lambda_3$, $\lambda_3\to\lambda_1$. The coupling constants are thereby associated with the 
three sectors, counting clockwise and starting with the sector to the right.

\section{Integrals}
\label{app:integrals}

In this appendix we collect the integral that are relevant for 
our three-loop calculation. 
In $D$-dimensional Euclidean space
the simple loop integral that involves two propagators
of massless fields with respective weights $\alpha$ and $\beta$
and external momentum $p^2=1$
can be expressed in terms of a $G$-function.
It is given by
\begin{equation}
G(\alpha,\beta)
=\frac{1}{(2\pi)^D}\int\frac{\de^Dk}{k^{2\alpha}(k-p)^{2\beta}}\Big|_{p^2=1}
=\frac{\Gamma(\tfrac{D}{2}-\alpha)\Gamma(\tfrac{D}{2}-\beta)\Gamma(\alpha+\beta-\tfrac{D}{2})}{(4\pi)^{\frac{D}{2}}\Gamma(\alpha)\Gamma(\beta)\Gamma(D-\alpha-\beta)}\pnt
\end{equation}
Likewise, the $G$-functions for integrals with one momentum or
two momenta in the numerators can respectively be written as
\begin{equation}
\begin{aligned}
G_1(\alpha,\beta)
&=\frac{1}{2}(-G(\alpha,\beta-1)+G(\alpha-1,\beta)+G(\alpha,\beta))\col\\
G_2(\alpha,\beta)
&=\frac{1}{2}(-G(\alpha,\beta-1)-G(\alpha-1,\beta)+G(\alpha,\beta))
\pnt
\end{aligned}
\end{equation}
To three loops, we need the following integrals and their overall
UV divergences
\begin{equation}\label{Integralexpr}
\begin{gathered}
\begin{aligned}
I_1&=
\settoheight{\eqoff}{$\times$}%
\setlength{\eqoff}{0.5\eqoff}%
\addtolength{\eqoff}{-4.5\unitlength}%
\raisebox{\eqoff}{%
\fmfframe(-1,-3)(-1,-3){%
\begin{fmfchar*}(15,15)
  \fmfleft{in}
  \fmfright{out1}
\fmf{phantom}{in,v1}
\fmf{phantom}{out,v2}
\fmfforce{(0,0.5h)}{in}
\fmfforce{(w,0.5h)}{out}
\fmffixed{(0.75w,0)}{v1,v2}
\fmf{plain,right=0.5}{v1,v2}
\fmf{plain,left=0.5}{v1,v2}
\end{fmfchar*}}}
=G(1,1)
\col\quad
&\mathcal{I}_1&=\frac{1}{(4\pi)^2}\frac{1}{\varepsilon} \col\\
I_2&=
\settoheight{\eqoff}{$\times$}%
\setlength{\eqoff}{0.5\eqoff}%
\addtolength{\eqoff}{-6.75\unitlength}%
\raisebox{\eqoff}{%
\fmfframe(-0.5,-5.5)(1,4){%
\begin{fmfchar*}(15,15)
  \fmfleft{in}
  \fmfright{out1}
\fmf{phantom}{in,v1}
\fmf{phantom}{out,v2}
\fmfforce{(0,0h)}{in}
\fmfforce{(w,0h)}{out}
\fmffixed{(0,0.75h)}{v2,v3}
\fmfpoly{phantom}{v1,v2,v3}
\fmf{plain}{v1,v2}
\fmf{plain}{v1,v3}
\fmf{plain,right=0.25}{v2,v3}
\fmf{plain,left=0.25}{v2,v3}
\end{fmfchar*}}}
=G(1,1)G(3-\tfrac{D}{2},1)
\col\quad
&\mathcal{I}_2&=\frac{1}{(4\pi)^4}\Big(-\frac{1}{2\varepsilon^2}+\frac{1}{2\varepsilon}\Big) 
\col\\
I_3&=
\settoheight{\eqoff}{$\times$}%
\setlength{\eqoff}{0.5\eqoff}%
\addtolength{\eqoff}{-6.75\unitlength}%
\raisebox{\eqoff}{%
\fmfframe(-0.5,-5.5)(9,4){%
\begin{fmfchar*}(15,15)
  \fmfleft{in}
  \fmfright{out1}
\fmf{phantom}{in,v1}
\fmf{phantom}{out,v2}
\fmfforce{(0,0h)}{in}
\fmfforce{(w,0h)}{out}
\fmffixed{(0,0.75h)}{v2,v3}
\fmfpoly{phantom}{v1,v2,v3}
\fmfpoly{phantom}{v3,v2,v4}
\fmf{plain}{v1,v2}
\fmf{plain}{v1,v3}
\fmf{plain}{v2,v3}
\fmf{plain}{v3,v4}
\fmf{plain,right=0.25}{v2,v4}
\fmf{plain,left=0.25}{v2,v4}
\end{fmfchar*}}}
=G(1,1)G(3-\tfrac{D}{2},1)G(5-D,1)
\col
&\mathcal{I}_3&=\frac{1}{(4\pi)^6}\Big(\frac{1}{6\varepsilon^3}-\frac{1}{2\varepsilon^2}+\frac{2}{3\varepsilon}\Big)
\col\\
I_{3\mathbf{t}}&=
\settoheight{\eqoff}{$\times$}%
\setlength{\eqoff}{0.5\eqoff}%
\addtolength{\eqoff}{-9\unitlength}%
\raisebox{\eqoff}{%
\fmfframe(-0.5,-5.5)(9,8.5){%
\begin{fmfchar*}(15,15)
  \fmfleft{in}
  \fmfright{out1}
\fmf{phantom}{in,v1}
\fmf{phantom}{out,v2}
\fmfforce{(0,0h)}{in}
\fmfforce{(w,0h)}{out}
\fmffixed{(0,0.75h)}{v2,v3}
\fmfpoly{phantom}{v1,v2,v3}
\fmfpoly{phantom}{v3,v2,v4}
\fmf{plain}{v1,v2}
\fmf{plain}{v1,v3}
\fmf{plain}{v2,v3}
\fmf{plain}{v3,v4}
\fmf{plain}{v2,v4}
\fmf{plain,right=1}{v1,v4}
\end{fmfchar*}}}
=I_{2\mathbf{t}}G(5-D,1)
\col\quad
&\mathcal{I}_{3\mathbf{t}}&=\frac{1}{(4\pi)^6}\frac{1}{\varepsilon}2\zeta(3)
\col\\
I_{3\mathbf{b}}&=
\settoheight{\eqoff}{$\times$}%
\setlength{\eqoff}{0.5\eqoff}%
\addtolength{\eqoff}{-6.75\unitlength}%
\raisebox{\eqoff}{%
\fmfframe(-0.5,-5.5)(9,4){%
\begin{fmfchar*}(15,15)
  \fmfleft{in}
  \fmfright{out1}
\fmf{phantom}{in,v1}
\fmf{phantom}{out,v2}
\fmfforce{(0,0h)}{in}
\fmfforce{(w,0h)}{out}
\fmffixed{(0,0.75h)}{v2,v3}
\fmfpoly{phantom}{v1,v2,v3}
\fmfpoly{phantom}{v3,v2,v4}
\fmf{plain}{v1,v2}
\fmf{plain}{v1,v3}
\fmf{plain}{v2,v4}
\fmf{plain}{v3,v4}
\fmf{plain,right=0.25}{v2,v3}
\fmf{plain,left=0.25}{v2,v3}
\end{fmfchar*}}}
\col\quad
&\mathcal{I}_{3\mathbf{b}}&=\frac{1}{(4\pi)^6}
\Big(\frac{1}{3\varepsilon^3}-\frac{2}{3\varepsilon^2}+\frac{1}{3\varepsilon}\Big)\col
&\\
I_{3\mathbf{bb}}&=
\settoheight{\eqoff}{$\times$}%
\setlength{\eqoff}{0.5\eqoff}%
\addtolength{\eqoff}{-6.75\unitlength}%
\raisebox{\eqoff}{%
\fmfframe(-0.5,-5.5)(9,4){%
\begin{fmfchar*}(15,15)
  \fmfleft{in}
  \fmfright{out1}
\fmf{phantom}{in,v1}
\fmf{phantom}{out,v2}
\fmfforce{(0,0h)}{in}
\fmfforce{(w,0h)}{out}
\fmffixed{(0,0.75h)}{v2,v3}
\fmfpoly{phantom}{v1,v2,v3}
\fmfpoly{phantom}{v3,v2,v4}
\fmf{plain,right=0.25}{v1,v2}
\fmf{plain,left=0.25}{v1,v2}
\fmf{plain}{v1,v3}
\fmf{plain}{v3,v4}
\fmf{plain,right=0.25}{v2,v4}
\fmf{plain,left=0.25}{v2,v4}
\end{fmfchar*}}}
=G(1,1)^2G(3-\tfrac{D}{2},3-\tfrac{D}{2})
\col\quad
&\mathcal{I}_{3\mathbf{bb}}&=\frac{1}{(4\pi)^6}
\Big(\frac{1}{3\varepsilon^3}-\frac{1}{3\varepsilon^2}-\frac{1}{3\varepsilon}\Big)
\col\\
\end{aligned}\\
\begin{aligned}
\hphantom{I_{3\mathbf{bb}}}&\hfill\\[-\baselineskip]
I_{32\mathbf{t}}&=
\settoheight{\eqoff}{$\times$}%
\setlength{\eqoff}{0.5\eqoff}%
\addtolength{\eqoff}{-6.75\unitlength}%
\raisebox{\eqoff}{%
\fmfframe(-0.5,-3.5)(10.5,2){%
\begin{fmfchar*}(15,15)
  \fmfleft{in}
  \fmfright{out1}
\fmf{phantom}{in,v1}
\fmf{phantom}{out,v2}
\fmfforce{(0,0h)}{in}
\fmfforce{(w,0h)}{out}
\fmffixed{(0.75w,0)}{v1,v2}
\fmfpoly{phantom}{v1,v2,v3}
\fmfpoly{phantom}{v3,v2,v4}
\fmfpoly{phantom}{v4,v2,v5}
\fmf{plain}{v1,v2}
\fmf{derplain}{v3,v1}
\fmf{plain}{v2,v3}
\fmf{plain}{v2,v4}
\fmf{plain}{v3,v4}
\fmf{plain}{v2,v5}
\fmf{derplain}{v4,v5}
\end{fmfchar*}}}
=G_1(2,1)G_1(4-\tfrac{D}{2},1)G_2(6-\tfrac{D}{2},1)
\col\qquad
\mathcal{I}_{32\mathbf{t}}=\frac{1}{(4\pi)^6}
\Big(-\frac{1}{3\varepsilon}\Big)
\col\qquad\quad\!\!\!\!
\end{aligned}
\end{gathered}
\end{equation}
where $\mathcal{I}=\Kop\Rop(I)$ denotes the pole part 
of the respective integral $I$.
It is extracted by $\Kop$ after the subdivergences have been removed 
by the operation $\Rop$. 
The integral $I_{2\mathbf{t}}$ that appears as substructure in 
$I_{3\mathbf{t}}$ and in the final expression for the two-loop chiral self 
energy \eqref{ctwoloopse} is finite and given by
\begin{equation}
\begin{aligned}\label{I2t}
I_{2\mathbf{t}}=
\settoheight{\eqoff}{$\times$}%
\setlength{\eqoff}{0.5\eqoff}%
\addtolength{\eqoff}{-6.75\unitlength}%
\raisebox{\eqoff}{%
\fmfframe(-0.5,-5.5)(9,4){%
\begin{fmfchar*}(15,15)
  \fmfleft{in}
  \fmfright{out1}
\fmf{phantom}{in,v1}
\fmf{phantom}{out,v2}
\fmfforce{(0,0h)}{in}
\fmfforce{(w,0h)}{out}
\fmffixed{(0,0.75h)}{v2,v3}
\fmfpoly{phantom}{v1,v2,v3}
\fmfpoly{phantom}{v3,v2,v4}
\fmf{plain}{v1,v2}
\fmf{plain}{v1,v3}
\fmf{plain}{v2,v3}
\fmf{plain}{v3,v4}
\fmf{plain}{v2,v4}
\end{fmfchar*}}}
=\frac{2}{D-4}G(1,1)(G(1,2)+G(3-\tfrac{D}{2},2))
=\frac{1}{(4\pi)^4}6\zeta(3)+\mathcal{O}(\epsilon)\pnt
\end{aligned}
\end{equation}

\section{Similarity transformations}
\label{app:strafo}

The representation of the dilatation operator is not unique, but 
it may be transformed by a change of the basis of operators that 
does not alter its eigenvalues. In this appendix, we work out the most general 
transformation that preserves the structural constraints coming from 
the underlying Feynman graphs. 
We include non-unitary transformations that allow us to remove the 
anti-Hermitean contributions in the three-loop dilatation 
operator \eqref{D3}.
 
The similarity transformations can be realized as
\begin{equation}\label{strafo}
\mathcal{D}'=\e^{-\chi}\mathcal{D}\e^{\chi}=\mathcal{D}+\delta\mathcal{D}
\col
\end{equation}
where $\chi$ is a linear combination of flavor operations.
We demand that the transformation preserves the structural constraints
coming from the underlying Feynman diagrams, i.e.\
the transformation must not increase the maximum range and the maximum power in 
$\rho$ and $\hat\rho$ found in the dilatation operator 
at a given order in $\bar g$. 
This is guaranteed if the weak coupling expansion of $\chi$ only contains 
those flavor operations that can be associated with Feynman diagrams at the 
considered order. It does not matter whether 
these Feynman diagrams have an overall UV divergence or are finite.
In addition to the chiral functions we therefore also have to consider 
operators in flavor space that are generated by the finite Feynman diagrams 
involving gauge interactions only. Their elementary building block is given
by
\begin{equation}\label{gaugeid}
\begin{aligned}
\settoheight{\eqoff}{$\times$}%
\setlength{\eqoff}{0.5\eqoff}%
\addtolength{\eqoff}{-12\unitlength}%
\raisebox{\eqoff}{%
\fmfframe(-1.5,2)(-12,2){%
\begin{fmfchar*}(20,20)
\idonei[phantom]{dots}{dots}{photon}{dots}{dots}
\fmfiv{label=$\scriptstyle i$,l.a=-90,l.dist=4}{vloc(__v3)}
\fmfiv{label=$\scriptstyle j$,l.a=-90,l.dist=4}{vloc(__v4)}
\end{fmfchar*}}}
=(\Lam_t)_{ij}
\col\qquad
(\Lam_t)_{ij}=
\begin{cases}
\rho\unitmatrix & (\Phi,*)\cup(*,\Phi)\cup(\tilde Q^{\dot I},Q_{\dot J}) \\
\hat\rho \unitmatrix & (\hat\Phi,*)\cup(*,\hat\Phi)\cup(Q_{\dot I},\tilde Q^{\dot J})\\
\end{cases}
\col
\end{aligned}
\end{equation}
where the subscript $t$ of $\Lambda_t$ is there to remind us that the 
underlying diagram is a $t$-channel when regarding the fields at 
positions $(i,j)$ as incoming.
In analogy to the chiral functions \eqref{chifuncdef}, we 
introduce the operators
\begin{equation}
\Lam_t(a_1,\dots,a_n)=\sum_{r=0}^{L-1}\prod_{i=1}^n(\Lam_t)_{r+a_i\,r+a_i+1}
\col\qquad a_1\le\dots\le a_n
\end{equation}
that are generated by Feynman diagrams involving $n$ gauge fields.
The identity operation is given by $\Lam_t()=\chi()$.
Note that we can impose the order $a_1\le\dots\le a_n$ on the list of arguments,
since the operators $(\Lam_t)_{ij}$, $(\Lam_t)_{kl}$ commute for any
$i$, $j$, $k$, $l$. 
If $\hat\rho=\rho$ these operators reduce to the identity and
cause no effect in the transformation \eqref{strafo}. This is the reason why 
in $\mathcal{N}=4$ SYM theory they need not be considered when constructing 
the similarity transformations.

Based on the aforementioned considerations, 
the most general ansatz for $\chi$ that leads to similarity transformations
up to three loops is given by
\begin{equation}\label{stansatz}
\begin{aligned}
\chi&=\bar g^2\big(\delta_1\chiop(1)
+\delta_{11}\Lam_t(1)\big)
+\bar g^4\big(
\delta_{21}\chiimp{1}{1}{1}
+\delta_{22}\chiimp{1}{2}{1}
+\delta_{23}\chiimp{2}{1}{1}
+\delta_{24}\chiimp{2}{2}{1}\\
&\phantom{{}={}\bar g^2\big(\delta_1\chiop(1)
+\delta_{11}\Lam_t(1)\big)+\bar g^4\big(}
+\hat\delta_{21}\chiimp[\tilde]{1}{1}{1}
+\hat\delta_{22}\chiimp[\tilde]{1}{2}{1}
+\hat{\delta}_{23}\chiimp[\tilde]{2}{1}{1}
+\hat\delta_{24}\chiimp[\tilde]{2}{2}{1}\\
&\phantom{{}={}\bar g^2\big(\delta_1\chiop(1)
+\delta_{11}\Lam_t(1)\big)+\bar g^4\big(}
+\delta_{25}\chi(1,2)
+\delta_{26}\chi(2,1)
+\delta_{27}\Lam_t(1)
\big)
\pnt
\end{aligned}
\end{equation}
Chiral functions that involve the flavor trace operator 
are not included, since they do not contribute
in the closed chiral subsector we are interested in. 
Furthermore, it is not necessary to consider at two loops
the operators $\Lam_t(1,2)$ and combinations of $(\Lam_t)_{ij}$ with the 
flavor operations of the chiral functions. They act identical to the 
operators already present in the above ansatz apart from additional 
dependences on $\rho$ and $\hat\rho$ that appear as prefactors.
Their contributions can be absorbed into the 
coefficients $\delta_{2i}$, $\hat\delta_{2i}$, $i=1,\dots,4$ 
and $\delta_{27}$ by allowing them to be 
complex linear functions of 
$\rho$, and $\hat\rho$. 
They should then fulfill
\begin{equation}\label{deltacond}
\begin{aligned}
\hat\delta_{2i}(\rho,\hat\rho)&=\delta_{2i}(\hat\rho,\rho)
\col\qquad i=1,\dots,4
\col\qquad\\
\delta_{27}(\rho,\hat\rho)&=\delta_{27}(\hat\rho,\rho)
\col
\end{aligned}
\end{equation}
while the remaining coefficients
$\delta_1$, $\delta_{11}$, $\delta_{25}$, $\delta_{26}$ are complex constants.

The one-loop contribution in \eqref{stansatz} coming with $\chi(1)$
is essentially the one-loop dilatation operator $\mathcal{D}_1$ itself as given 
in \eqref{D1D2}. It therefore commutes with $\mathcal{D}_1$ and
does not generate a transformation of the two-loop dilatation operator 
$\mathcal{D}_2$. However, the one-loop term
involving $\Lam_t(1)$ leads to a transformation of $\mathcal{D}_2$. 
We do not want this to happen, since $\mathcal{D}_1$ and also 
$\mathcal{D}_2$ are already Hermitean and in a minimal form.
Hence, from now on we set $\delta_{11}=0$.
The three-loop dilatation operator $\mathcal{D}_3$ is then modified by the 
following terms
\begin{equation}\label{deltaD3ansatz}
\footnotesize
\begin{aligned}
\delta\mathcal{D}_3&=
-\comm{\chi_1}{\mathcal{D}_2}-\comm{\chi_2}{\mathcal{D}_1}\\
&=2\big(
\epsilon_{2}\comm{\chiop(1)}{\chiop(1,2)}
+\epsilon_{\bar 2}\comm{\chiop(1)}{\chiop(2,1)}\\
&\phantom{{}={}2\big(}
+\epsilon_{2c}\comm{\smash[t]{\chiimp{1}{1}{1}}}{\smash[t]{\chiimp{1}{2}{1}}}
+(\epsilon_{2b}+\epsilon_{2c})\comm{\smash[t]{\chiimp{1}{1}{1}}}{\smash[t]{\chiimp{2}{1}{1}}}
+\epsilon_{2a}\comm{\smash[t]{\chiimp{1}{1}{1}}}{\smash[t]{\chiimp{2}{2}{1}}}\\
&\phantom{{}={}2\big(}
+\epsilon_{2b}\comm{\smash[t]{\chiimp{1}{2}{1}}}{\smash[t]{\chiimp{2}{1}{1}}}
+(\epsilon_{2a}-\epsilon_{2c})\comm{\smash[t]{\chiimp{1}{2}{1}}}{\smash[t]{\chiimp{2}{2}{1}}}
+(\epsilon_{2a}-\epsilon_{2b}-\epsilon_{2c})\comm{\smash[t]{\chiimp{2}{1}{1}}}{\smash[t]{\chiimp{2}{2}{1}}}\\
&\phantom{{}={}2\big(}
+\hat\epsilon_{2c}\comm{\smash[t]{\chiimp[\tilde]{1}{1}{1}}}{\smash[t]{\chiimp[\tilde]{1}{2}{1}}}
+(\hat\epsilon_{2b}+\hat\epsilon_{2c})\comm{\smash[t]{\chiimp[\tilde]{1}{1}{1}}}{\smash[t]{\chiimp[\tilde]{2}{1}{1}}}
+\hat\epsilon_{2a}\comm{\smash[t]{\chiimp[\tilde]{1}{1}{1}}}{\smash[t]{\chiimp[\tilde]{2}{2}{1}}}\\
&\phantom{{}={}2\big(}
+\hat\epsilon_{2b}\comm{\smash[t]{\chiimp[\tilde]{1}{2}{1}}}{\smash[t]{\chiimp[\tilde]{2}{1}{1}}}
+(\hat\epsilon_{2a}-\hat\epsilon_{2c})\comm{\smash[t]{\chiimp[\tilde]{1}{2}{1}}}{\smash[t]{\chiimp[\tilde]{2}{2}{1}}}
+(\hat\epsilon_{2a}-\hat\epsilon_{2b}-\hat\epsilon_{2c})\comm{\smash[t]{\chiimp[\tilde]{2}{1}{1}}}{\smash[t]{\chiimp[\tilde]{2}{2}{1}}}\\
&\phantom{{}={}2\big(}
+\mu\comm{\smash[t]{\chiimp{1}{1}{1}}}{\smash[t]{\chiimp[\tilde]{1}{1}{1}}}
+(\mu+\hat\epsilon_{2c})\comm{\smash[t]{\chiimp{1}{1}{1}}}{\smash[t]{\chiimp[\tilde]{1}{2}{1}}}
+(\mu+\hat\epsilon_{2b}+\hat\epsilon_{2c})\comm{\smash[t]{\chiimp{1}{1}{1}}}{\smash[t]{\chiimp[\tilde]{2}{1}{1}}}\\
&\phantom{{}={}2\big(}
+(\mu+\hat\epsilon_{2a})\comm{\smash[t]{\chiimp{1}{1}{1}}}{\smash[t]{\chiimp[\tilde]{2}{2}{1}}}
+(\mu-\epsilon_{2c})\comm{\smash[t]{\chiimp{1}{2}{1}}}{\smash[t]{\chiimp[\tilde]{1}{1}{1}}}
+(\mu-\epsilon_{2c}+\hat\epsilon_{2c})\comm{\smash[t]{\chiimp{1}{2}{1}}}{\smash[t]{\chiimp[\tilde]{1}{2}{1}}}\\
&\phantom{{}={}2\big(}
+(\mu-\epsilon_{2c}+\hat\epsilon_{2b}+\hat\epsilon_{2c})\comm{\smash[t]{\chiimp{1}{2}{1}}}{\smash[t]{\chiimp[\tilde]{2}{1}{1}}}
+(\mu-\epsilon_{2c}+\hat\epsilon_{2a})\comm{\smash[t]{\chiimp{1}{2}{1}}}{\smash[t]{\chiimp[\tilde]{2}{2}{1}}}\\
&\phantom{{}={}2\big(}
+(\mu-\epsilon_{2b}-\epsilon_{2c})\comm{\smash[t]{\chiimp{2}{1}{1}}}{\smash[t]{\chiimp[\tilde]{1}{1}{1}}}
+(\mu-\epsilon_{2b}-\epsilon_{2c}+\hat\epsilon_{2c})\comm{\smash[t]{\chiimp{2}{1}{1}}}{\smash[t]{\chiimp[\tilde]{1}{2}{1}}}\\
&\phantom{{}={}2\big(}
+(\mu-\epsilon_{2b}-\epsilon_{2c}+\hat\epsilon_{2b}+\hat\epsilon_{2c})\comm{\smash[t]{\chiimp{2}{1}{1}}}{\smash[t]{\chiimp[\tilde]{2}{1}{1}}}
+(\mu-\epsilon_{2b}-\epsilon_{2c}+\hat\epsilon_{2a})\comm{\smash[t]{\chiimp{2}{1}{1}}}{\smash[t]{\chiimp[\tilde]{2}{2}{1}}}\\
&\phantom{{}={}2\big(}
+(\mu-\epsilon_{2a})\comm{\smash[t]{\chiimp{2}{2}{1}}}{\smash[t]{\chiimp[\tilde]{1}{1}{1}}}
+(\mu-\epsilon_{2a}+\hat\epsilon_{2c})\comm{\smash[t]{\chiimp{2}{2}{1}}}{\smash[t]{\chiimp[\tilde]{1}{2}{1}}}\\
&\phantom{{}={}2\big(}
+(\mu-\epsilon_{2a}+\hat\epsilon_{2b}+\hat\epsilon_{2c})\comm{\smash[t]{\chiimp{2}{2}{1}}}{\smash[t]{\chiimp[\tilde]{2}{1}{1}}}
+(\mu-\epsilon_{2a}+\hat\epsilon_{2a})\comm{\smash[t]{\chiimp{2}{2}{1}}}{\smash[t]{\chiimp[\tilde]{2}{2}{1}}}\\
&\phantom{{}={}2\big(}
+\nu\comm{\Lam_t(1)}{\chi(1)}
\big)
\col
\end{aligned}
\end{equation}
where $\chi_1$, $\chi_2$ are the respective one- and two-loop contributions
in \eqref{stansatz}, and we have introduced the new parameters
\begin{equation}\label{epsilons}
\begin{aligned}
\epsilon_{2}&=\delta_1-\delta_{25}\col\\
\epsilon_{\bar 2}&=\delta_1-\delta_{26}\col\\
\end{aligned}
\qquad
\begin{aligned}
\epsilon_{2a}&=\delta_{21}-\delta_{24}\col\qquad
&\hat\epsilon_{2a}&=\hat\delta_{21}-\hat\delta_{24}\col\\
\epsilon_{2b}&=\delta_{22}-\delta_{23}\col\qquad
&\hat\epsilon_{2b}&=\hat\delta_{22}-\hat\delta_{23}\col\\
\epsilon_{2c}&=\delta_{21}-\delta_{22}\col\qquad
&\hat\epsilon_{2c}&=\hat\delta_{21}-\hat\delta_{22}\col\\
\end{aligned}
\qquad
\begin{aligned}
\mu&=\delta_{21}-\hat\delta_{21}\col\\
\nu&=\delta_{27}
\pnt
\end{aligned}
\end{equation}
The coefficient $\delta_1$ is absorbed into 
the redefinitions of $\delta_{25}$ and $\delta_{26}$. 
We can hence set $\delta_1=0$ such that 
$\delta\mathcal{D}_3$ is entirely generated by the commutator 
of $\chi_2$ with $\mathcal{D}_1$.
The individual commutators of chiral functions 
that appear in \eqref{deltaD3ansatz} are evaluated to
\begin{equation}
\footnotesize
\begin{aligned}
\comm{\chiop(1)}{\chiop(1,2)}
&=\chiop(2,1,2)+\chiop(1,3,2)-\chiop(2,1,3)-\chiop(1,2,1)\col\\
\comm{\chiop(1)}{\chiop(2,1)}
&=\chiop(1,3,2)+\chiop(1,2,1)-\chiop(2,1,2)-\chiop(2,1,3)
\pnt
\end{aligned}
\end{equation}
The commutators of chiral functions with fixed incoming and outgoing flavor
arrangements are given by
\begin{equation}
\footnotesize
\begin{aligned}
\comm{\smash[t]{\chiimp{1}{1}{1}}}{\smash[t]{\chiimp{1}{2}{1}}}
&=\chiimp{1}{2}{2,1}+\hat\rho\chiimp{1}{2}{1}\col\\
\comm{\smash[t]{\chiimp{1}{1}{1}}}{\smash[t]{\chiimp{2}{1}{1}}}
&=-\hat\rho\chiimp{2}{1}{1}-\chiimp{2}{1}{1,2}\col\\
\comm{\smash[t]{\chiimp{1}{1}{1}}}{\smash[t]{\chiimp{2}{2}{1}}}
&=\chiimp{2}{2}{2,1}-\chiimp{2}{2}{1,2}\col\\
\comm{\smash[t]{\chiimp{1}{2}{1}}}{\smash[t]{\chiimp{2}{1}{1}}}
&=\rho\chiimp{1}{1}{1}-\hat\rho\chiimp{2}{2}{1}\col\\
\comm{\smash[t]{\chiimp{1}{2}{1}}}{\smash[t]{\chiimp{2}{2}{1}}}
&=\chiimp{2}{3}{2,1}+\rho\chiimp{1}{2}{1}\col\\
\comm{\smash[t]{\chiimp{2}{1}{1}}}{\smash[t]{\chiimp{2}{2}{1}}}
&=-\rho\chiimp{2}{1}{1}-\chiimp{3}{2}{1,2}\col\\
\end{aligned}
\qquad
\begin{aligned}
\comm{\smash[t]{\chiimp[\tilde]{1}{1}{1}}}{\smash[t]{\chiimp[\tilde]{1}{2}{1}}}
&=\chiimp[\tilde]{1}{2}{2,1}+\rho\chiimp[\tilde]{1}{2}{1}\col\\
\comm{\smash[t]{\chiimp[\tilde]{1}{1}{1}}}{\smash[t]{\chiimp[\tilde]{2}{1}{1}}}
&=-\rho\chiimp[\tilde]{2}{1}{1}-\chiimp[\tilde]{2}{1}{1,2}\col\\
\comm{\smash[t]{\chiimp[\tilde]{1}{1}{1}}}{\smash[t]{\chiimp[\tilde]{2}{2}{1}}}
&=\chiimp[\tilde]{2}{2}{2,1}-\chiimp[\tilde]{2}{2}{1,2}\col\\
\comm{\smash[t]{\chiimp[\tilde]{1}{2}{1}}}{\smash[t]{\chiimp[\tilde]{2}{1}{1}}}
&=\hat\rho\chiimp[\tilde]{1}{1}{1}-\rho\chiimp[\tilde]{2}{2}{1}\col\\
\comm{\smash[t]{\chiimp[\tilde]{1}{2}{1}}}{\smash[t]{\chiimp[\tilde]{2}{2}{1}}}
&=\chiimp[\tilde]{2}{3}{2,1}+\hat\rho\chiimp[\tilde]{1}{2}{1}\col\\
\comm{\smash[t]{\chiimp[\tilde]{2}{1}{1}}}{\smash[t]{\chiimp[\tilde]{2}{2}{1}}}
&=-\hat\rho\chiimp[\tilde]{2}{1}{1}-\chiimp[\tilde]{3}{2}{1,2}\col\\
\end{aligned}
\end{equation}
and by
\begin{equation}
\footnotesize
\begin{aligned}
\comm{\smash[t]{\chiimp{1}{1}{1}}}{\smash[t]{\chiimp[\tilde]{1}{1}{1}}}
&=0\col\\
\comm{\smash[t]{\chiimp{1}{1}{1}}}{\smash[t]{\chiimp[\tilde]{1}{2}{1}}}
&=\chiimp{1,2}{1,3}{1,2}\col\\
\comm{\smash[t]{\chiimp{1}{1}{1}}}{\smash[t]{\chiimp[\tilde]{2}{1}{1}}}
&=-\chiimp{1,3}{1,2}{2,1}\col\\
\comm{\smash[t]{\chiimp{1}{1}{1}}}{\smash[t]{\chiimp[\tilde]{2}{2}{1}}}
&=\chiimp{1,3}{1,3}{1,2}-\chiimp{1,3}{1,3}{2,1}\col\\
\comm{\smash[t]{\chiimp{1}{2}{1}}}{\smash[t]{\chiimp[\tilde]{1}{1}{1}}}
&=-\chiimp[\tilde]{1,2}{1,3}{1,2}\col\\
\comm{\smash[t]{\chiimp{1}{2}{1}}}{\smash[t]{\chiimp[\tilde]{1}{2}{1}}}
&=\chiimp{1,2}{2,3}{1,2}-\chiimp[\tilde]{1,2}{2,3}{1,2}\col\\
\comm{\smash[t]{\chiimp{1}{2}{1}}}{\smash[t]{\chiimp[\tilde]{2}{1}{1}}}
&=0\col\\
\comm{\smash[t]{\chiimp{1}{2}{1}}}{\smash[t]{\chiimp[\tilde]{2}{2}{1}}}
&=\chiimp{1,3}{2,3}{1,2}\col\\
\end{aligned}
\qquad%
\begin{aligned}
\comm{\smash[t]{\chiimp{2}{1}{1}}}{\smash[t]{\chiimp[\tilde]{1}{1}{1}}}
&=\chiimp[\tilde]{1,3}{1,2}{2,1}\col\\
\comm{\smash[t]{\chiimp{2}{1}{1}}}{\smash[t]{\chiimp[\tilde]{1}{2}{1}}}
&=0\col\\
\comm{\smash[t]{\chiimp{2}{1}{1}}}{\smash[t]{\chiimp[\tilde]{2}{1}{1}}}
&=\chiimp{2,3}{1,2}{2,1}-\chiimp[\tilde]{2,3}{1,2}{2,1}\col\\
\comm{\smash[t]{\chiimp{2}{1}{1}}}{\smash[t]{\chiimp[\tilde]{2}{2}{1}}}
&=-\chiimp{2,3}{1,3}{2,1}\col\\
\comm{\smash[t]{\chiimp{2}{2}{1}}}{\smash[t]{\chiimp[\tilde]{1}{1}{1}}}
&=\chiimp[\tilde]{1,3}{1,3}{2,1}-\chiimp[\tilde]{1,3}{1,3}{1,2}\col\\
\comm{\smash[t]{\chiimp{2}{2}{1}}}{\smash[t]{\chiimp[\tilde]{1}{2}{1}}}
&=-\chiimp[\tilde]{1,3}{2,3}{1,2}\col\\
\comm{\smash[t]{\chiimp{2}{2}{1}}}{\smash[t]{\chiimp[\tilde]{2}{1}{1}}}
&=\chiimp[\tilde]{2,3}{1,3}{2,1}\col\\
\comm{\smash[t]{\chiimp{2}{2}{1}}}{\smash[t]{\chiimp[\tilde]{2}{2}{1}}}
&=0
\pnt
\end{aligned}
\end{equation}
Furthermore, $\Lambda_t(1)$ does not alter the positions of the 
impurities but only produces a factor of either $\rho$ or $\hat\rho$. 
Thus, its commutators with 
the chiral functions with fixed positions of the incoming and outgoing 
impurities read
\begin{equation}
\begin{aligned}
\comm{\Lam_t(1)}{\chiimp{i_1\dots,i_I}{o_1,\dots,o_I}{a_1,\dots,a_n}}
&=(\rho-\hat\rho)\sum_{k=1}^I(-1)^{k}(i_k-o_k)\chiimp{i_1\dots,i_I}{o_1,\dots,o_I}{a_1,\dots,a_n}\col\\
\comm{\Lam_t(1)}{\chiimp[\tilde]{i_1\dots,i_I}{o_1,\dots,o_I}{a_1,\dots,a_n}}
&=-(\rho-\hat\rho)\sum_{k=1}^I(-1)^{k}(i_k-o_k)\chiimp[\tilde]{i_1\dots,i_I}{o_1,\dots,o_I}{a_1,\dots,a_n}\pnt
\end{aligned}
\end{equation}

Inserting the above expressions for the commutators into 
\eqref{deltaD3ansatz} and adding the result to the three-loop dilatation 
operator $\mathcal{D}_3$ given in \eqref{D3}, we obtain the transformed  
expression
\begin{equation}
\footnotesize
\begin{aligned}
\mathcal{D}'_3
&=
-4(\chiop(1,2,3)+\chi(3,2,1))
+8(\rho+\hat\rho)(\chiop(1,2)+\chiop(2,1))\\
&\phantom{{}={}}
-2(2+\epsilon_2-\epsilon_{\bar 2})\chi(1,2,1)
-2(2-\epsilon_2+\epsilon_{\bar 2})\chi(2,1,2)
-2(2(\rho+\hat\rho)^2-(\rho-\hat\rho)^2\zeta(3))\chiop(1)\\
&\phantom{{}={}}
-2(1-\epsilon_2-\epsilon_{\bar 2})(\chiop(1,3,2)-\chiop(2,1,3))
-4(\hat\rho\chiimp{*,*}{*,*}(1,3)+\rho\chiimp[\tilde]{*,*}{*,*}(1,3))\\
&\phantom{{}={}}
+2(\rho\epsilon_{2b}-(1-\hat\rho^2)\zeta(3))\chiimp{1}{1}{1}
+2(\hat\rho\hat\epsilon_{2b}-(1-\rho^2)\zeta(3))\chiimp[\tilde]{1}{1}{1}\\
&\phantom{{}={}}
+2(\rho\epsilon_{2a}-(\rho-\hat\rho)(\epsilon_{2c}-\nu)-(1-\hat\rho^2)\zeta(3))\chiimp{1}{2}{1}\\
&\phantom{{}={}}
+2(\hat\rho\hat\epsilon_{2a}+(\rho-\hat\rho)(\hat\epsilon_{2c}-\nu)-(1-\rho^2)\zeta(3))\chiimp[\tilde]{1}{2}{1}\\
&\phantom{{}={}}
-2(\rho\epsilon_{2a}-(\rho-\hat\rho)(\epsilon_{2b}+\epsilon_{2c}-\nu)+(1-\rho^2)\zeta(3))\chiimp{2}{1}{1}\\
&\phantom{{}={}}
-2(\hat\rho\hat\epsilon_{2a}+(\rho-\hat\rho)(\hat\epsilon_{2b}+\hat\epsilon_{2c}-\nu)+(1-\hat\rho^2)\zeta(3))\chiimp[\tilde]{2}{1}{1}
\\
&\phantom{{}={}}
-2(\hat\rho\epsilon_{2b}+(1-\rho^2)\zeta(3))\chiimp{2}{2}{1}
-2(\rho\hat\epsilon_{2b}+(1-\hat\rho^2)\zeta(3))\chiimp[\tilde]{2}{2}{1}\\
&\phantom{{}={}}
-2(\epsilon_{2b}+\epsilon_{2c}+(\rho-\hat\rho))\chiimp{2}{1}{1,2}
-2(\hat\epsilon_{2b}+\hat\epsilon_{2c}-(\rho-\hat\rho))\chiimp[\tilde]{2}{1}{1,2}\\
&\phantom{{}={}}
+2\epsilon_{2c}\chiimp{2}{1}{2,1}+2\hat\epsilon_{2c}\chiimp[\tilde]{2}{1}{2,1}\\
&\phantom{{}={}}
-2(\epsilon_{2a}+(\rho-\hat\rho))(\chiimp{2}{2}{1,2}-\chiimp{2}{2}{2,1})
-2(\hat\epsilon_{2a}-(\rho-\hat\rho))(\chiimp[\tilde]{2}{2}{1,2}-\chiimp[\tilde]{2}{2}{2,1})\\
&\phantom{{}={}}
-2(\epsilon_{2a}-\epsilon_{2b}-\epsilon_{2c})\chiimp{3}{2}{1,2}
-2(\hat\epsilon_{2a}-\hat\epsilon_{2b}-\hat\epsilon_{2c})\chiimp[\tilde]{3}{2}{1,2}\\
&\phantom{{}={}}
+2(\epsilon_{2a}-\epsilon_{2c}+(\rho-\hat\rho)\zeta(3))\chiimp{2}{3}{2,1}
+2(\hat\epsilon_{2a}-\hat\epsilon_{2c}-(\rho-\hat\rho)\zeta(3))\chiimp[\tilde]{2}{3}{2,1}\\
&\phantom{{}={}}
+2(\mu+\hat\epsilon_{2c}+(\rho-\hat\rho)\zeta(3))\chiimp{1,2}{1,3}{1,2}
-2(\mu-\epsilon_{2c}+(\rho-\hat\rho)\zeta(3))\chiimp[\tilde]{1,2}{1,3}{1,2}\\
&\phantom{{}={}}
-2(\mu+\hat\epsilon_{2b}+\epsilon_{2c})\chiimp{1,3}{1,2}{2,1}
+2(\mu-\epsilon_{2b}-\epsilon_{2c})\chiimp[\tilde]{1,3}{1,2}{2,1}\\
&\phantom{{}={}}
+2(\mu-\epsilon_{2c}+\hat\epsilon_{2c}+(\rho-\hat\rho)\zeta(3))(\chiimp{1,2}{2,3}{1,2}-\chiimp[\tilde]{1,2}{2,3}{1,2})\\
&\phantom{{}={}}
-2(\mu-\epsilon_{2b}-\epsilon_{2c}+\hat\epsilon_{2b}+\hat\epsilon_{2c}+(\rho-\hat\rho)\zeta(3))(\chiimp{2,3}{1,2}{2,1}-\chiimp[\tilde]{2,3}{1,2}{2,1})\\
&\phantom{{}={}}
+2(\mu+\hat\epsilon_{2a})(\chiimp{1,3}{1,3}{1,2}-\chiimp{1,3}{1,3}{2,1})
-2(\mu-\epsilon_{2a})(\chiimp[\tilde]{1,3}{1,3}{1,2}-\chiimp[\tilde]{1,3}{1,3}{2,1})\\
&\phantom{{}={}}
+2(\mu-\epsilon_{2c}+\hat\epsilon_{2a})\chiimp{1,3}{2,3}{1,2}
-2(\mu-\epsilon_{2a}+\hat\epsilon_{2c})\chiimp[\tilde]{1,3}{2,3}{1,2}\\
&\phantom{{}={}}
-2(\mu-\epsilon_{2b}-\epsilon_{2c}-\hat\epsilon_{2a}-(\rho-\hat\rho)\zeta(3))\chiimp{2,3}{1,3}{2,1}\\
&\phantom{{}={}}
+2(\mu-\epsilon_{2a}+\hat\epsilon_{2b}+\hat\epsilon_{2c}-(\rho-\hat\rho)\zeta(3))\chiimp[\tilde]{2,3}{1,3}{2,1}
\end{aligned}
\end{equation}

A particular choice of the parameters of the transformation
then allows us to simplify the expression of $\mathcal{D}_3$ 
as obtained from Feynman diagrams and given in \eqref{D3}. 
The anti-Hermitean terms can be completely removed
with the following choice of the parameters
\begin{equation}
\begin{aligned}
\re\epsilon_2=\re\epsilon_{\bar 2}=1
\end{aligned}
\end{equation}
in analogy to the case of $\mathcal{N}=4$ SYM theory and moreover with
\begin{equation}
\begin{aligned}
\re\epsilon_{2a}&=\re\mu\col\quad
&\re\hat\epsilon_{2a}&=-\re\mu\col\\
\re\epsilon_{2b}&=-2\re\epsilon_{2c}+\re\mu\col\quad
&\re\hat\epsilon_{2b}&=-2\re\hat\epsilon_{2c}-\re\mu\col\\
\im\epsilon_{2b}&=0\col\quad
&\im\hat\epsilon_{2b}&=0\col
\end{aligned}
\qquad
\begin{aligned}
\re\mu&=-(\rho-\hat\rho)\zeta(3)\col\\
\re\nu&=(\rho+\hat\rho)\zeta(3)
\pnt
\end{aligned}
\end{equation}
Since $\hat\mu=-\mu$, $\hat\nu=\nu$  according to \eqref{epsilons}, 
the above choices respect 
the constraints \eqref{deltacond}. With these choices and 
furthermore by setting
\begin{equation}
\begin{aligned}
\im\epsilon_2=\im\epsilon_{\bar 2}=0
\col
\end{aligned}
\qquad
\begin{aligned}
\im\epsilon_{2a}&=0\col\quad
&\im\hat\epsilon_{2a}&=0\col\\
\re\epsilon_{2c}&=0\col\quad
&\re\hat\epsilon_{2c}&=0\col\\
\re\epsilon_{2c}&=0\col\quad
&\re\hat\epsilon_{2c}&=0\col
\end{aligned}
\qquad
\begin{aligned}
\im\mu&=0\col\\
\im\nu&=0\col
\end{aligned}
\end{equation}
one obtains the simplified and Hermitean result \eqref{D3red}
that is presented in section \ref{sec:result}.
The identified transformation explicitly reads
\begin{equation}\label{stmin}
\begin{aligned}
\chi&=
\bar g^4\big(
(\rho-\hat\rho)\zeta(3)(\chiimp{2}{1}{1}
+\chiimp{2}{2}{1}
+\chiimp[\tilde]{1}{1}{1}
+\chiimp[\tilde]{1}{2}{1})
+\delta_{21}\chi(1)
+\chi(1,2)
+\chi(2,1)\\
&\phantom{{}={}\bar g^4\big(}
+(\rho+\hat\rho)\zeta(3)\Lam_t(1)
\big)
\col
\end{aligned}
\end{equation}
where the parameter $\delta_{21}$ can be chosen at will, since 
in \eqref{deltaD3ansatz} the respective
term drops out when taking the commutator with $\mathcal{D}_1$.

Applying the transformation \eqref{stmin} to the momentum eigenstates 
\eqref{momeigenstates} and using
\eqref{chiimpphaseshifts}, \eqref{chisumphaseshifts}, we obtain
for the terms relevant to three loops
\begin{equation}\label{strafomomeigenstates}
\begin{aligned}
\psi'(p')&=\e^{-\chi}\psi(p)=[1+\bar g^4N_{2,L,p}(\rho,\hat\rho)]\psi(p+i\bar g^4(\rho^2-\hat\rho^2)\zeta(3))\col\\
\tilde\psi'(p')&=e^{-\chi}\,\tilde\psi(p)=[1+\bar g^4N_{2,L,p}(\hat\rho,\rho)]\tilde\psi(p-i\bar g^4(\rho^2-\hat\rho^2)\zeta(3))\pnt\\
\end{aligned}
\end{equation}
The momentum acquires a constant imaginary two-loop shift, and the 
normalization is corrected by
\begin{equation}
N_{2,L,p}(\rho,\hat\rho)=(\delta_{21}+\cos p)\big[4\sin^2\tfrac{p}{2}+(\rho^{\frac{1}{2}}-\hat\rho^{\frac{1}{2}})^2\big]
-(\rho-\hat\rho)\zeta(3)\big[1-\e^{ip}\hat\rho-(\rho-L\hat\rho)\big]
\pnt
\end{equation}
Note that the rational term already appears in the $\mathcal{N}=4$ case, 
where it removes the anti-Hermitean contribution $2(\chiop(2,1,3)-\chiop(1,3,2))$ in the third line of \eqref{D3}.


\unitlength=0.875mm

\end{fmffile}

\bibliographystyle{JHEP}
\bibliography{references}

\end{document}